\documentclass[twocolumn]{aastex63}
\usepackage{url, amssymb, amsmath, graphicx, epsf, fancyhdr, wrapfig, threeparttable}
\usepackage[caption=false]{subfig}

\usepackage{lineno}

\newcommand{\fermi}{{\it Fermi}}

\newcommand{\reffig}[1]{Figure~\ref{fig:#1}}
\newcommand{\reftab}[1]{Table~\ref{tab:#1}}
\newcommand{\refeq}[1]{Eq.~\ref{eq:#1}}
\newcommand{\refsec}[1]{Section~\ref{sec:#1}}
\newcommand{\refapp}[1]{Appendix~\ref{app:#1}}
\newcommand{\p}{\prime}
\newcommand{\mrm}{\mathrm}

\newcommand{\SP}{SP$_\mrm{E54}$}
\newcommand{\MP}{MP$_\mrm{E54.5}$}

\newcommand{\soft}[1]{\textsc{#1}}

\graphicspath{{./}{plots_publication/}}

\received{}
\revised{}
\accepted{}
\submitjournal{ApJ}

\shorttitle{Lepto-hadronic models for energetic GRBs}
\shortauthors{Rudolph et al.}

\begin{document}

\title{Multi-collision internal shock lepto-hadronic models for energetic GRBs}

\correspondingauthor{Annika Rudolph}
\email{annika.lena.rudolph@nbi.ku.dk}

\author[0000-0003-2040-788X]{Annika Rudolph}
\affil{Niels Bohr International Academy and DARK, Niels Bohr Institute, University of Copenhagen \\
Blegdamsvej 17, 2100, Copenhagen, Denmark}

\affil{Deutsches Elektronen-Synchrotron DESY \\
Platanenallee 6, 15738 Zeuthen, Germany}

\author[0000-0001-6640-0179]{Maria Petropoulou}
\affil{Department of Physics, National and Kapodistrian University of Athens \\
University Campus Zografos, GR 15783, Athens, Greece}

\affil{Institute of Accelerating Systems \& Applications \\
University Campus Zografos, GR 15783, Athens, Greece} 

\author[0000-0001-6536-0320]{\v Zeljka Bo\v snjak}
\affil{Faculty of Electrical Engineering and Computing , University of Zagreb \\
Unska ul. 3 , 10000 Zagreb, Croatia\\
}

\author[0000-0001-7062-0289]{Walter Winter}
\affil{Deutsches Elektronen-Synchrotron DESY \\
Platanenallee 6, 15738 Zeuthen, Germany}

\begin{abstract}
For a sub-population of energetic Gamma-Ray Bursts (GRBs), a moderate baryonic loading may suffice to 
power Ultra-High-Energy Cosmic Rays (UHECRs). Motivated by this, we study the radiative signatures of cosmic-ray protons in the prompt phase of energetic GRBs. Our framework is the internal shock model with multi-collision descriptions of the relativistic ejecta (with different emission regions along the jet), plus time-dependent calculations of photon and neutrino spectra. Our GRB prototypes are motivated by {\em Fermi}-LAT detected GRBs (including GRB~221009A) for which further, owing to the large energy flux, neutrino non-observation of single events may pose a strong limit on the baryonic loading.
We study the feedback of protons
on electromagnetic spectra in synchrotron- and inverse Compton-dominated scenarios to identify the multi-wavelength signatures, to constrain the maximally allowed baryonic loading, and to point out the differences between hadronic and inverse Compton signatures. We find that hadronic signatures appear as correlated flux increases in the optical-UV to soft X-ray and GeV to TeV gamma-ray ranges in the synchrotron scenarios, whereas they are difficult to identify in inverse Compton-dominated scenarios.
We demonstrate that baryonic loadings around 10, which satisfy the UHECR energetic requirements, do not distort the predicted photon spectra in the {\em Fermi}-GBM range and are consistent with constraints from neutrino data if the collision radii are large enough (i.e., the time variability is not too short). It therefore seems plausible that under the condition of large dissipation radii a
population of energetic GRBs can be the origin of the UHECRs. 

\end{abstract}


\section{Introduction} \label{sec:intro}
Gamma-ray bursts, as the most luminous 
extragalactic energy sources in the Universe, were proposed early on as promising candidates for the acceleration of ultra-high energy cosmic rays  
\citep[UHECRs;][]{Milgrom:1995um, Waxman:1995vg, Vietri:1995hs, Waxman:1997ti, Bhattacharjee:1999mup, Dermer:2006bb}. 
The exact mechanism for  accelerating UHECRs in the GRB outflow to rigidities up to  $1.6-3.5 \cdot 10^{18} \, \mathrm{V}$~\citep{Heinze:2019jou} has not been pinned down yet, with magnetic reconnection \citep[e.g.][]{Giannios:2010cv} and shock acceleration \citep[e.g.][]{Murase:2005hy, Dermer:2010iz} posing the two leading scenarios.
The presence of very energetic protons and nuclei in the GRB outflow 
is accompanied by the production of High-Energy (HE) neutrinos, following the interactions of cosmic rays with the prompt emission photon field \citep{Waxman:1997ti, Dermer:2006bb, Murase:2005hy, Hummer:2011ms, Asano:2013jea, Zhang:2012qy, Petropoulou:2014awa, Biehl:2017zlw, Pitik:2021xhb}. 

The leading models for the GRB prompt emission 
involve a dissipative fireball expanding relativistically (Lorentz factor $\Gamma \geq$ 100), where the energy dissipation occurs in internal shocks when portions of the jet are moving outwards with  
varying Lorentz factors \citep{Rees:1994nw, Daigne:1998xc, Kobayashi:1997jk}, or via magnetic reconnection in Poynting-flux dominated outflows \citep{Meszaros:1996ww, Spruit:2000zm, Drenkhahn:2001ue, 2002A&A...391.1141D, Lyutikov:2003ih, Kumar:2015cka,Giannios:2018nin}. 
In both cases a fraction of the thermal electrons in the outflow is accelerated to relativistic velocities and their subsequent cooling typically produces the synchrotron spectrum observed as a prompt gamma-ray burst emission at sub-MeV energies. Energy dissipation in the sub-photospheric region resulting in electron heating can also lead to the significant contribution from photospheric photons to the observed non-thermal spectrum \citep[e.g.][]{Giannios:2011ex, Rees:2004gt, Peer:2005qoc, Toma:2010xw}. 

If accelerated protons carry a much larger amount of the dissipated energy than electrons, then the photon spectrum below and above the MeV energy range may show additional components related to photohadronic interactions and/or photodisintegration of nuclei~\citep[see e.g.][]{Asano:2008tc, Murase:2010va}. The possibility that the sub-MeV prompt emission spectrum is explained by radiation from secondary leptons produced in photohadronic interactions in proton-dominated outflows has also been explored \citep[e.g.][]{Murase:2011cx, Petropoulou:2014sja}. 

While the Gamma-Ray Burst Monitor (GBM) on board the {\it Fermi} satellite remains the most prolific instrument for the detection of GRBs in the sub-MeV to MeV energy range, the Large Area Telescope offers complementary spectral information at energies above 100 MeV~\citep{Ajello:2019zki}\footnote{Note that {\it Fermi} LAT probes primarily 
most energetic events 
observed by the  Gamma-Ray Burst Monitor (GBM), 
though LAT-detected GRBs cover a large range of peak flux and fluence values when compared to the GBM population.}. 
For instance, LAT observations of the prompt phase of several individual GRBs have revealed the presence of a hard power-law component (with photon index less than 2) extending from the lower GBM energies to the GeV band \citep[e.g.][]{Fermi-LAT:2010err, Zhang:2010ey, Guiriec:2015ppa, Tang:2021wsb}.
Among the proposed scenarios offered as an explanation for this additional spectral component, those involving UHECRs \citep[e.g.][]{Asano:2009ta} are attractive for multi-messenger connections. The possibility that UHECRs are accelerated in 
GRBs was also studied through the
direct cross-correlations between the LAT HE ($\gtrsim$ 1 GeV) events and UHECRs detected by the Telescope Array (TA) and the Pierre Auger Observatory (PAO) \citep{Alvarez:2016otl}.  
However, no statistically significant correlation was found in comparisons to date. The detectable neutrino fluence predicted from the simplest theoretical models was neither confirmed in the analysis of the IceCube data for individual bright gamma-ray bursts \citep{Gao:2013fra}, nor identified in the model-independent stacking analysis searches performed on large samples of GRBs with different spectra 
\citep{IceCube:2012qza,IceCube:2016ipa,Aartsen:2017wea,IceCube:2022rlk}. The recent detections of very-high energy ($>$ 100 GeV) photons from GRBs \citep{MAGIC:2019lau, Abdalla:2019dlr, GRB221009A_LHAASO_GCN32677} have prompted the searches for the associated multi-messenger counterparts, but no neutrino detection from such GRBs has been confirmed so far. 

The null results from multi-messenger searches of GRBs so far do not necessarily reject the hypothesis that UHECRs are accelerated in GRB outflows. In fact, several theoretical predictions about UHECRs, HE gamma-rays and neutrinos during the GRB prompt phase rely on one-zone models, where the emission from the whole burst is assumed to be produced in one region of the outflow, see e.g. \citet{Waxman:1997ti,Murase:2005hy,Hummer:2011ms,Asano:2012jr, He:2012tq,Baerwald:2014zga,Biehl:2017zlw,Pitik:2021xhb}. Since both the neutrino and UHECR production are sensitive to the locally produced photon spectrum, it is necessary to account for the temporal evolution of the relativistic outflow and evolution of the physical conditions in the emission regions. 
In a series of more recent works \citep{Bustamante:2014oka,Bustamante:2016wpu,Rudolph:2019ccl,Heinze:2020zqb}, neutrino and cosmic-ray emission from multiple emission regions were considered. It was demonstrated that different messengers originate from different shock collision radii and that the neutrino fluence is dominated by the collisions close to the photosphere, leading to lower predicted neutrino fluences than the one-zone models. However, the local photon spectrum in these works was motivated from observations and not self-consistently computed. While \citet{Globus:2014fka} considered a multi-zone model for UHECR nuclei with target photons self-consistently produced by leptonic processes, they did not account for the feedback from hadronic processes on the electromagnetic spectrum. 

In this paper we present a fully self-consistent lepto-hadronic radiative model for multiple internal shocks occurring at different locations above the GRB photosphere. Our model accounts for the different physical conditions of each collision, while taking into account the feedback of high-energy protons on the locally produced photon spectrum. We focus on energetic {\em Fermi}-LAT detected GRBs (i.e. $E_{\gamma,\mathrm{iso}} \gtrsim 10^{54} \, \mathrm{erg}$). These GRBs can provide the necessary energy output per GRB in order to power the UHECRs, while only a moderate baryonic loading (defined as the energy injected into non-thermal protons versus electrons) is required and even energy equipartition might be possible. The latter property is especially attractive because it can mitigate the energetic problems that models (for typical GRBs) with high baryonic loadings face when it comes to the afterglow emission. Given the moderate dissipation efficiency of internal shocks (typically less than 10\%, see e.g. \citet{Panaitescu:1999dw, Beloborodov:2000nn, Bosnjak:2008bd}), a high baryonic loading implies a high outflow kinetic luminosity, a large fraction of which will end up powering the afterglow.
This may be in tension with afterglow observations \citep[see e.g.][]{Beniamini:2016hzc}, although the implications of the afterglow measurements for the prompt emission efficiency depend on the afterglow model.
Nonetheless, if scenarios with high baryonic loadings for ``ordinary'' GRBs are disfavored,  this raises the question of why energetic GRBs should be more efficient UHECR accelerators than ``ordinary'' GRBs. An answer to that question either points towards a different population of bursts or to ``friendlier'' conditions for UHECR acceleration in energetic bursts, see Sec.~\ref{sec:discussion} for a deeper discussion.

The paper is structured as follows. In Section~\ref{sec:ismodel} we present the implementation of the internal shock scenario for the GRB prompt emission and the parameters of the model describing the physical conditions in the shocked regions. For the numerical modeling of the cooling of the injected distributions of particles we used the time-dependent code \textsc{AM$^3$} \citep{Gao:2016uld}. We introduce two model GRBs (prototypes), a single-pulse burst and a multi-peaked event (inspired by GRB~170214A) in Section~\ref{sec:prototypes}. We explore first the leptonic model using as test-bed the single-pulse GRB, and present our results in Section~\ref{sec:leptonicmodel}. In Section~\ref{sec:leptohadronicmodel} we examine the lepto-hadronic models for both prototypes. We investigate different baryonic loadings and the impact of the variability timescales for single-peaked and multi-peaked synthetic GRBs. A multi-collision GRB model offers the possibility to investigate the time-dependence for the observation of different messengers (photons, neutrinos, UHECRs). In Section~\ref{sec:multimess} we discuss possible ways to discriminate between different parameter regions using current multi-messenger observations. We continue the discussion on other aspects of our models, such as UHECR composition and future directions in Section~\ref{sec:discussion}. We summarize our work and present our conclusions in Section~\ref{sec:summary}.   

\section{A multi-collision internal shock radiation model}
\label{sec:ismodel}
In the internal shock model the GRB prompt emission is generated in interactions between fast and slow parts of the relativistic ejecta \citep{Rees:1994nw, Kobayashi:1997jk, Daigne:1998xc}. We follow the formalism presented in \cite{Daigne:1998xc} and discretize the outflow as a series of plasma layers (called `shells') that propagate at different velocities. As fast shells catch up with slower ones they collide and energy is dissipated. The overall fireball emission is obtained by adding the contributions of all single collisions. 

The relativistic outflow is discretised as shells of source-frame width $\Delta = c \mrm{d}\tau$, where $\mrm{d}\tau$ is calculated from the number of initial shells $N^\mrm{ini}_\mrm{shells}$ and the engine active time $t_\mrm{eng}$ as $\mrm{d}\tau = t_\mrm{eng} / N_\mrm{shells}^\mrm{ini}$. Each shell is further characterised by its mass $M$ and Lorentz factor $\Gamma$ and has a volume $V= 4 \pi R^2 \Delta $ at a distance $R$ from the central emitter \footnote{We use three different frames of reference in this paper: The source (or engine) frame, the plasma comoving frame and the observers frame. Quantities in those frames will be denoted as $X$, $X^\p$ and $X_\mrm{obs}$ respectively.}. We emphasise that the number of initial shells (and the resulting number of collisions) in our model are a discretisation choice and the results are independent of it as long as there are enough shells to adequately resolve the fireball evolution.

\subsection{Two-shell collision}

We first recapitulate the formulas describing the collision between two shells.
A collision (labelled with a subscript `$\mrm{Coll}$') between a fast (subscript `f') and a slow (`s') shell at a radius $R_\mrm{Coll}$ and time $T_\mrm{Coll}$ (in the source frame) creates a new merged (subscript `m') shell that continues in the fireball. 
From energy and momentum conservation the merged shell mass and Lorentz factor can be calculated as 
\begin{eqnarray}
    M_\mrm{m} & = M_\mrm{f} + M_\mrm{s} \, ,
    \label{eq:m_m} \\
    \Gamma_\mrm{m} &= \sqrt{\frac{M_\mrm{f} \Gamma_\mrm{f} + M_\mrm{s} \Gamma_\mrm{s}}{M_\mrm{f}/ \Gamma_\mrm{f} + M_\mrm{s}/ \Gamma_\mrm{s}}} \, .
    \label{eq:gamma_m}
\end{eqnarray}

The collision time as measured in the observer's frame is given by 
\begin{equation}
T_\mrm{Coll, obs} = (1+z)(T_\mrm{Coll} - \frac{R_\mrm{Coll}}{ c}) \,
\label{eq:t_obs}
\end{equation}
where $z$ is the redshift of the burst. Note that this also corresponds to the earliest time at which photons of a collision may be observed.

We specify next the plasma conditions in the shocked plasma of the merged shell. We assume the Lorentz factor of the emission region is the same as that of the shocked plasma region, which is given by
\begin{equation}
\Gamma_\mrm{em} = \sqrt{\Gamma_\mrm{s} \Gamma_\mrm{f}} \, .
\end{equation}
This formula is obtained by assuming that most of the energy is dissipated as the less massive shell has swept up a mass comparable to its own. The mass density of the plasma is calculated as 
\begin{equation}
\rho^\p = \frac{M_\mrm{m}}{V_\mrm{Coll}^\p} = \frac{M_\mrm{m}}{4 \pi R_\mrm{Coll}^2 \Gamma_\mrm{em} c \, \mrm{d}\tau} \, ,
\end{equation}
where we used the comoving width of the shell $\Delta_\mrm{Coll}^\p = \Gamma_\mrm{em} c \, \mrm{d}\tau $. 
Energy conservation gives the dissipated energy as 
\begin{equation}
E_\mrm{diss} = (M_\mrm{f}\Gamma_\mrm{f}+ M_\mrm{s} \Gamma_\mrm{s} - M_\mrm{m} \Gamma_\mrm{m})c^2 \, .
\end{equation}
From the comoving dissipated energy $E_\mrm{diss}^\p = E_\mrm{diss}/ \Gamma_\mrm{em}$ we define the comoving energy density as

\begin{equation}
u^{\p}_{\rm diss} \equiv \frac{E_\mrm{diss}^\p}{V^\p_{\rm Coll}} \, .
\end{equation}
We further define the dissipated energy per unit mass $\epsilon^\p_\mrm{diss}$ as

\begin{equation}
    \epsilon^\p_\mrm{diss} = \frac{E_\mrm{diss}^\p}{M_\mrm{m}} \, .
\end{equation}

The characteristic timescale of the system (also called `dynamical timescale') is identified as the shell expansion time 
\begin{equation}
t^\p_\mrm{dyn} = \frac{R_\mrm{Coll}}{\Gamma_\mrm{em} c} \, . 
\label{eq:t_ex}
\end{equation}

We also specify the fraction of energy that is transferred to the different particle species, quantified by the microphysics parameters $\epsilon_\mrm{i}$. Under the assumption that the observed prompt emission is dominated by emission of non-thermal electrons, it is convenient to relate all quantities to the fraction of energy transferred to non-thermal electrons $\epsilon_e$. We thus define $f_\mrm{p/e} = \epsilon_\mrm{p} / \epsilon_\mrm{e}$ (where $ \epsilon_\mrm{p}$ is the fraction of energy transferred to non-thermal protons), $f_\mathrm{TH/e} = \epsilon_\mathrm{TH} / \epsilon_\mrm{e}$ (where $ \epsilon_\mathrm{TH}$ is the fraction of energy transferred to thermal particles) and  $f_\mrm{B/e} = \epsilon_\mrm{B} / \epsilon_\mrm{e}$ (where $ \epsilon_\mrm{B}$ is the fraction of energy transferred to the magnetic field). Note that by this definition, $f_\mathrm{TH/e}$ accounts for both electrons and protons and that the parameter, although set to 0 in the following, may in reality non-negligible. The comoving magnetic field strength can then be expressed in terms of the comoving non-thermal electron energy density $u^\p_\mrm{ele,NT} = \epsilon_\mrm{e} u^\p_{\rm diss}  $ as follows:
\begin{equation}
B^\p = \sqrt{8 \pi f_\mrm{B/e} u^\p_\mrm{ele,NT}} \, .
\end{equation}

\subsection{Fireball energy normalisation} 
The initial kinetic energy of the fireball can be written as
\begin{eqnarray}
    E_\mrm{kin, ini} & = &  \varepsilon^{-1} E_{\rm diss} = \varepsilon^{-1}\left(E^{\rm tot}_\mathrm{e, NT} + E^{\rm tot}_\mathrm{p, NT} + E_{\rm B} + E^{\rm tot}_\mathrm{TH} \right) \nonumber \\ 
    & = & \varepsilon^{-1} E_\mathrm{e, NT}^\mathrm{tot}(1 + f_\mrm{p/e} + f_\mrm{B/e} + f_\mrm{TH/e}),
    \label{eq:ekin_initial}
\end{eqnarray}
where $\varepsilon$ is the fireball (dissipation) efficiency defined as
\begin{equation}
\varepsilon \equiv 1  - \frac{\sum_\mrm{i} M_\mrm{i, fin}\Gamma_\mrm{i, fin}}{ \sum_\mrm{i} M_\mrm{i, ini}\Gamma_\mrm{i, ini}}. 
\end{equation}
In the following we will assume that the outflow is launched with a constant wind luminosity, which implies that all initial shells carry the same initial energy $E_\mrm{i, ini} = \mrm{d}\tau/t_\mrm{eng} \cdot E_\mrm{kin, ini}$.

Assuming that the observed sub-MeV prompt spectrum is predominately produced by leptonic processes, it is convenient to normalise the initial fireball kinetic energy to the total energy transferred to non-thermal electrons $E_\mathrm{e, NT}^\mathrm{tot}$ that is needed to produce a given isotropic-equivalent energy in gamma rays $E_\mrm{\gamma, iso}$ (the isotropic equivalent energy emitted in gamma-rays in the energy range of 1 - 100 keV). With this normalization we obtain $\epsilon_\mrm{e}$ for each set of ($f_\mrm{p/e}$, $f_\mrm{B/e}$, $f_\mrm{TH/e}$) as
\begin{equation}
\epsilon_\mrm{e} = (1 + f_\mrm{p/e} + f_\mrm{B/e} + f_\mrm{TH/e})^{-1} \, .
\label{eq:epsilon_e}
\end{equation}

\subsection{The deceleration radius}
\label{sec:deceleration_radius}
The deceleration radius $R_\mathrm{dec}$ which marks the end of the prompt emission phase of the fireball is reached when the initial fireball kinetic energy is equal to the energy of the heated downstream plasma (given by $E_\mrm{SW} = \Gamma^2 M_\mrm{SW}(R) c^2$, where $M_\mrm{SW}(R)$ is the swept-up mass at a radius $R$). We adjust Eq.~1 from \citet{Rees:1992ek} for a wind-like medium and find 

\begin{equation}
    R_\mrm{dec} = \frac{1}{4 \pi \rho_0}
    \sum_i 
    \frac{M_\mrm{i, ini}}{\Gamma_\mrm{i, ini}} \, , 
\end{equation}
where $i$ sums over all initial shells contained in the fireball. We calculate $\rho_0$ as $\rho_0 = \frac{\dot{M}}{4 \pi v_w}$ with typical values of the mass ejection rate $\dot{M}$ and wind velocity $v_w$ that are given by $\dot{M} = 10^{-5} M_\odot / \, \mrm{yr}$ and $v_w = 10^3$~km/s. Note that a deviation from those values results in a different deceleration radius.

The initial masses can be re-expressed through the initial kinetic energy of the fireball:
\begin{equation}
    \sum_i\frac{M_\mrm{i, ini}}{\Gamma_\mrm{i, ini}} = \frac{\mrm{d}\tau}{t_\mrm{eng} c^2} E_\mrm{kin, ini} \sum_i\frac{1}{\Gamma_{\mathrm{i, ini}}^2} \, ,
\end{equation}
where $i$ again sums over the initial shells of the fireball.  

For the calculating of the deceleration radius we invoke two sets of parameters: \textit{(a)} The \textit{equipartition} case with $f_\mrm{p/e} = f_\mrm{B/e} = f_\mrm{TH/e} = 1$ (where thus non-thermal electrons, protons, the magnetic field and thermal particles receive equal parts of energy) and \textit{(b)} the \textit{hadronic} case with $f_\mrm{p/e} = 30$,  $f_\mrm{B/e} =1$ and $f_\mrm{TH/e} = 0$ (where thus the majority of the energy is transferred to non-thermal protons). The corresponding radii will be labelled $R_\mrm{dec, equi}$ and  $R_\mrm{dec, had}$.

\subsection{Non-thermal particle distributions and radiative calculations}
\label{sec:model_description_injection}
From the internal shock model that specifies the conditions in the shocked plasma we can obtain the parameters for the non-thermal particle distributions that are an input for the radiation modeling.
Note that we do not account for the effect of photons emitted from earlier collisions.

\subsubsection{Injected particle distributions}
We first describe how we calculate the injected distributions of primary particles, considering a species $i$ with mass $m_\mrm{i}$, Lorentz factor $\gamma^\p_\mrm{i}$ and charge number $Z_\mrm{i} $. As we consider no heavy nuclei, $Z_\mrm{i} = 1$. 
We assume that particles are accelerated at a very thin region close to the shock front before being injected into the (homogeneous) radiation zone of the shocked plasma. In the radiation zone of a single collision we do not account for any spatial dependence of quantities in the shocked plasma, such as magnetic field decay away from the shock \citep[e.g.][]{Lemoine:2012yw}. 

The injection rate of particle species $i$ into the radiation zone is given by
\begin{equation}
\frac{\mrm{d} n^\p_\mrm{i}}{\mrm{d}t^\p \mrm{d}\gamma^\p_\mrm{i}} = \dot{n}_\mrm{0,i} \gamma_\mrm{i}^{\p -p_{\rm i}} e^{-\gamma^\p_\mrm{i} / \gamma^\p_\mrm{i, max}}
\end{equation}
for Lorentz factors above a minimum value of $\gamma^\p_\mrm{i, min}$. The maximum Lorentz factor, $\gamma^\p_\mrm{i,max}$, at any given time is determined by the balance between the radiative loss timescale and the acceleration timescale. The loss timescales for electrons are due to synchrotron, inverse Compton and adiabatic cooling, while for protons they are due to photo-pion, photo-pair, synchrotron and adiabatic cooling.
In general, the acceleration time can be written as
\begin{equation}
t_\mrm{acc}^\p =  \eta_\mrm{acc} \frac{E_\mrm{i}^\p}{c B^\p e} \, ,
\label{eq:tacc}
\end{equation}
where $\eta_\mrm{acc}$ specifies the acceleration efficiency. Considering the fastest acceleration time we set $\eta_\mrm{acc} = 1$ throughout this paper.
The power-law index  for electrons $p_\mrm{e}$ is considered to be a free parameter determined by prompt observations. For protons, however, the index cannot be usually inferred by electromagnetic observations alone. So, our benchmark choice is $p_\mrm{p} = 2$, but we discuss the impact of other indices in Section~\ref{sec:multimess}. 

If $\epsilon_\mrm{i}$ defines the fraction of energy transferred to the species $i$ (that can be calculated from the $f_\mrm{i/e}$ values and $\epsilon_e$ through \refeq{epsilon_e}), the number fraction of accelerated particles $\zeta_\mrm{i}$ and the minimum Lorentz factor $\gamma^\p_\mrm{i,min}$ are related through
\begin{equation}
\gamma^\p_\mrm{i, min} = \frac{p_\mrm{i} -2 }{p_\mrm{i} - 1}\frac{\epsilon_\mrm{i}}{\zeta_\mrm{i}} \frac{m_\mrm{p}}{m_\mrm{i}} \frac{u^\p_ \mrm{diss}}{\rho^\p c^2} \, .
\label{eq:zeta_gammamin}
\end{equation}
This relationship is obtained by setting the total particle density $n^\p_\mrm{i}$ (integrated over $\gamma^\p_\mrm{i}$) equal to $\zeta_\mrm{i} \rho^\p/ m_\mrm{p} $, assuming that the shocked plasma is composed of ionized hydrogen. 
For both electrons and protons we assume $\zeta_\mrm{i} \propto \epsilon_\mrm{diss}^\p$ which results in a constant $\gamma^\p_\mrm{i, min}$ throughout the fireball evolution. As suggested in \citet{Asano:2008tc}, the minimum Lorentz factor in mildly relativistic shocks should be of order unity and following their approach we set $\gamma^\p_\mrm{p, min} = 10$. On the other hand, $\gamma^\p_\mrm{e, min}$ will be adjusted to fit the peak of the spectrum in the observer's frame for each scenario. In Appendix~\ref{app:modelling-choices} we further explore the additional scenario of $\zeta_\mrm{e} = {\rm const.} $ where $\gamma^\p_\mrm{e, min}$ evolves proportional to $\epsilon_\mrm{diss}^\p$ throughout the fireball evolution \citep[see also][]{Bosnjak:2014hya}.

The normalisation $\dot{n}_\mrm{0,i}$ of the injected particle distribution is given by  
\begin{equation}
\epsilon_\mrm{i} u^\p_ \mrm{diss}  = m_\mrm{i} c^2  \int_{\hat{t^\p}=0}^{\hat{t^\p} = t^\p_\mathrm{inj}}  \mrm{d}\hat{t^\p}  \, \dot{n}_\mrm{0,i} \int_{\gamma^\p_\mrm{i, min}}^{\infty} 
\mrm{d}\gamma^\p_\mrm{i} \, \gamma_\mrm{i}^{\p -p_\mrm{i} + 1} .
\label{eq:injection_normalisation}
\end{equation}
Here, $t^\p_\mathrm{inj}$ is the timescale over which particles are injected in the radiation zone. Our fiducial assumption is continuous injection of accelerated particles over the dynamical timescale, i.e. $t^\p_\mathrm{inj} = t^\p_\mathrm{dyn}$. In Appendix \ref{app:modelling-choices} we further explore an acceleration timescale much shorter than $t^\p_\mathrm{dyn}$, labelled as $\delta t^\p_\mathrm{inj} \rightarrow 0$. In the numerical calculations this is approximated by an injection over a timescale much shorter than $t_\mrm{dyn}^\p$; here, we choose $t^\p_\mathrm{inj} = 0.01 t^\p_\mathrm{dyn} $.

\subsubsection{Numerical treatment}
The numerical radiation modeling is performed with the time-dependent code \textsc{AM$^3$} \citep{Gao:2016uld} which follows the coupled evolution of photons, electrons, positrons, muons, pions, protons, neutrons and neutrinos. The software includes all relevant non-thermal processes such as synchrotron and synchrotron self-absorption, inverse Compton scatterings, photo-pair and photo-pion production, $\gamma \gamma $-annihilation, adiabatic cooling and escape. Secondary particles produced in these interactions are added to the overall particle distributions and undergo the same processes as primaries.

The numerical treatment is described in detail in \cite{Gao:2016uld}. Here, we briefly list the modifications to the original version of the code:
\begin{itemize}
    \item Adiabatic cooling of charged particles is implemented with a cooling rate $\mrm{d}\gamma^\p_\mrm{i}/\mrm{d}t^\p = -\gamma^\p_\mrm{i}/t^\p_\mrm{dyn}$.
    \item Muons and pions are treated as separate species that may be subject to synchrotron and adiabatic cooling prior to their decay and emit synchrotron radiation.
    \item Quantum synchrotron radiation is implemented following \cite{Brainerd_1987}.
    \item The treatment of photo-pair production has been updated and now follows \cite{Kelner:2008ke}.
\end{itemize}
For each collision we then compute the temporal and spectral evolution of the comoving particle energy density of each particle species, $u^\p_{ \mrm{i}}(E^\p, t^\p)$. We point out that feedback between different collisions is not accounted for.

\subsubsection{Calculation of emitted spectra}
While charged particles are assumed to be confined by the magnetic fields, neutral particles escape at a rate $t_\mathrm{esc}^{\p -1} = 1 / t^\p_\mrm{dyn}$ .
The differential spectrum of particles that have escaped until a time $t^\p$ can be calculated from the time-dependent comoving density of the relevant particle species, 
$u^\p_\mrm{i} (E^\p, t^\p) = E_\mrm{i}^\p \mrm{d}N_\mrm{i} /\mrm{d}E_\mrm{i}^\p \mrm{d}V^\p$, as 
\begin{equation}
u^\p_\mrm{esc, i} (E^\p, t^\p)= \frac{1}{t^\p_\mrm{esc}} \int_{\hat{t^\p} = 0}^{\hat{t^\p} = t^\p} u^\p_\mrm{i}(E^\p, \hat{t^\p}) \mrm{d}\hat{t^\p} \, .
\label{eq:escaped_single_spec}
\end{equation}
To compute the differential emitted spectrum, $u^\p_\mrm{em} (E^\p)$, we follow the system's evolution until $t^\p = t^\p_\mrm{inj} + t^\p_\mrm{dyn}$, at which point all primary particles have either escaped or cooled. The particles remaining in the system at this point are then added to the spectra of escaped particles. 

\subsubsection{Conversion into observed quantities}
For the calculation of time-dependent quantities (light curves and time-resolved spectra), we take into account the curvature of the emitting surface following \cite{Granot:1998ep}.
We however introduce a simplified approach where the emission from the thin shell is assumed to be released at a single time $T_\mrm{Coll}$ and from a single shell surface of radius $R_\mrm{Coll}$ (thus the integral over $R$ in \cite{Granot:1998ep} is a $\delta$-function for each collision). For calculations that invoke only a single radiation zone, this over-simplification would severely impact the observed profile and not reproduce the Fast-Rise-Exponential-Decay (FRED) \citep[see e.g.][for the case of an infinitesimally thin shell]{Granot:1998ep, Salafia:2016wru}. 
When computing the emission from a large number of collisions though, as in this work (see next section), the emission profile is dominated by the fireball evolution and our simplified approach will not impact the predictions significantly.

For a given collision (with radius $R_\mrm{Coll}$, emission time $T_\mrm{Coll}$, Lorentz factor $\Gamma_\mrm{em}$ and comoving width $\Delta^\p_\mrm{Coll}$) 
we calculate the energy flux, $F(T_{\rm obs})$, integrated over the full energy range (in erg s$^{-1}$ cm$^{-2}$) as 

\begin{equation}
F (T_\mrm{obs})  = \frac{1}{2 D^2} \int \mrm{d} E^\p \, \frac{R_\mrm{Coll} c }{(1+z) \Lambda^2} u^{\p}_\mrm{em i} (E^\p) \Delta_\mrm{Coll}^\p  \, ,
\label{eq:energyflux}
\end{equation}
where $D$ is the comoving distance and $\Lambda = \Gamma_\mrm{em} \left[1 - \frac{c \beta_\mrm{em }}{R_\mrm{Coll}} \left(T_\mrm{Coll} - \frac{T_\mrm{obs}}{1+z} \right) \right]$. 

For spectra integrated over the full duration of the burst we apply a simplified procedure. The (differential in energy) observed fluence $\mathcal{F}_E$ of a single collision is simply given by 
\begin{equation}
    E \mathcal{F}_{E} =u^{\p}_\mrm{em} (E^\p) V_\mrm{Coll}^\prime \frac{\Gamma_\mrm{em} E^\prime}{4\pi (1+z)D^2} \, .
\end{equation}
The full fireball emission is obtained by adding the contributions of all single collisions.

The effect of absorption due to interactions with the Extragalactic Background Light (EBL) are discussed in \refsec{constraints_em_cascade} and will be omitted in all other sections. We calculate it with the open-source \soft{gammapy}-package \citep{Deil:2017yey, Nigro:2019hqf}, selecting the model of \cite{Dominguez:2010bv}.

\section{Introducing two prototypes}\label{sec:prototypes}
For the purposes of
our study it is useful to introduce two model GRBs that differ mainly in their temporal properties. The first prototype is characterized by a smooth single-pulse (SP) light curve, while the second one has a multi-peaked (MP) light curve with short-timescale variability. Given that we will pay special attention to possible HE emission signatures, we loosely base the parameters of our prototypes on the properties of GRBs detected by \textit{Fermi}-LAT.
As pointed out in \cite{Ajello:2019zki}, these populate the upper range of the $E_\mrm{\gamma, iso}-$distribution. Consequently, we choose events with a high energy budget. The $E_\mathrm{\gamma, iso}$- $\Gamma$ correlation \citep[see e.g.][]{Liang:2009zi, Ghirlanda:2017opl} and the requirement of being optically thin to $\gamma \gamma$ pair production, further imply comparatively high Lorentz factors of the outflow. 

Leptonic scenarios will be illustrated with solely the simple, single-peaked burst. On the other hand, lepto-hadronic models (with different baryonic loadings $f_\mrm{p/e}$) will be discussed for both prototypes. 

\subsection{Model parameters}

\begin{figure}
    \centering
\includegraphics[width = 0.4\textwidth]{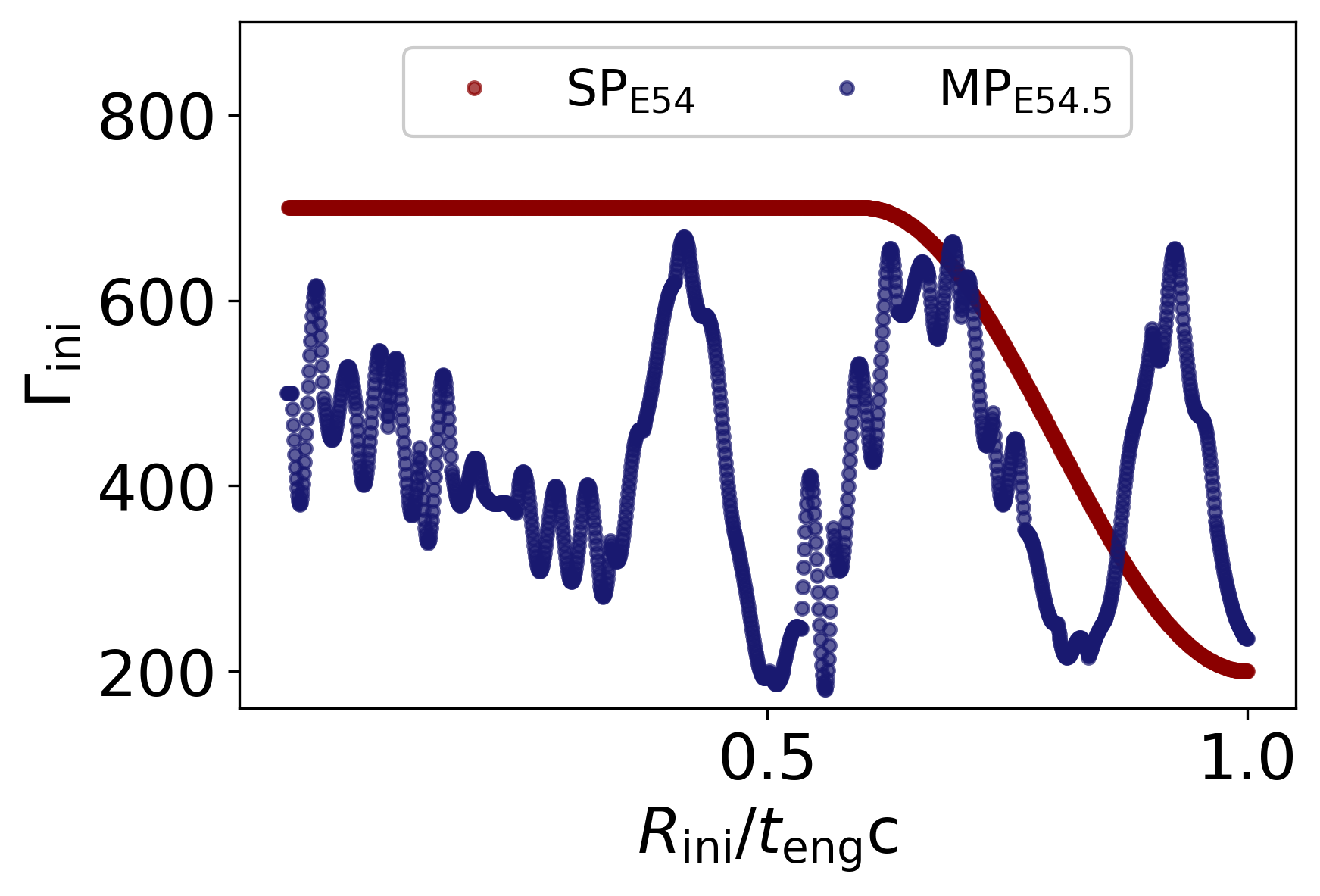}
    \caption{Initial Lorentz factor distribution for two GRB prototypes as a function of initial radius $R_\mrm{ini}$ (normalised to $t_\mrm{eng} c $).}
    \label{fig:initial_lorentz_multipulse}
\end{figure}

Before going into detail on the two prototypes let us collect the model parameters and assumptions for better overview. Each of our prototypes will have a targeted gamma-ray isotropic energy, observed duration and observed peak energy, while it is placed at an assumed redshift (see Table~\ref{tab:models_characteristics}). Targeted here refers to the order of magnitude value that we will aim to reproduce. To this end we impose \textit{internal shock model parameters} (characterising jet parametrisation through shells) and \textit{microphysical parameters} (characterising the conditions in the shocked plasma), collected in Table~\ref{tab:model_parameters}. The internal shock model parameters are complemented by the initial Lorentz factor distributions, as depicted in \reffig{initial_lorentz_multipulse}.

\textit{Internal shock model parameters.} We choose the engine active time $t_\mathrm{eng}$ that reproduces the targeted duration. Then, after selecting an initial Lorentz factor distribution of the shells (that is chosen to match a certain light curve profile), the initial energies of the shells are set such that the targeted $E_\mathrm{\gamma, iso}$ is approximately matched. For the latter we follow \refeq{ekin_initial}, 
assuming a set of microphysics parameters $f_\mathrm{i}$. 

\textit{Microphysical parameters.} In the spirit of a parameter study we will explore different combinations of $f_\mathrm{i}$. Namely, we will study a scenario with a strong and a weak magnetic field in the outflow. This is achieved by imposing
$f_\mathrm{B/e} = 1$ and $f_\mathrm{B/e} = 10^{-3}$, which correspond respectively to a synchrotron (SYN)-dominated scenario for electron cooling, and an inverse Compton (IC)-dominated scenario. 
Both regimes will be explored for leptonic ($f_\mrm{p/e} = 0$) and lepto-hadronic models ($f_\mrm{p/e} > 0$). The benchmark baryonic loading used in the latter is $f_\mrm{p/e} = 30$, but we will also explore other values in the range $0.3-100$. \\
To reproduce the sub-MeV peak energy that we use as benchmark we adjust the fraction of accelerated electrons $\zeta_\mathrm{e}$ for each parameter set. More specifically,
we set $E_\mathrm{peak, obs}$ as the synchrotron peak energy of electrons at the minimum Lorentz factor as defined in Eq.~\ref{eq:zeta_gammamin} and solve for $\zeta_\mathrm{e, 0}$. This gives $\zeta_\mathrm{e, 0}^\mathrm{k}$ for the $k-$th collision as 
\begin{equation}
    \zeta_\mathrm{e, 0}^\mathrm{k} = 6.2 \cdot 10^{-4} \left[ \frac{(\epsilon_\mathrm{B} u^{\prime k }_\mathrm{diss})^{1/2} \Gamma_\mrm{em}^\mrm{k}}{(1+z)E_\mathrm{peak, obs}} \right]^{1/2}  \frac{p_\mrm{e}-2}{p_\mrm{e}-1} \epsilon_\mathrm{e} \, . 
\end{equation}
The reported value of $\zeta_\mathrm{e, 0}$ is then the average over all collisions, weighted by the dissipated energy of the collision.  
For completeness we will further list 
the minimum Lorentz factor that corresponds to the average $\zeta_\mathrm{e, 0}$. The initial power-law index of electrons will be set to $p_\mathrm{e} = 2.5$ (to reproduce typical high-energy slopes of GRBs) and $p_\mathrm{e} = 3.0$ (to reproduce a steeper high-energy slope that was observed by \fermi-LAT for GRB~170414A, which inspired our second prototype). 
For simplicity a minimum Lorentz factor of 10  and $p_\mathrm{p} = 2.0 $ will be assumed in all cases for protons.

\begin{table*}
\caption{
Parameter values used for two GRB prototypes inspired by observations of energetic bursts detected by \textit{Fermi}-LAT.} 
\label{tab:models_characteristics}
\centering
\renewcommand{\arraystretch}{1.35}
\resizebox{2.\columnwidth}{!}{%
\begin{tabular}{c c c c c}
\toprule
Parameter & Symbol  & \SP & \multicolumn{2}{c}{\MP}\\ 
 & & & $\delta t_\mrm{var} = 1.13$ s & $\delta t_\mrm{var} = 0.11$ s \\ \hline 
 Targeted isotropic $\gamma$-ray energy (source frame) & $E_{\gamma, \mathrm{iso}}$ & $10^{54}$ erg & $ 2.9 \cdot 10^{54}$ erg & $ 2.9 \cdot 10^{54}$ erg \\
 Targeted peak energy (observer frame) & $E_\mathrm{peak, obs}$ & 400 keV & 566 keV  & 566 keV \\
 Targeted duration (observer frame) & $T_\mathrm{dur, obs}$ & 15 s & 106 s & 10.6 s \\
 Assumed redshift &  $z$ & 2 & 2 & 2 \\
\hline
\end{tabular}
}
\end{table*}

\begin{table*}
\caption{Fireball parameters and fiducial microphysics parameters used for the modeling of two energetic GRB prototypes.}
\label{tab:model_parameters}
\centering
\renewcommand{\arraystretch}{1.35}
\resizebox{2.0\columnwidth}{!}{%
\begin{threeparttable}
\begin{tabular}{l c c c c}
\toprule
  Parameter & Symbol & \SP & \MP \, ($\delta t_\mrm{var} = 1.13$ s) & \MP \, ($\delta t_\mrm{var} = 0.11$ s)\\ \hline
  Number of initial shells & $N_\mathrm{shells}^\mathrm{ini}$ & 1000 & 1297 & 1297\\
 Engine active time  & $t_\mathrm{eng}$ & 5 s & 34 s & 3.4 s \\ 
 Number of collisions & $N_\mathrm{coll}$ & 999 & 1139 & 1139\\
Total energy in non-thermal electrons & $E_\mathrm{e, NT}^\mathrm{tot}$ & $1.3 \cdot 10^{54}$ erg &  $3.5 \cdot 10^{54} \, \mathrm{erg}$ &  $3.5 \cdot 10^{54} \, \mathrm{erg}$  \\
 Average collision radius & $\langle R_\mathrm{Coll} \rangle$  & $1.9 \cdot 10^{16}$ cm & $2.4 \cdot 10^{16}$ cm & $2.4 \cdot 10^{15}$ cm \\ 
 Overall dissipation efficiency & $\varepsilon$ & 7.8\% &  2.98 \% &  2.98 \% \\ \hline
 Power-law index of non-thermal electrons & $p_\mrm{e}$ & 2.5 & 3.0 & 3.0\\
 Power-law index of  non-thermal protons & $p_\mrm{p}$ & 2.0 & 2.0 & 2.0 \\
 Minimum Lorentz factor of  non-thermal protons & $\gamma^\prime_\mrm{p, min}$ & 10 & 10 & 10\\
Relative fraction of energy transferred to thermal particles & $f_\mrm{TH/e}= \epsilon_\mrm{TH}/\epsilon_\mrm{e}$ & 0 & 0 & 0 \\
   \hline
  & \multicolumn{4}{c}{`SYN-dominated'} \\
Relative fraction of energy transferred to magnetic field & $f_\mrm{B/e}= \epsilon_\mrm{B}/\epsilon_\mrm{e}$ & 1 & 1 & 1 \\
 Relative fraction of energy transferred to protons & $f_\mrm{p/e}= \epsilon_\mrm{p}/\epsilon_\mrm{e}$ & $ \{ \mathbf{0} , 10,  \mathbf{30}, 100 \}$ & $ \{ \mathbf{0} , 3, 10 ,  \mathbf{30} \}$ & $ \{ \mathbf{0} , 0.3, 1 ,  \mathbf{3} \}$ \\
 Normalisation for number fraction of accelerated electrons & $\zeta_{\rm 0,e} $ [$10^{-4}$] & 18.7 &  21.6 & 119.7 \\
  Minimum Lorentz factor of non-thermal electrons & $\gamma_{\rm e, min} $ [$10^{4}$] & 1.2 &  1.5 & 0.2 \\
 & \multicolumn{4}{c}{`IC-dominated'} \\
  Relative fraction of energy transferred to magnetic field & $f_\mrm{B/e}= \epsilon_\mrm{B}/\epsilon_\mrm{e}$ & $10^{-3}$ & $10^{-3}$ & - \\ 
   Relative fraction of energy transferred to protons & $f_\mrm{p/e}= \epsilon_\mrm{p}/\epsilon_\mrm{e}$ & $ \{ \mathbf{0} , 10,  \mathbf{30}, 100 \}$ & $ \{ \mathbf{0} , 3, 10 ,  \mathbf{30} \}$ & - \\
  Normalisation for number fraction of accelerated electrons & $\zeta_{\rm 0,e} $ [$10^{-4}$] & 3.3 & 3.8 & - \\
Minimum Lorentz factor of non-thermal electrons & $\gamma_{\rm e, min}^\p $ & 6.5 &  8.6 & - \\

\hline
\end{tabular}
\begin{tablenotes}
\item \textit{Notes. --} For $f_\mrm{p/e}$ we list all parameter values explored in this work and mark in bold those used as benchmark values for the leptonic and lepto-hadronic models. The variability timescale in the source frame is given by $\delta t_\mrm{var}  = t_\mrm{eng} / N_\mrm{osc}$, where $N_\mrm{osc} = 30$ is the number of short timescale oscillations in the initial Lorentz factor distribution, the average collision radius is obtained by weighing the distribution of $R_\mathrm{Coll}$ with their respective dissipated energy $E_\mrm{diss}$. The number fraction of accelerated electrons in a collision in a collision can be calculated as $\zeta_e = \mrm{min} \left(1, \zeta_{\rm 0,e} \frac{\epsilon^\p_\mrm{diss}}{100 \, \mrm{MeV/proton}} \right)$.
\end{tablenotes}
\end{threeparttable}
}
\end{table*}

\subsection{Single-pulse burst (SP$_\mrm {E54}$)}\label{sec:single-pulse}

\begin{figure*}
    \centering
    \centering
\includegraphics[width = 0.49\textwidth]{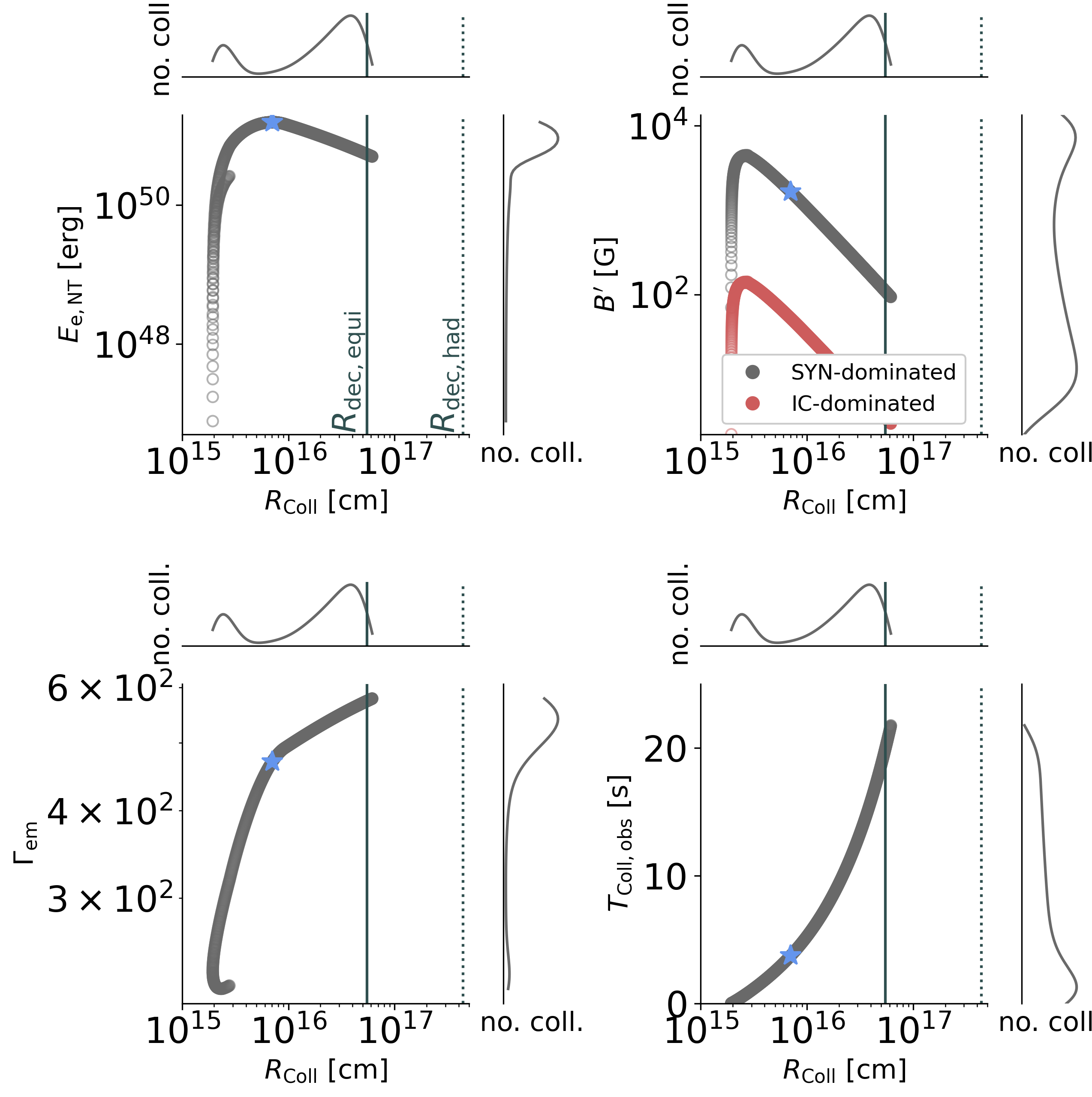}
\includegraphics[width = 0.49 \textwidth]{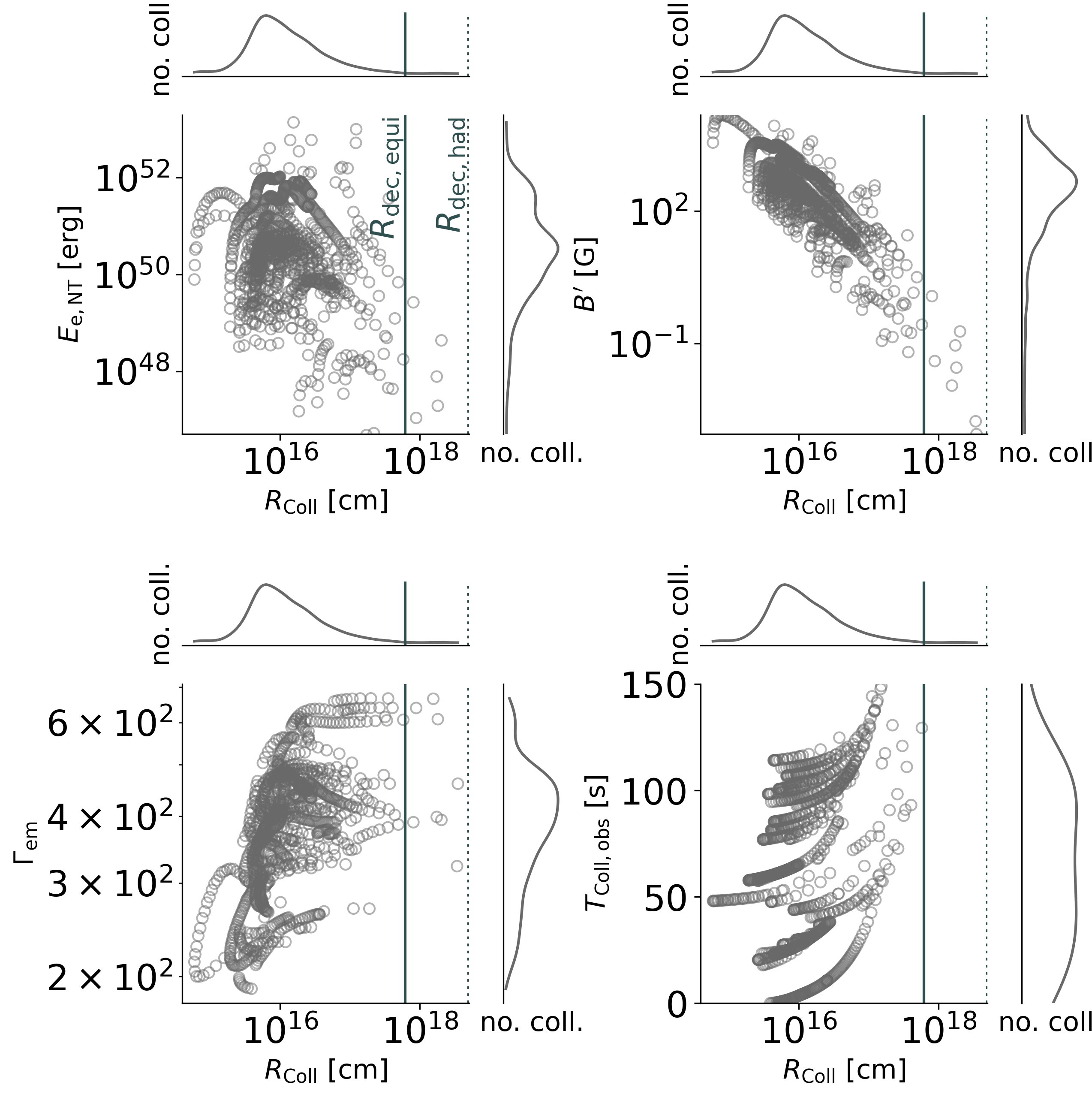}
    \caption{Evolution of outflow parameters as a function of collision radius for the \textit{(left)} \SP \, and \textit{(right)} \MP \, prototypes: 
    the energy carried by primary non-thermal electrons ($E_{\rm e, NT}$), 
    the Lorentz factor of the emitting shocked plasma ($\Gamma_\mrm{em}$), the comoving magnetic field ($B^\prime$) for the SYN- and IC-dominated scenarios, and the collision time in the observer's frame ($T_\mrm{Coll, obs}$).
    Vertical lines indicate the deceleration radius for two sets of microphysical parameters:  $f_\mrm{p/e} = f_\mrm{B/e} = f_\mrm{TH/e} = 1$ (an equipartition (\textit{equi}) scenario) and $f_\mrm{p/e} = 30$,  $f_\mrm{B/e} =1$ and $f_\mrm{TH/e} = 0$ (a hadronic (\textit{had}) scenario); see \refsec{deceleration_radius} for details.
    For the \SP \, scenario on the left we mark the representative collision introduced in \refsec{single-pulse} with a $\star$ symbol and show magnetic fields for both the IC-and the SYN-dominated scenario. For the \MP \, scenario on the right we only display the magnetic fields for the SYN-dominated scenario. For both prototypes, side panels show one-dimensional density distributions of the collisions.
    }
    \label{fig:ediss_rcoll}
\end{figure*}

Our first prototype is a single-pulse burst without short-timescale variability in the light curve, referred to as \SP \, (the subscript refers to the isotropic-equivalent energy of the GRB). This type of simple example has been studied in the past \citep[see for example][]{Bosnjak:2008bd, Globus:2014fka}. This light curve is representative of a single pulse in the GRB time evolution and the simple temporal and spatial structure are useful to study the effects of different emission zones and their contribution to the overall emitted spectrum.

The burst  
is generated from a smooth initial Lorentz factor distribution that is defined as 
$\Gamma (R_\mathrm{ini}) = \Gamma_{\max}$, for $R_\mrm{ini}/c < 0.6 \, t_\mrm{eng}$ and $\Gamma(R_\mathrm{ini}) = (\Gamma_\mrm{max}+\Gamma_\mrm{min})/2 - (\Gamma_\mrm{max}-\Gamma_\mrm{min})\cos\left(\pi \frac{R_\mathrm{ini}}{0.4 c t_\mrm{eng}}\right)/2$, for $ 0.6  \, t_\mrm{eng} \le R_\mrm{ini}/c <  t_\mrm{eng}$ \citep{Daigne:1998xc};
here $R_\mrm{ini}$ is the initial shell radius and $t_{\rm eng}$ is the overall engine activity time. We choose $\Gamma_\mathrm{min}=200$ and $\Gamma_\mathrm{max} = 700 $ as benchmark values, but we also examine different values later in \refapp{modelling-choices}. The burst characteristics and input fireball parameters are summarised in Tables \ref{tab:models_characteristics} and \ref{tab:model_parameters}, respectively. 

The emission from the full burst is the superposition of the emission from all individual collisions. As these occur at different locations in the outflow, the physical conditions of the emitting plasma (e.g. magnetic fields, energy densities, and so on) will be different. This is exemplified in \reffig{ediss_rcoll} (left panel) where we plot the total non-thermal electron energy, the magnetic field strength, the bulk Lorentz factor of the emission region, and the observed time of a collision as a function of collision radius (each point represents a collision). 
 
Because of the different physical conditions involved in each collision, it is difficult to examine in detail the role of the various physical processes at work by studying the emission of the full burst. For this reason, we will use a single \textit{representative} collision.
This is defined as the collision with the maximum dissipated energy $E_\mrm{diss}$ and is indicated with a star ($\star$) symbol in \reffig{ediss_rcoll}.
We consider it representative in the sense that it will be the one dominating the observed emission, and its spectrum will be somewhat similar to the overall burst spectrum (this will become clearer in the next section).  
By comparing the modeling results for the representative collision to those obtained from the full fireball evolution, we will discuss the limitations of simplified one-zone models where a single radiation zone is used for describing the full burst. 

\subsection{Multi-pulse burst with short-time variability (MP$_\mrm {E54.5}$)}
Most GRB light curves have
complex structures consisting of many pulses and exhibiting variability on short timescales. These properties cannot be captured by the single-pulse model described in the previous section. We therefore introduce a second prototype burst with a multi-pulse light curve and short-timescale variability, referred to as MP$_\mrm {E54.5}$. 

The prototype is inspired by GRB~170214A, a very energetic burst that was observed by both \textit{Fermi}-GBM and \textit{Fermi}-LAT. The onset of the LAT detection happened during the prompt phase, although delayed approximately 60~s with respect to the GBM trigger. Given that a similar short-time variability was observed in the GBM and LAT energy bands, an internal (prompt-phase) origin for the HE emission was proposed in \citet{Tang:2017mmr}. In terms of energetics and (preliminary) intrinsic peak energy, the burst is also similar to the recently observed GRB~221009A~\citep{GRB221009A_GBM_GCN32642, GRB221009A_KONUS_GCN32668, GRB221009A_AGILE_GCN32650}.  

To produce an analogous burst (i.e. with similar 
$E_\mathrm{iso}$, $T_\mathrm{90}$, $E_\mathrm{peak}$, high-energy photon index $\beta$ as summarised in \reftab{models_characteristics}) we impose the fireball parameters indicated in \reftab{model_parameters}. For the conversion to observed quantities we adopt the same redshift as for \SP, i.e. $z = 2 $.  
The observed light curve of the burst is composed of a few long-timescale pulses with superimposed variability on short timescales. This type of pattern needs to be represented in the initial Lorentz factor distribution (which in the internal shock model is directly reflected on the light curve structure \citep[see e.g. ][]{Bustamante:2016wpu}). Our chosen initial Lorentz factor distribution of the outflow is displayed in \reffig{initial_lorentz_multipulse}. 

In this case the distribution of plasma parameters as a function of collision radius deviates from the simple behaviour displayed in \reffig{ediss_rcoll}. This is exemplified in \reffig{ediss_rcoll} (right), where we plot the same quantities as in \reffig{ediss_rcoll}, with side panels showing the projected one-dimensional density distributions. 
We observe that the bulk of collisions occurs at $R_\mrm{Coll} \sim 10^{16}$~cm (similar to \SP). This is further reflected in the average dissipation radius $\langle R_\mrm{Coll} \rangle$, obtained by weighing the distribution of $R_\mrm{Coll}$ with the dissipated energy per collision. For \MP \, this weighted average is computed as $\langle R_\mrm{Coll} \rangle = 2.4 \cdot 10^{16}$~cm (compared to $\langle R_\mrm{Coll} \rangle = 1.9 \cdot 10^{16}$~cm for \SP). This large typical collision radius is mainly driven by the adopted long duration and the relatively large Lorentz factors. 
Although the $E_\mrm{iso}$ - $\Gamma$ correlation may point to high dissipation radii for energetic bursts, typical quoted collision radii of GRBs in the internal shock model are $R_\mrm{Coll} \sim 10^{11}$~cm to $R_\mrm{Coll} \sim 10^{15}$~cm, which are significantly smaller than the values for \MP \, and \SP. 

In a one-zone internal shock model where a single collision is considered representative for the complete burst, the source-frame variability timescale $\delta t_\mrm{var}$ can be used to estimate  the  collision radius through $R_\mrm{Coll} = 2 \Gamma_\mrm{em}^2 c \delta t_\mrm{var}$. Similarly, in multi-collision models the variability timescale can be used as a proxy of the collision radius; here we obtain $R_\mrm{Coll} \simeq 1.7 \cdot 10^{16} \, \mathrm{cm}$ for \MP\ using $\delta t_\mrm{var}=1.13 \, \mathrm{s}$, in rough consistency with the $\langle R_\mrm{Coll} \rangle = 2.4 \cdot 10^{16}$~cm obtained earlier (see also Table~\ref{tab:model_parameters}). \\
For the purpose of studying an event with smaller dissipation radius we introduce a modified version of \MP \, that has the same Lorentz factor distribution which is however ejected over a smaller engine active time $t_\mrm{eng}$. This yields a shorter variability timescale $\delta t_\mrm{var}$. Here $\delta t_\mrm{var} \equiv t_\mrm{eng}/N_\mrm{osc}$, where $N_\mrm{osc} = 30$ is the number of short-timescale oscillations introduced in the initial Lorentz factor distribution (see \reffig{initial_lorentz_multipulse}). The corresponding parameters are also listed in \reftab{models_characteristics}. The distribution of collisions is similar to \reffig{ediss_rcoll} except for being shifted to smaller radii by a factor 10. The typical collision radius in this case is $\langle R_\mrm{Coll} \rangle \sim 2.4 \cdot 10^{15}$~cm.

\section{Leptonic modeling of \SP}
\label{sec:leptonicmodel}
We begin our investigation with a leptonic radiation model for \SP, as this is the simplest scenario and is widely used under the name ``synchrotron self-Compton model''. The structure is further representative for a single pulse in a light curve composed of several pulses. We show results for the SYN- and the IC-dominated scenario introduced previously (see also Table~\ref{tab:model_parameters}).
To obtain a better understanding of the role of physical processes in shaping the overall spectrum, 
we commence with the presentation of results for the representative collision.  We then illustrate how the full spectrum is built up from the contributions of single collisions.
Finally we examine the predicted emission from the full burst on the relative contributions of different radiative processes (e.g. synchrotron and inverse Compton radiation from different particle populations) and the time at which they become observable.

\begin{figure*}
    \centering
    \includegraphics[width = 0.99 \textwidth, trim = 10 20 0 0]{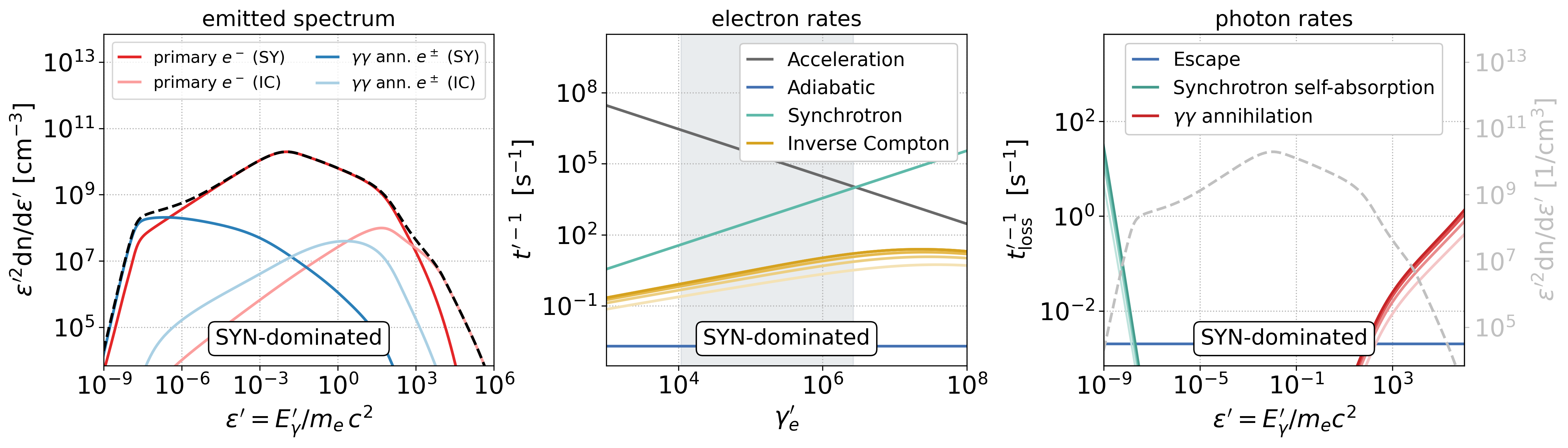}
    \\
    \vspace{3mm}
    \includegraphics[width = 0.99 \textwidth, trim = 10 20 0 0]{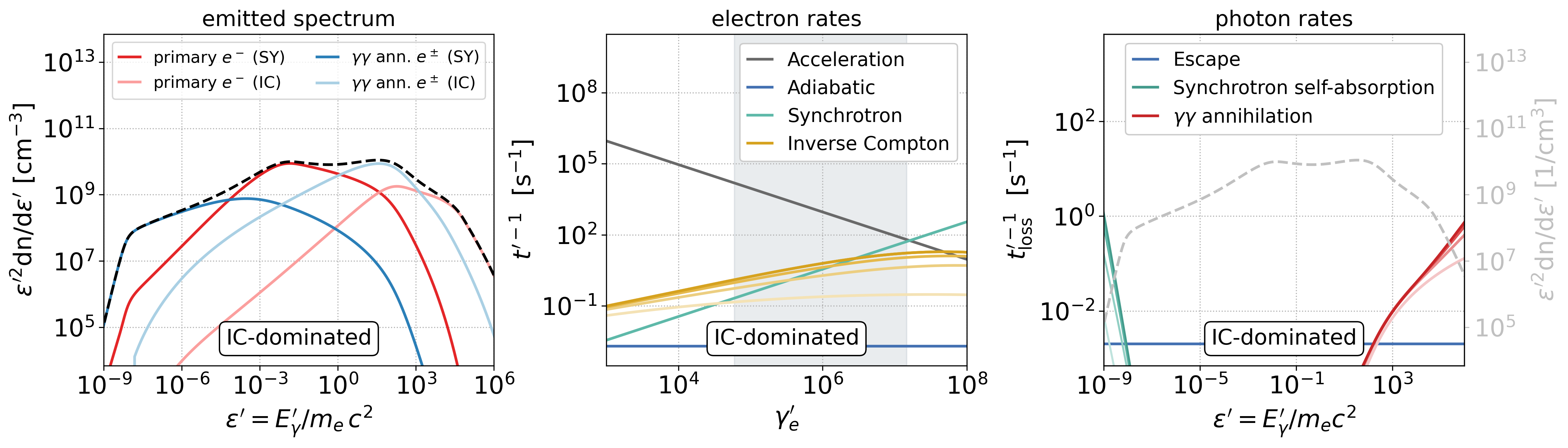}
    \caption{Examining the physical processes at play in the leptonic scenario 
    for the representative collision of the \SP \, prototype. We show the results for the SYN-dominated scenario (top row) and the IC-dominated scenario  (bottom row). \textit{Left:} decomposed emitted spectrum with synchrotron (SY) and inverse Compton (IC) contributions of primary electrons and secondary leptons produced in $\gamma \gamma$-annihilation, shown with different colors (see inset legends). \textit{Middle:} loss/acceleration rates of electrons as a function of electron Lorentz factor with the energy range of injected primaries shown as a grey shaded band. \textit{Right:} loss rates of photons with the emitted spectrum overplotted with a dashed grey line and units shown on the right $y$-axis. For time-dependent quantities we show the results for $t^\p \in \{0.25, 0.5, 0.75, 1.0\} t^\p_\mrm{dyn}$ where light (dark) colors correspond to early (late) times. 
    All quantities are in the comoving frame of the shocked plasma.
    }
    \label{fig:hl_coll345_leptonic_decomposed}
\end{figure*}

\subsection{Results for the representative collision}
For a better understanding of the physical processes at play we first present the modeling results for the representative collision (that corresponds to the point where the dissipated energy is maximal) for a SYN-dominated and a IC-dominated scenario. The collision parameters can be inferred from \reffig{ediss_rcoll} where the representative collision is marked with a star. The emitted (decomposed) spectrum and the loss rates for leptons and photons are shown in \reffig{hl_coll345_leptonic_decomposed}.

The comoving emitted photon spectra (left plots of \reffig{hl_coll345_leptonic_decomposed}) are decomposed into different components based on the radiating
particle species (primary electrons or secondary lepton pairs from $\gamma \gamma$-annihilation) and on the radiation process (synchrotron or inverse Compton). The inverse Compton component of a particle distribution is defined as the emission produced by those particles scattering 
off the complete photon distribution. 
In the SYN-dominated case (of strong magnetic fields) the primary synchrotron component dominates the overall broadband spectrum with contributions of secondaries being small appearing at low energies and at the high end of the spectrum.
On the other hand, in the IC-dominated case (of weaker magnetic fields) inverse Compton radiation of both primary and secondary leptons significantly reshapes the  broadband spectrum by increasing the radiative output at high photon energies (i.e. for $E'_\gamma > m_{\rm e}c^2$). Because electrons cool more efficiently via inverse Compton scatterings, the peak photon energy density of the primary synchrotron component decreases.

The plots in the middle panels of \reffig{hl_coll345_leptonic_decomposed} show the loss rates of electrons. 
The energy range of the injected primary distribution is overplotted for reference (grey band). 
Loss rates of time-dependent processes are shown at $t^\p \in \{0.25, 0.5, 0.75 \} t^\p_\mrm{dyn}$ with lighter (darker) colors referring to early (late) times. 
Naturally, the weaker magnetic field for the IC-dominated case demands a larger minimum electron Lorentz factor to reproduce the same synchrotron peak energy (recall that the synchrotron peak energy $E^\p_\mrm{syn} \propto \gamma^{\p 2}_\mrm{e, min} B^\p$). 
Moreover, because the synchrotron cooling rate scales as $B^{\p - 2}$, electrons can be accelerated to higher Lorentz factors for weaker magnetic fields (see Eq.~\ref{eq:tacc}). Consequently, the primary electron distribution is shifted to higher energies. 
While the inverse Compton cooling rate is
sub-dominant at all times and energy ranges for the SYN-dominated case, it overcomes the synchrotron cooling rate
for low- to medium-energy electrons for the IC-dominated case. 
Indeed, the electrons around $\gamma^\p_\mrm{e, min}$ cool predominantly through inverse Compton scatterings for $t^\p \gtrsim 0.25 t^\p_\mrm{dyn}$. As these tend to occur in the
Klein-Nishina regime, the 
dependence of the cooling rate on the electron Lorentz factor differs from the synchrotron one. As pointed out in earlier works \citep[e.g.][]{Daigne:2010fb}, this can lead to a synchrotron spectral shape that
differs from the classical synchrotron fast- or slow-cooling predictions: the resulting spectral index is steeper, while energy extraction from electrons is still efficient.

Finally, on the right-hand side panels we show the photon loss rates as a function of photon energy. For reference, we also indicate the emitted photon spectrum (with units shown on the right $y$-axis). The photon fields are shaped by synchrotron self-absorption (SSA) at the lowest energies and $\gamma \gamma$-annihilation at the highest energies. The SSA dimensionless energy here is at $\sim ~10^{-8}$, which in the observer's frame (assuming $\Gamma \simeq 500 $) corresponds to energies of roughly 1~eV that are observed in the optical band. On the other hand, $\gamma \gamma$-annihilation dominates over escape at a dimensionless energy of $\sim ~10^2$ that is approximately 10~GeV in the observer's frame. Both loss rates evolve little with time. The position of the spectral breaks can be inferred from the intersection of the respective loss rates with the escape rate. We point to \refapp{buildup_spectra} for the time-evolution of photon spectra in the collision, which further illustrates e.g. the suppression of high-energy photons due to $\gamma \gamma$-annihilation.
\begin{figure}
    \centering
\includegraphics[width = 0.47\textwidth]{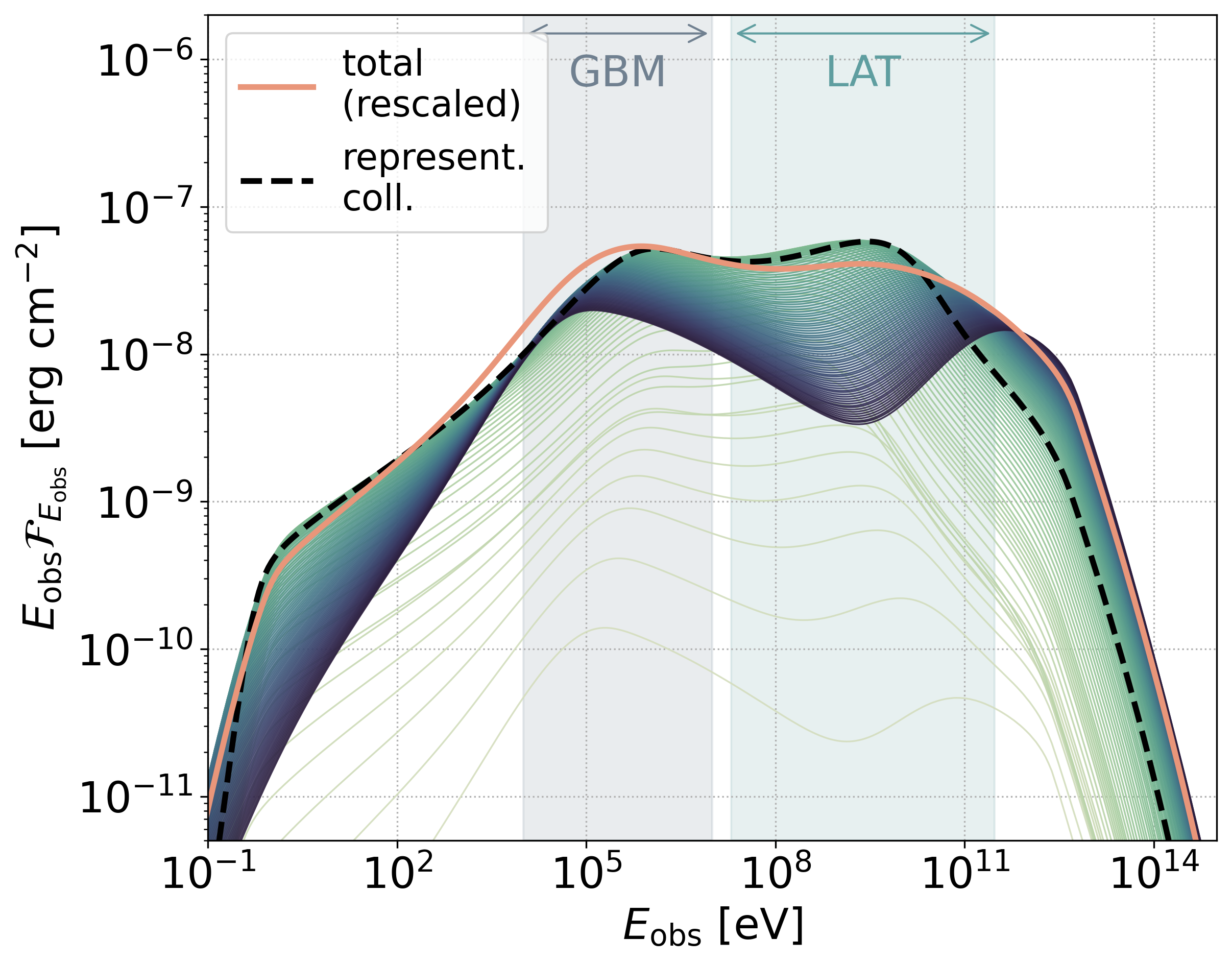}
    \caption{Decomposition of the full burst photon spectrum of \SP into the contributions of single collisions (IC-dominated).
    The thick orange line corresponds to the overall time-integrated spectrum that is obtained by summing 
    the contributions of all single collisions (rescaled by the ratio of the energy dissipated in the representative collision and the total dissipated energy $E^\mathrm{*}_\mathrm{diss}/ E^\mathrm{tot}_\mathrm{diss}$ for easier comparison). Thin colored lines show the spectra of every 10th collision, with light green (dark blue) curves corresponding to early (late) collisions. The spectrum of the representative collision is indicated with a dashed black line. Shaded bands indicate the \textit{Fermi}-GBM and LAT bands.
    All quantities are in the observer's frame assuming a redshift of $z=2$. EBL attenuation is not included here.}
    \label{fig:single_collisions_full_burst}
\end{figure}

\subsection{The build-up of the full burst spectrum}
From the spectrum of the representative collision we now move to the full burst spectrum. This is obtained by summing the contributions of all single collisions.
This is illustrated in  \reffig{single_collisions_full_burst}, where we show the time-integrated, full spectrum (rescaled to approximately match the fluence of single collision spectra) of the IC-dominated scenario of \SP. Thin curves show the contribution of every 10th collision and the colour-coding indicates the observed time of the collision: light green spectra correspond to early collisions, dark blue indicate collisions to later times. The spectrum of the representative collision discussed in the last section is indicated with a dashed line. 

Although the sub-MeV synchrotron peak energy is by construction comparable in all single-collision spectra (recall that we chose $\zeta_\mrm{e} = \mrm{const.}$, which results in an almost constant $\gamma_\mrm{e, min}^\p$ throughout the fireball evolution), the shape and fluence varies for the different collisions. For example, the HE ($>1$~GeV) emission is largely powered by late collisions. This may be explained by differences
in $\gamma \gamma$-absorption opacity in the region of the emitting plasma:
close to the source, high densities increase the optical depth to $\gamma \gamma$-absorption, effectively hindering HE photons from escaping. At larger distances from the central engine, lower densities enable the escape of photons of higher energies.
On the other hand, the low-energy spectrum in the eV range is mostly shaped by early collisions. The reason for this can be understood from the same reasoning: the high $\gamma \gamma$-absorption efficiency in early collisions results in a large number of secondary lepton pairs that contribute through synchrotron radiation at low energies (see also \reffig{hl_coll345_leptonic_decomposed}). 

Comparing the rescaled overall spectrum to the spectrum of the representative collision we find that the sub-MeV synchrotron peak of the total spectrum is broadened by the contributions of many collisions. The structures at higher energies are equally washed out, and the overall spectrum extends to higher energies than the one of the representative collision due to contributions from collisions occurring at even larger distances. 
These results highlight the importance of moving from single-collision to multi-collision models for an accurate description of the broadband photon spectrum.

\subsection{Full-burst decomposed spectra and light curves}
\begin{figure*}
    \captionsetup[subfigure]{labelformat=empty}
\includegraphics[width = 0.47 \textwidth]{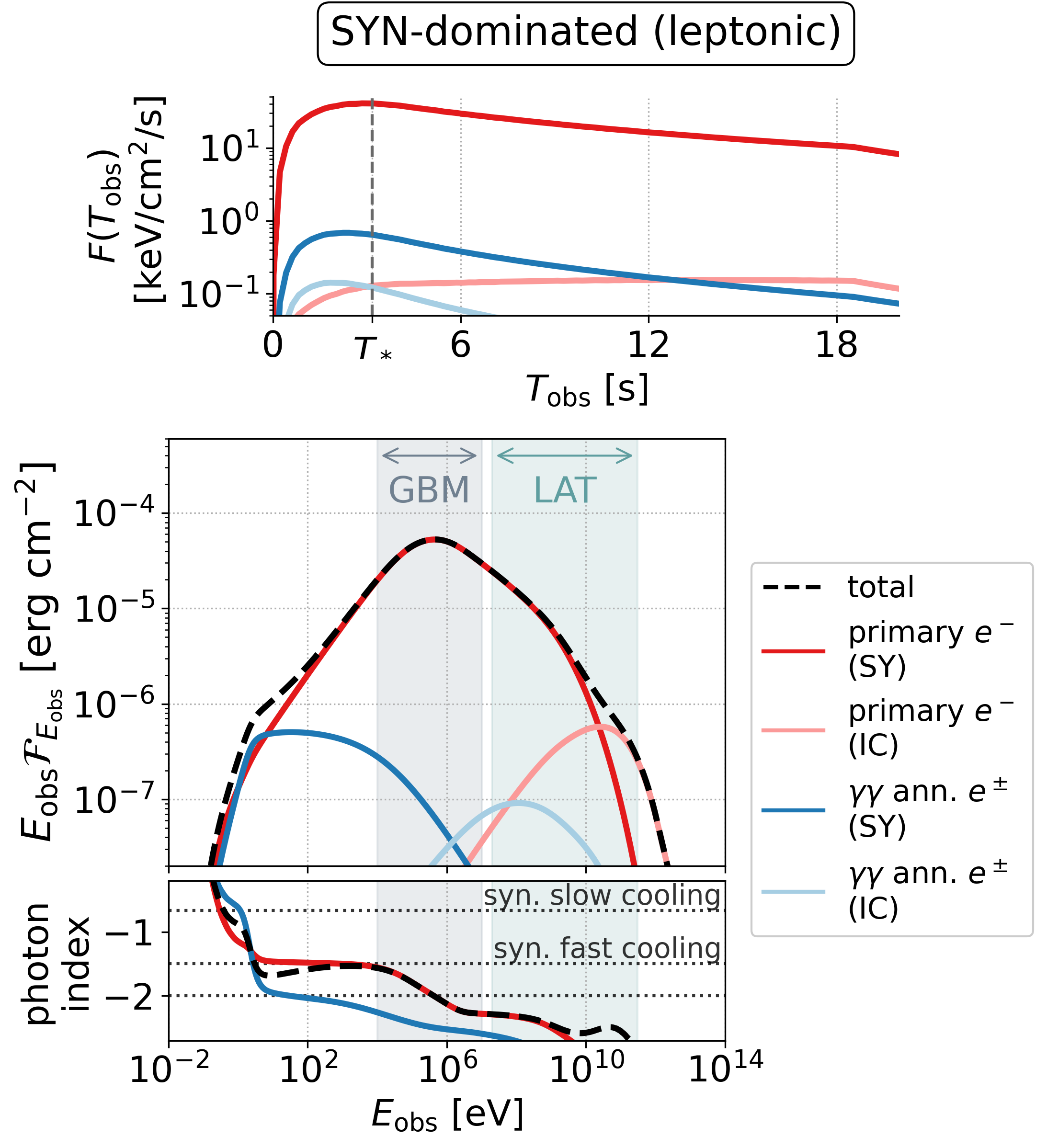}
    \hspace{0.8cm}
    \includegraphics[width = 0.47 \textwidth]{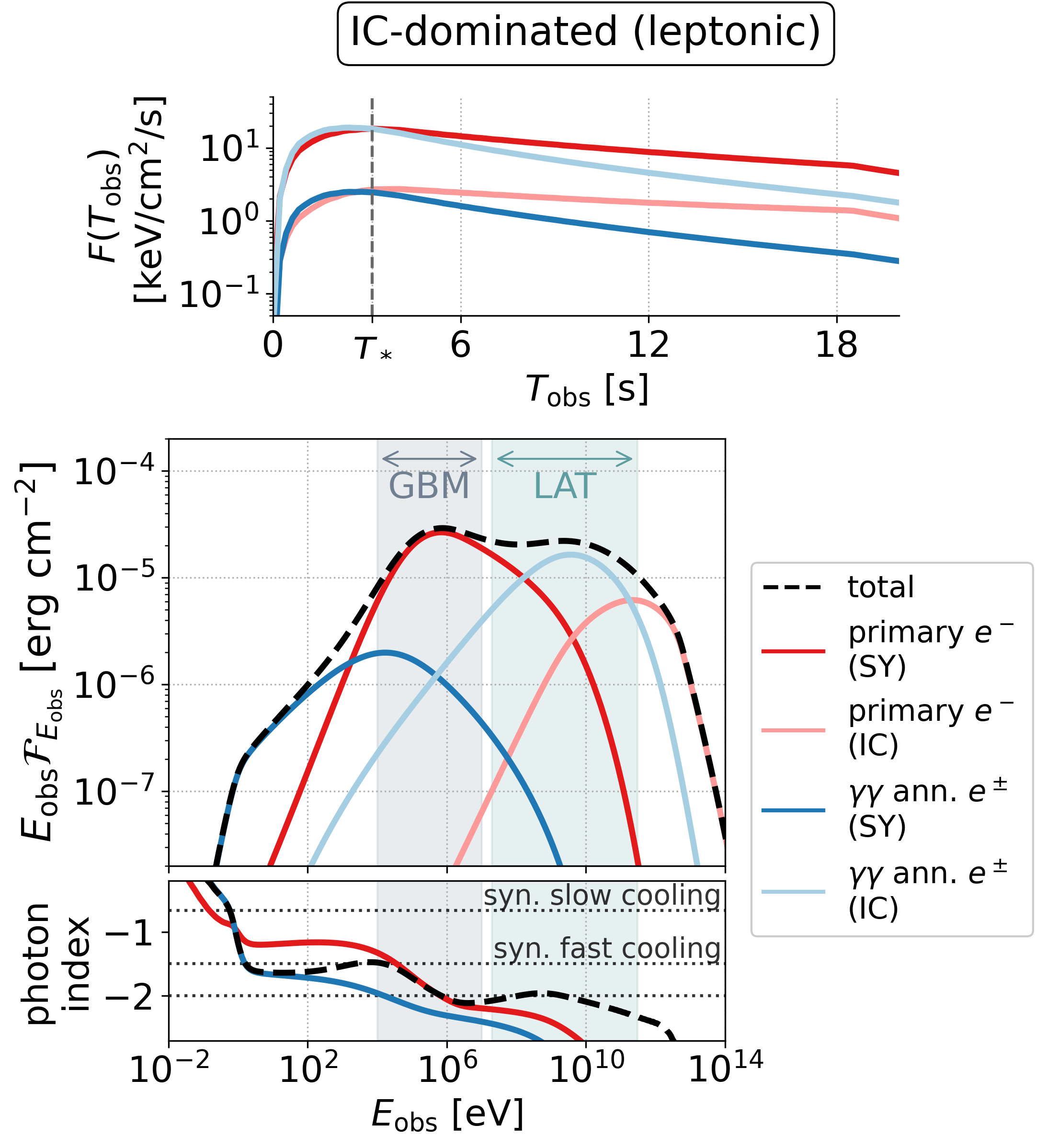}
    \caption{Full-burst decomposed light curve, spectra and photon indices for the \SP leptonic model, examining \textit{(left)} the SYN-dominated scenario and \textit{(right)} the IC-dominated scenario. We show the synchrotron (SY) and inverse Compton (IC) of primary electrons and secondary lepton pairs from $\gamma \gamma$-annihilation. In the upper plots we show the fluence as a function of observed time for the different components. In the lower plots, shaded bands indicate the \fermi-GBM and LAT bands. For the photon indices in the lower panels, we only show the SY contributions of primary and secondary leptons. We further indicate the photon indices for the synchrotron slow-cooling (-2/3) and fast-cooling (-3/2) regimes as well as a photon index of $-2$ (that marks peaks of $E_\mrm{obs} \mathcal{F}_\mrm{E_{obs}}$) as dotted lines.}
    \label{fig:hl_spectra_leptonic_decomposed}
\end{figure*}

We finally proceed to evaluate the full burst results for the SYN- and IC-dominated scenarios introduced before. In \reffig{hl_spectra_leptonic_decomposed} we display the full burst spectra and light curves, decomposed in a similar manner as the representative collision before. 

The contributions of the single emission processes to the full burst spectra are qualitatively similar to those of the representative collision for both scenarios (compare to \reffig{hl_coll345_leptonic_decomposed}); we point out that the superposition of inverse Compton emission from many collisions leads to a relatively flat (i.e. $\propto E_{\rm obs}^{-2}$) HE spectrum for the IC-dominated scenario. As for the representative collision, the synchrotron peak is decreased in fluence. This underlines that by examining the representative collision we can indeed gain some understanding for the processes that shape the full spectrum.

The lower panels of \reffig{hl_spectra_leptonic_decomposed} show the photon index as a function of observed photon energy. In the SYN-dominated case
the photon index below the synchrotron peak mostly equals the fast-cooling synchrotron prediction of $-3/2$ and is softened by contributions of secondary lepton pairs only around the synchrotron self-absorption break at $\sim$eV energies. 
In the IC-dominated case the low-energy photon index (in the range $10-10^4$~eV) is 
determined by the synchrotron emission of both primary and secondary electrons. The primary electron synchrotron spectrum has a photon index of $\approx -1$, because electrons radiating at these energies cool mainly via inverse Compton scatterings in the Klein-Nishina regime \citep{Nakar:2009er, Daigne:2010fb, Duran:2012ww}. Still, secondary electrons cool mostly via inverse Compton scatterings in the Thomson regime, thus resulting in a photon index $\approx -1.5$. As a result, the combined synchrotron spectrum has a photon index which is $\lesssim -1.5$ at low energies. Our results highlight the importance of a complete radiation treatment invoking also secondary particles and their emission, which can reshape the photon spectra. 

Besides the question \textit{at which energy} different radiation processes contribute, one may ask \textit{at which time} these processes dominate the overall emission.
To investigate this aspect we show the time-dependent energy fluxes in the upper panels of \reffig{hl_spectra_leptonic_decomposed} for the SYN-dominated and IC-dominated scenarios.  
By construction, the flux of the primary synchrotron emission peaks at the observed time of the representative collision (which is when the dissipated energy is maximal). 
The flux created by secondary leptons (both synchrotron and inverse Compton emission) peaks at an earlier time. We recall that the flux at these early times is produced in collisions close to the source (see \reffig{ediss_rcoll}), where the densities and consequently the optical depth to $\gamma \gamma$-absorption are high. The inverse Compton emission of primary electrons peaks latest. This is most evident for  the SYN-dominated scenario, even though the actual contribution of this component to the overall spectrum is negligible at all times. 
For the IC-dominated scenario the inverse Compton emission of primary electrons peaks around the time of the representative collision $T_*$, while inverse Compton emission of secondary leptons peaks before $T_*$. As they peak at different energies, an energy-dependent shift in peak time can thus be expected: the HE emission in the LAT range peaks \textit{before} the keV synchrotron peak, the flux at the highest energies peaks \textit{after} the keV synchrotron peak~\citep[see also][]{Asano:2012jr}. We point out that the early peak due to secondary leptons, was not present in \cite{Bosnjak:2008bd} who considered scenarios of low $\gamma \gamma$-opacity.

\section{Lepto-hadronic models}
\label{sec:leptohadronicmodel}
Hadronic signatures have been explored as the potential origin of the HE component that was in some cases observed by the \textit{Fermi}-LAT \citep{Asano:2012jr, Wang:2018xkp}. In the same spirit as those studies, we investigate lepto-hadronic scenarios paying special attention to HE emission. In contrast to before, this section will include results for \textit{both} prototypes, \SP \, and \MP.

In the leptonic study (see previous section), no HE enhancement was found for the SYN-dominated scenario. In contrast to this, IC-dominated scenarios showed an increased HE emission, due to inverse Compton scatterings of primary and secondary leptons. Motivated by these findings, we again select the SYN- and IC-dominated scenarios to investigate lepto-hadronic models. 

As before, we will first show the results for the representative collision before moving to the emission from the full burst. While our fiducial choice of baryonic loading is $f_\mrm{p/e} = 30$, we will further explore $f_\mrm{p/e} \in \{0, 10, 30, 100 \}$ and discuss the impact of the variability timescale for \MP.

\subsection{\SP: Results for the representative collision}
\begin{figure*}
    \centering\includegraphics[width = 0.7 \textwidth]{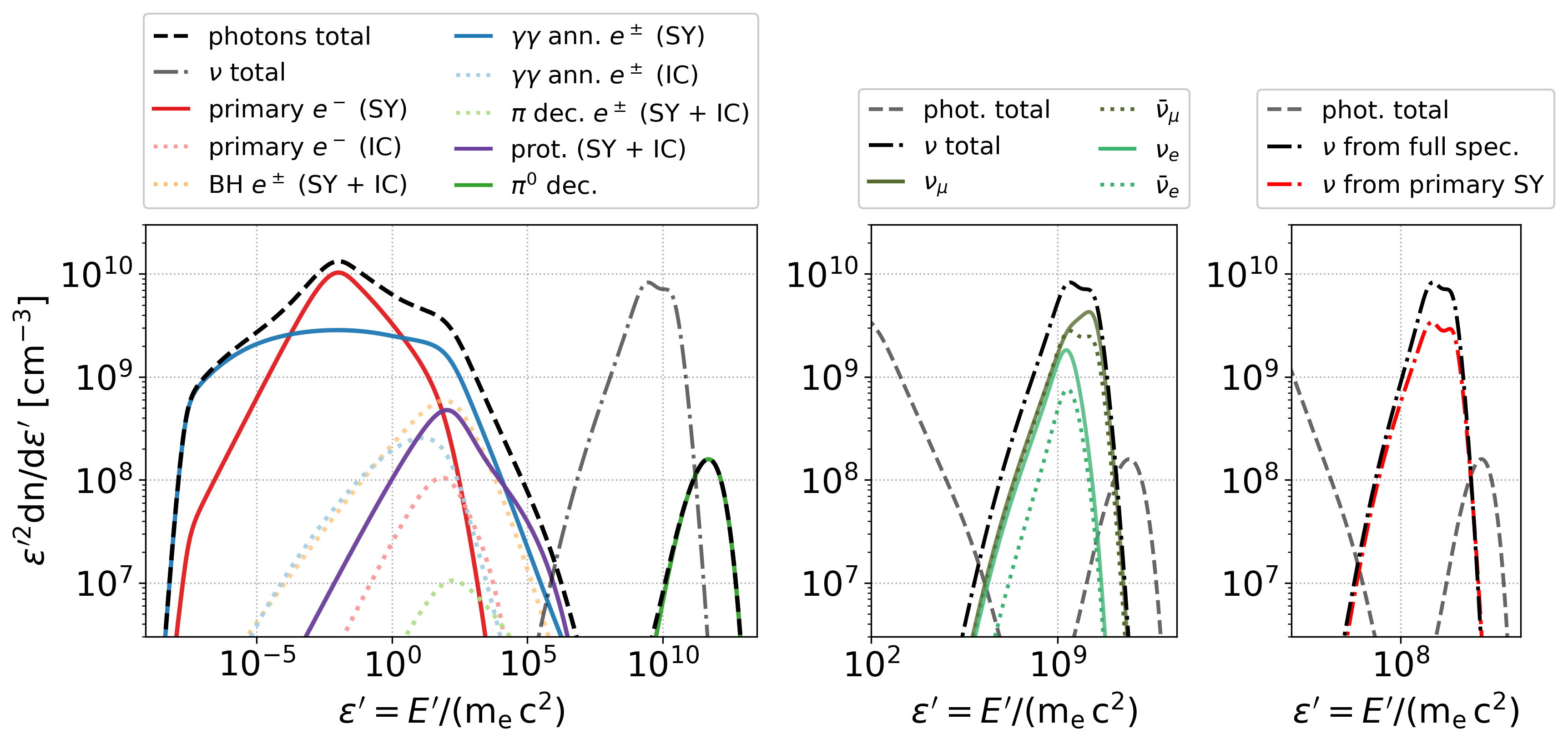}
    \\
    \vspace{3mm}\includegraphics[width = 0.7 \textwidth]{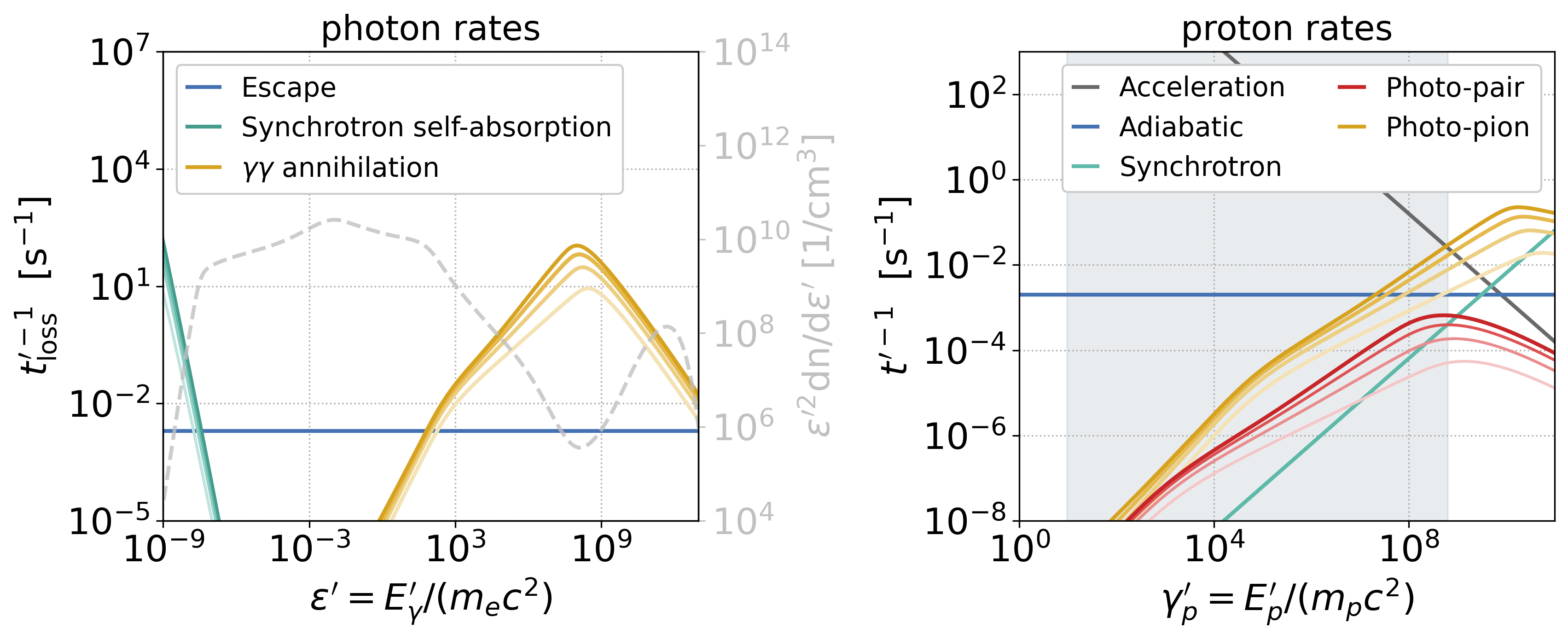}
    \caption{Examining the physical processes at play in a lepto-hadronic scenario for the representative collision of \SP \, (SYN-dominated, $f_\mrm{p/e} = 30$).
    Upper panels: decomposed comoving spectrum of photons \textit{(left)} and neutrinos \textit{(middle, right)}. The red dashed-dotted in the right plot corresponds to the neutrino spectrum from the interactions of protons with the primary electron synchrotron component only. \textit{Lower left panel:} Loss rates of photons with the emitted spectrum over-plotted with a dashed grey line and units shown on the right $y$-axis. For time-dependent quantities  we show the results for $t^\p \in \{0.25, 0.5, 0.75, 1.0\} t^\p_\mrm{dyn}$ with progressively darker colors. \textit{Lower right panel:} Loss/acceleration rates of protons as a function of the proton Lorentz factor with the energy range of injected protons shown as a grey shaded band. 
    }
    \label{fig:hl_coll345_hadronic_decomposed}
\end{figure*}
The comoving photon and neutrino energy densities for the SYN-dominated case are shown in the upper panels of \reffig{hl_coll345_hadronic_decomposed}. The left-hand side plot shows the broadband photon spectrum and its decomposition into various components:
primary electrons, $e^\pm$ from  photo-pair production (labelled as Bethe-Heitler, BH), secondary $e^\pm$ from $\gamma \gamma$-annihilation, $e^\pm$ from pion decays, primary protons, and photons from neutral pion decays. We indicate the synchrotron (SY) and inverse Compton (IC) emission from charged particles. Dominant contributions to the overall spectrum are shown as solid lines while sub-dominant contributions are plotted with dotted lines. 

For the chosen baryonic loading, the synchrotron emission of secondary pairs from $\gamma \gamma$ annihilation follows a $n_{\rm e}'(\gamma'_e) \propto \gamma_e^{'-3}$ distribution. This creates a broad flat spectrum that dominates over all other hadronic-related contributions. Similar features were reported in earlier works on hadronic emission models for GRBs \citep{Asano:2014nba, Petropoulou:2014awa, Wang:2018xkp}. Contrary to the corresponding leptonic SYN-dominated scenario, the injection rate of pairs from $\gamma \gamma$ annihilation is higher in this case, because of the high luminosity of VHE photons. These photons are mainly produced from $\pi^0$ decays. The fact that the pion bump at $\varepsilon' \sim 10^{12}$ is lower in normalization than the all-flavour neutrino bump indicates the amount of attenuation; this can be also seen by the attenuation rate shown in the panel with the photon rates.

The all-flavour neutrino spectra are shown together with the photon spectra in the upper left panel, and per-flavour ($\nu_\mu$, $\bar{\nu}_\mu$,$\nu_e$, $\bar{\nu}_e$) in the upper middle panel. The neutrino peak energy and spectral shape differ by species and depend on the parent~\citep{Baerwald:2010fk}, with $\nu_\mu$ having the highest peak energies. The $\bar{\nu}_\mu$ spectrum extends to equal energies as the $\nu_\mu$ spectrum, but peaks at a lower energy. The ${\nu}_{\rm e}$ and $\bar{\nu}_{\rm e}$ spectra are similar to each other and have peak energies similar to the $\bar{\nu}_\mu$ spectrum, although not extending to similarly high energies. Interestingly, the neutrino spectra peak at much lower energy than the $\pi^0$-decay photons. We attribute this difference to two effects:
\begin{itemize}
\item \textit{Cooling of intermediate pions and muons.} Charged intermediate secondaries are subject to adiabatic and synchrotron cooling. For the high magnetic fields of the SYN-dominated scenario, both species cool via synchrotron radiation before decaying. This introduces a parent-dependent cooling break in the neutrino spectra, see e.g. \citet{Waxman:1998yy,Lipari:2007su,Baerwald:2010fk,Baerwald:2011ee, Tamborra:2015qza, Bustamante:2020bxp}, and reduces their energy with respect to the parent proton energy. The longer decay time of muons and their larger synchrotron cooling rate enhance this effect on the spectra of electron neutrinos (produced in muon decays). This effect and its manifestation throughout the fireball evolution are discussed in detail in \refapp{pion_muon_cooling_neutrinos}.
\item \textit{Attenuation of photons vs. neutrinos.} Neutrinos as free-streaming particles reflect the in-source distribution throughout the complete evolution. This is different for photons, which are subject to $\gamma \gamma$-annihilation. In a single collision, HE photons above $\sim$GeV can escape more easily at small $t^\prime$ when the radiation densities are still low (see rates in the lower panels of \reffig{hl_coll345_hadronic_decomposed}). At these early $t^\p$, the low radiation densities also result in a low pion-production efficiency. This in turn enables large proton maximal energies, which evolve to lower energies as the densities increase (see proton spectra and photo-pion rates in \reffig{hl_coll345_hadronic_decomposed}). Thus, the escaping VHE photon spectrum is dominated by early in-source spectra with high maximal proton energies, whereas the neutrino spectra capture the complete evolution up to late $t^\prime$ when the maximal proton energies are lower. For further illustration of this, we also point to \refapp{buildup_spectra} and \reffig{cooling_impact_on_neutrinos}, where we show the time evolution of single collision spectra.
\end{itemize}

It is also interesting to note that the neutrino production can be higher (and the spectra peak at higher energies) than in the simplest one zone models. 
First of all, cooling effects on the secondaries are not as strong as for models with smaller collision radii, which means that the cooling breaks are at comparatively high energies. 
Then, consider that the spectral index below the spectral peak in the synchrotron fast cooling regime (-3/2) is softer than the one typically inferred from observations (-1). This implies that the dominant energy relevant for the pion production, i.e., the energy where the photon number density peaks, can be found below the photon break at $\epsilon'_{\gamma,\mathrm{eff}} = \mathrm{max} \left[ y_\Delta \, m_p/(20 \, E'_{\nu,\mathrm{peak}}),\epsilon'_{\gamma,\mathrm{min}} \right]$ (following case~2 in \citet{Fiorillo:2021hty}). Here $y_\Delta \simeq 0.2-0.5 \, \mathrm{GeV}$ is a suitable choice for the pitch angle-averaged cross section (see Fig.~4 in \citet{Hummer:2010vx}) and $\epsilon'_{\gamma,\mathrm{min}}$ corresponds to the synchrotron self-absorption cutoff in our case.\footnote{Compared to \citet{Fiorillo:2021hty}, the neutrino peak neutrino energy has been used instead of the maximal proton energy, since cooling effects of the secondaries can be included in that way.} Here we are limited by the first term, as can be seen in \reffig{hl_coll345_hadronic_decomposed}, lower right panel:  the photo-pion rate still increases monotonously even beyond the maximal proton energy, and the synchrotron self-absorption induced break lies significantly beyond. In order to compare to \reffig{hl_coll345_hadronic_decomposed}, upper left panel, we can derive the corresponding parameter $\varepsilon'_{\gamma,\mathrm{eff}} \simeq 3\, \cdot 10^{-5}$ (for $\varepsilon'_{\nu,\mathrm{peak}} \simeq 10^9$ and $y_\Delta \simeq 0.5 \, \mathrm{GeV}$); this value is in the region where the blue (electromagnetic cascade)  and red (primary synchrotron) curves  intersect -- about two orders of magnitude in energy below the peak. Thus there are two effects which enhance the neutrino production: a) the photon number density at that point is about an order of magnitude higher than for a spectral index -1, and b) photons from the secondary cascade enhance the neutrino production by a factor of a few. We show the latter effect in \reffig{hl_coll345_hadronic_decomposed}, upper right panel, where the contribution from the primary synchrotron spectrum is shown separately; here the effect is about an additional factor of 2-3 enhancement, where the additional impact of the secondary cascade depends on the hardness of the  primary spectrum and the baryonic loading.

\subsection{\SP: Full-burst decomposed spectra and light curves}
\begin{figure*}
    \centering
    \includegraphics[width = 0.49 \textwidth]{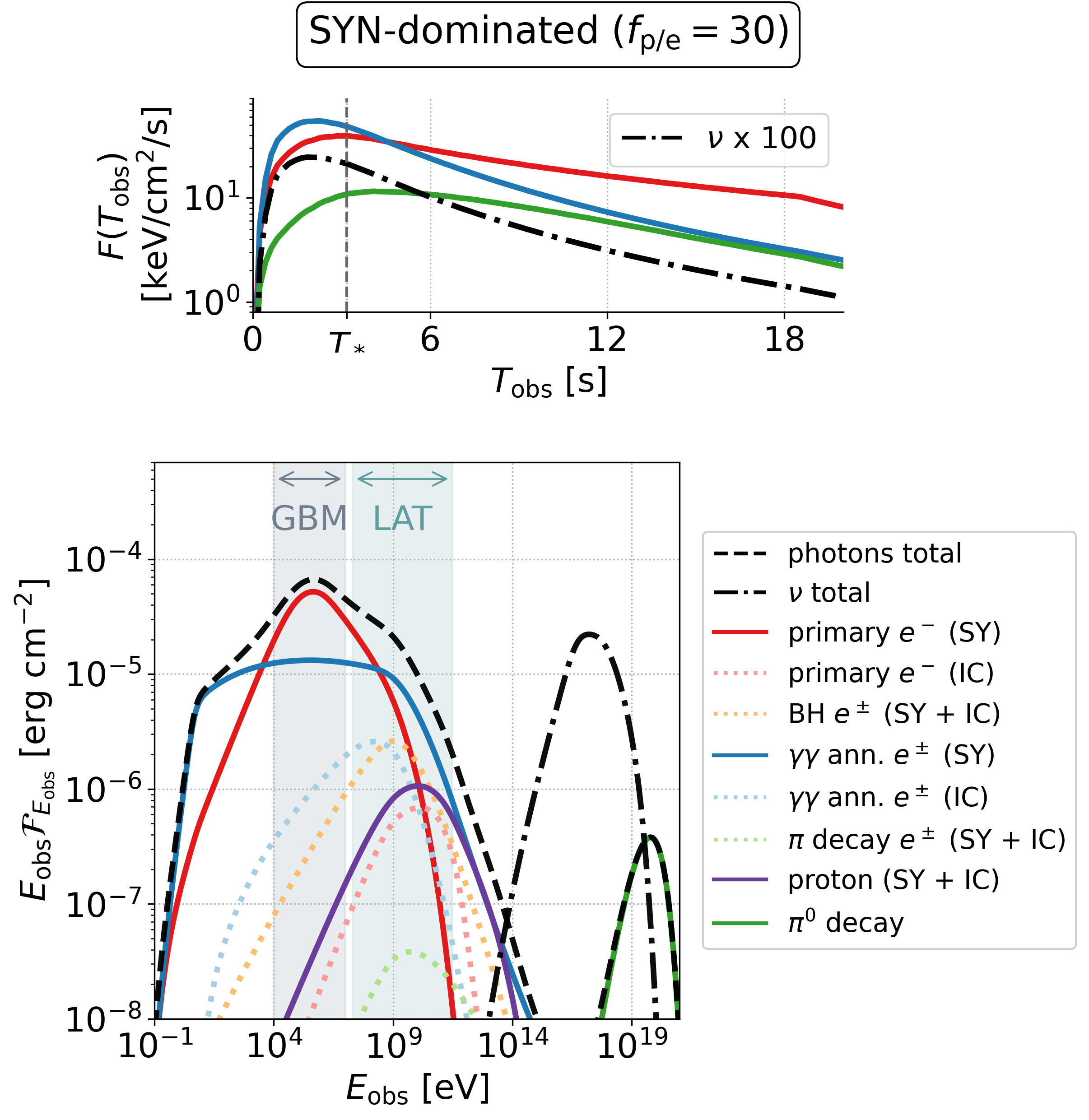}
    \includegraphics[width = 0.49 \textwidth]{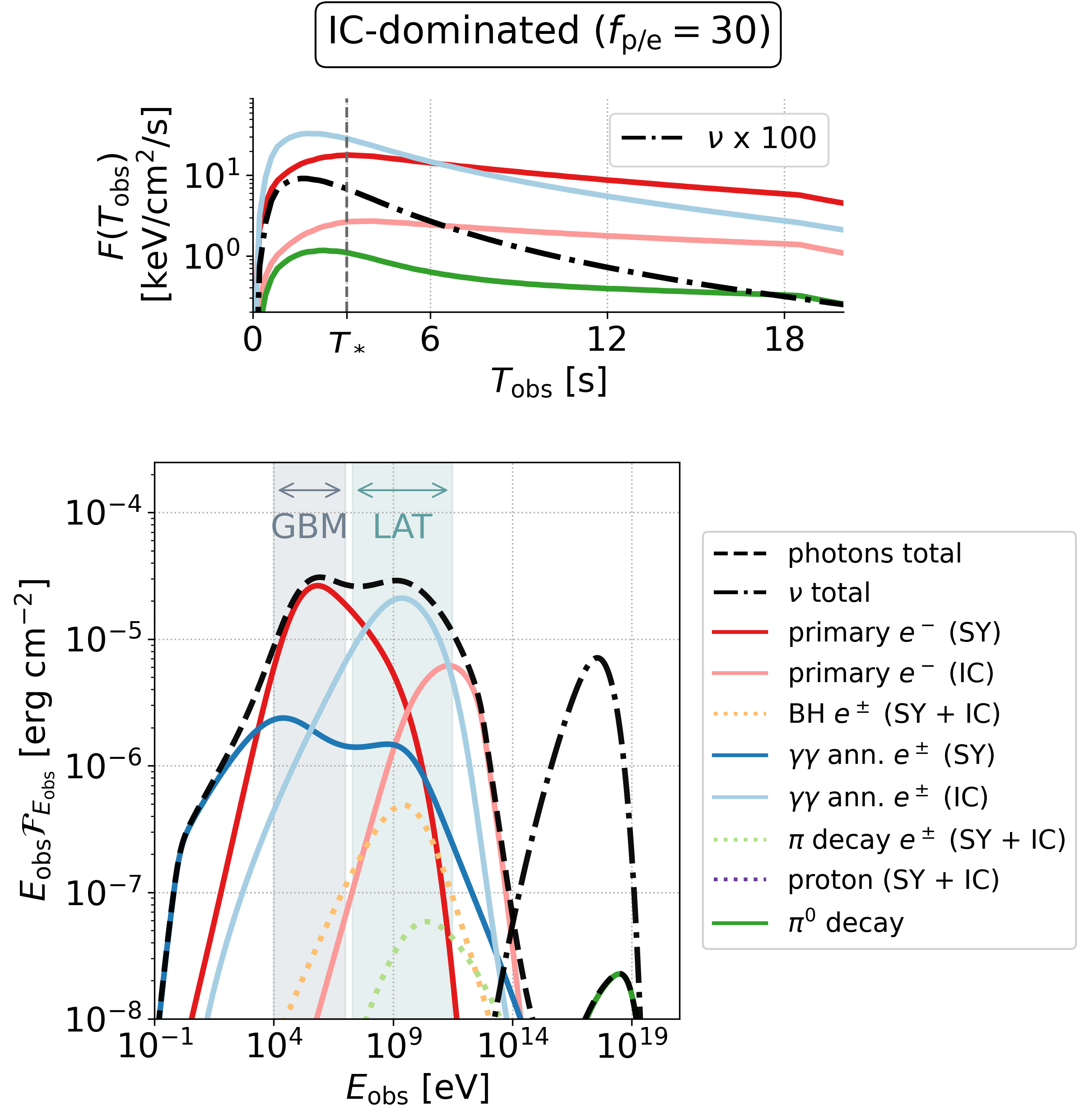}
    \caption{ 
    Full-burst decomposed light curves and spectra for the \SP lepto-hadronic model, examining \textit{(left)} the SYN-dominated scenario and \textit{(right)} the IC-dominated scenario with $f_\mrm{p/e} = 30$. Coloured lines show various contributions to the total spectrum which is plotted with dashed line (for details, see legends). The all-flavour neutrino fluences and fluxes are overplotted with dash-dotted black lines. For the light curves, the neutrino fluxes we scaled up by a factor 100 to match the same scale. 
     Shaded regions indicate the energy ranges of the \fermi-GBM and LAT detectors.
     In the energy flux light curves the dashed vertical line indicates the observed time of the  representative collision, marked with a star in \reffig{ediss_rcoll}.
     }
    \label{fig:leptohadronic-decomposed}
\end{figure*}

The simulated full-burst photon and all-flavour neutrino spectra, as well as the corresponding light curves for the SYN-dominated and IC-dominated scenarios are shown in the left and right panels of \reffig{leptohadronic-decomposed}, respectively. Coloured lines indicate the various components that make up the total spectrum (see inset legend for details). 

Again commencing with a discussion of the spectra, we find that the spectral features of the full-burst SYN-dominated spectrum are similar to those of the representative collision discussed in the previous section (see also \reffig{hl_coll345_hadronic_decomposed}). On the other hand, the neutrino peak properties relative to the photon peak are slightly different: First, the fluence of the neutrino peak relative to the photon peak fluence is $\sim 45$~\% higher for the representative collision than for the complete burst. Also, in the representative collision the neutrino peak energies are $\sim 25$~\% lower than for the full burst. Thus, if we scale up the neutrino spectra from the representative collision to the full burst we \textit{overestimate} the fluence, while we \textit{undererstimate} the peak energy. Both these effects increase the detection perspectives by instruments like IceCube, or, for non-detection, increase potential conflicts with neutrino limits.

In the IC-dominated case the broadband spectrum differs from the SYN-dominated case, and resembles that of the pure leptonic scenario (compare to  \reffig{hl_spectra_leptonic_decomposed}). Because of the lower $f_{\rm B/e}$ value, pairs injected by $\gamma \gamma$-annihilation are predominantly cooling via inverse Compton scatterings. As a result, the associated inverse Compton component is much brighter than their synchrotron component, and potentially modifies the spectrum in the \fermi-LAT energy range. For the selected parameters, the secondary inverse Compton emission again outshines the primary inverse Compton component. The VHE peak, which is associated with the $\pi^0$ decays, has a much lower peak fluence than in the SYN-dominated case. We attribute this to two effects: Firstly, a lower maximum proton energy (which can be inferred from the lower peak energy of the pion bump) results in a reduced pion production efficiency. This lower pion production efficiency is also reflected in the lower neutrino fluxes.
The lower maximum proton energy is driven by the slower acceleration in the weaker magnetic field, whereas the dominant loss processes are independent of the magnetic field (in contrast to electrons, where the weaker magnetic field for the IC-dominated scenario enables higher $\gamma_\mathrm{e, max}$). Second, the opacity to $\gamma \gamma$ annihilation is higher around the VHE peak due to the lower peak energy (compare to \reffig{hl_coll345_hadronic_decomposed} lower left). This is indicated by the higher difference in energy flux of neutrinos and $\gamma$-rays when compared to the SYN-dominated case.

The temporal evolution of the observed fluxes of various components in the SYN- and IC-dominated scenarios is shown in the upper panels of \reffig{leptohadronic-decomposed}.
It is useful to recall at this point that small $T_\mrm{obs}$ correspond to small collision radii $R_\mrm{Coll}$, small shell volumes, and high particle densities (see \reffig{ediss_rcoll}). Starting with the SYN-dominated, we find that the primary electron synchrotron flux peaks at $T_*$ as expected; the dissipated energy, a fraction of which is transferred to primary electrons, becomes maximal at this time. However, the synchrotron emission of secondary pairs from $\gamma \gamma$-annihilation and the neutrino emission reach their maximum flux at earlier times. This early emission originates closer to the central engine where radiation densities are higher. This naturally enhances the efficiency of density-dependent processes, such as $\gamma \gamma$-annihilation and photo-pion production. While the latter process is more efficient at earlier times, the photon flux from $\pi^0$ decays peak a little later, when the low-energy photon densities decrease, thus leading to a suppression of the in-source $\gamma \gamma$-annihilation rate. These results are in agreement with the findings of \cite{Bustamante:2016wpu}, where neutrinos were found to originate from small radii (where the optical thickness to photohadronic interactions is high) and VHE $\gamma$-rays from large radii (where the  $\gamma \gamma$ optical thickness is low). 

Similar trends are found in the IC-dominated case, except for earlier peak time of the $\pi^0$ photon flux.
We recall that $\pi^0$ flux depends on both the $\gamma \gamma$-annihilation and the pion production efficiency. In the SYN-dominated scenario the early flux is suppressed by $\gamma \gamma$-annihilation and thus peaks at later times. On the other hand in the IC-dominated scenario the pion production efficiency in late collisions is low, which suppresses the $\pi^0$ photon flux at late times.

\subsection{Investigating different baryonic loadings} 

\begin{figure*}
    \captionsetup[subfigure]{labelformat=empty}
    \centering
    \includegraphics[width = 0.49 \textwidth]{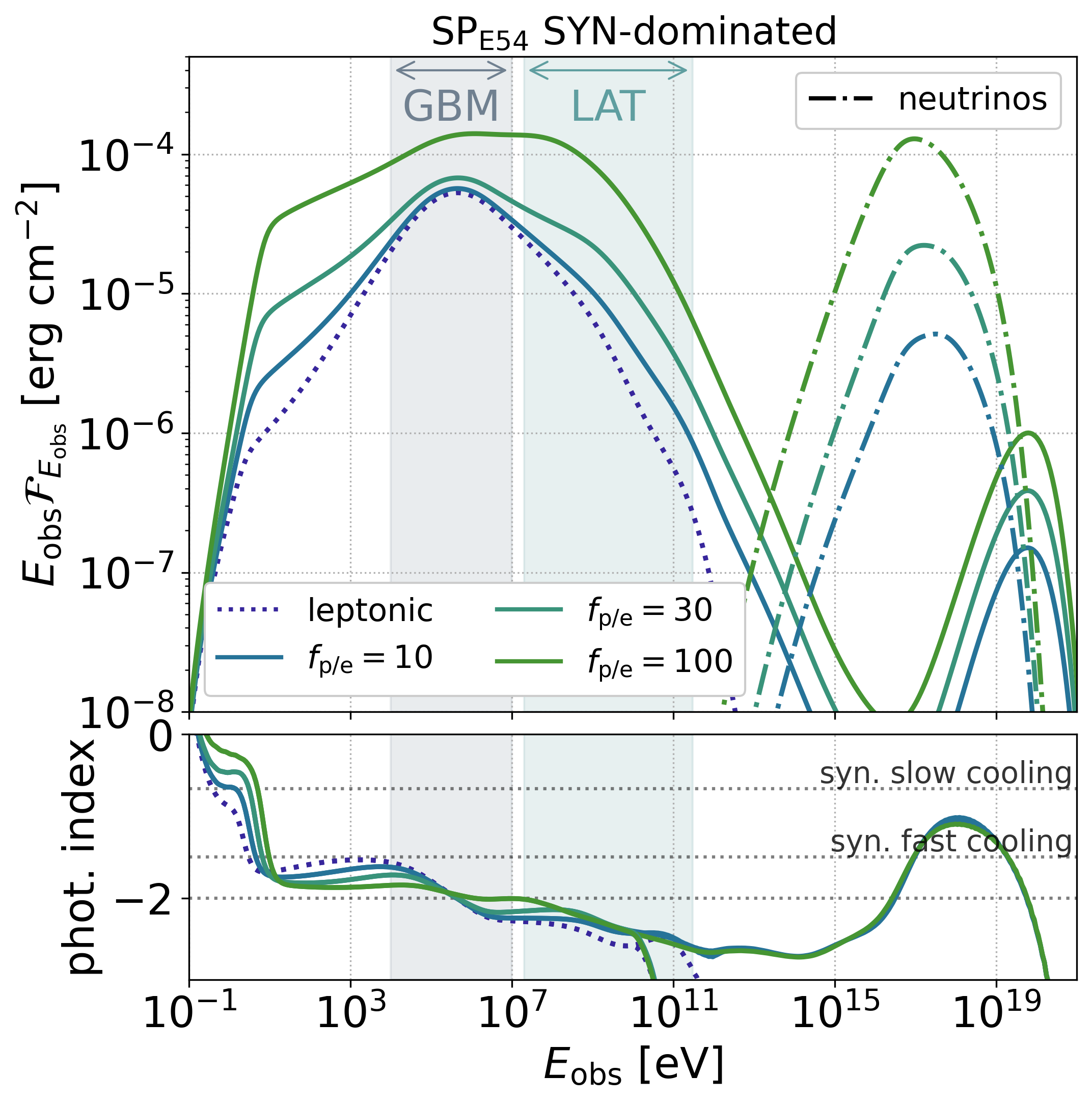}
    \hfill
    \includegraphics[width = 0.49 \textwidth]{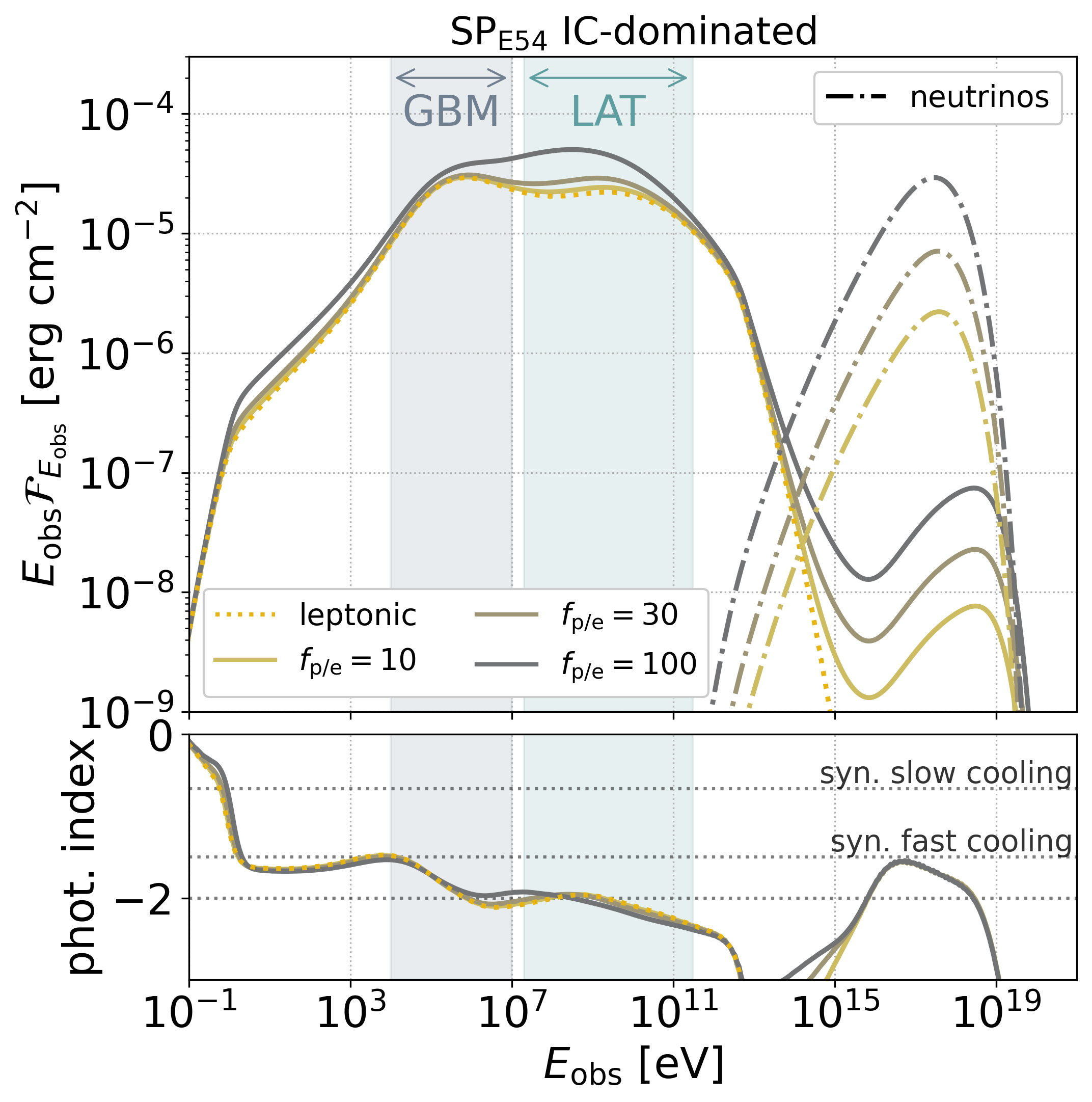}
    \\
    \vspace{3mm}
    \includegraphics[width = 0.49 \textwidth]{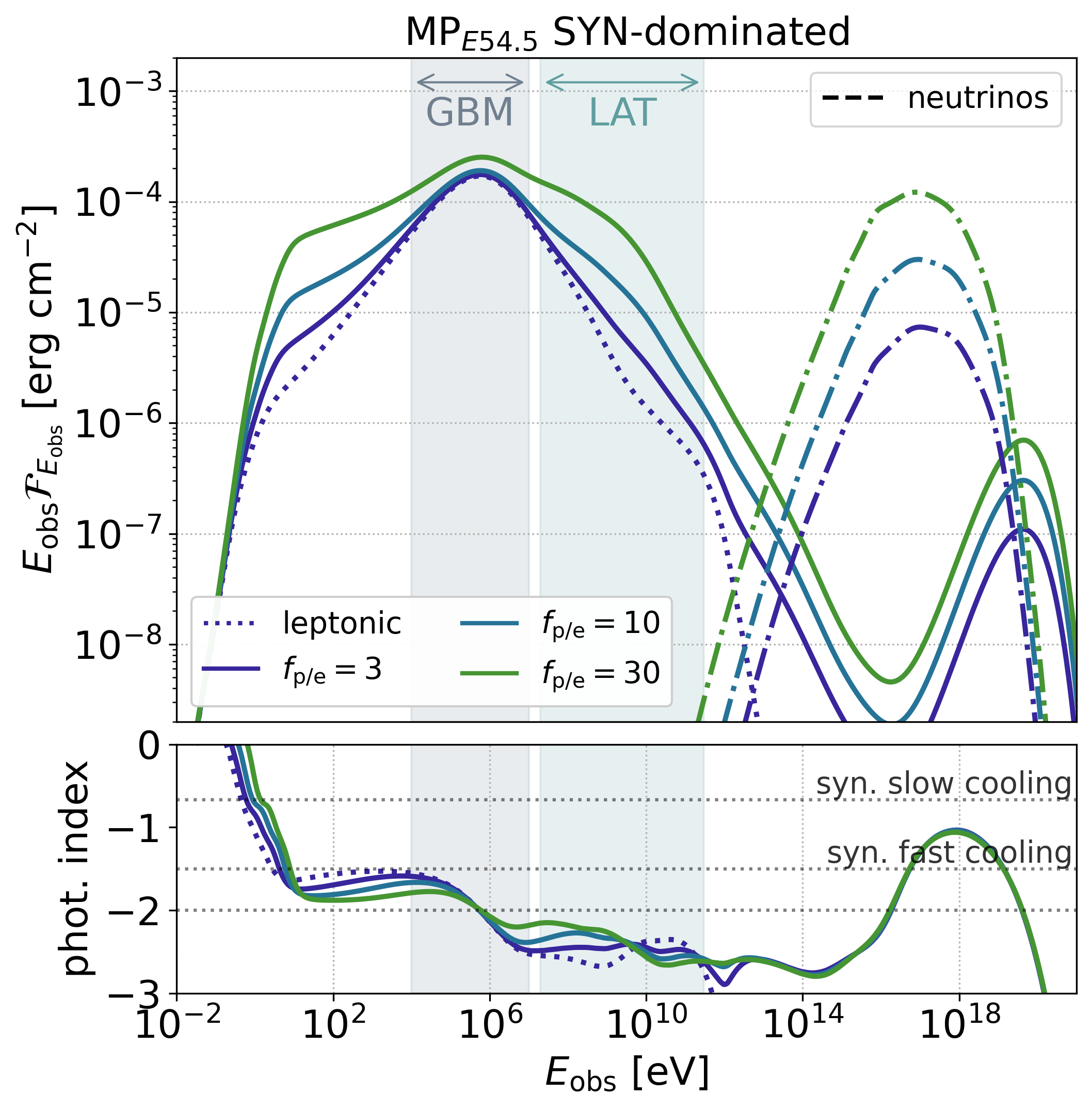}
    \hfill
    \includegraphics[width = 0.49 \textwidth]{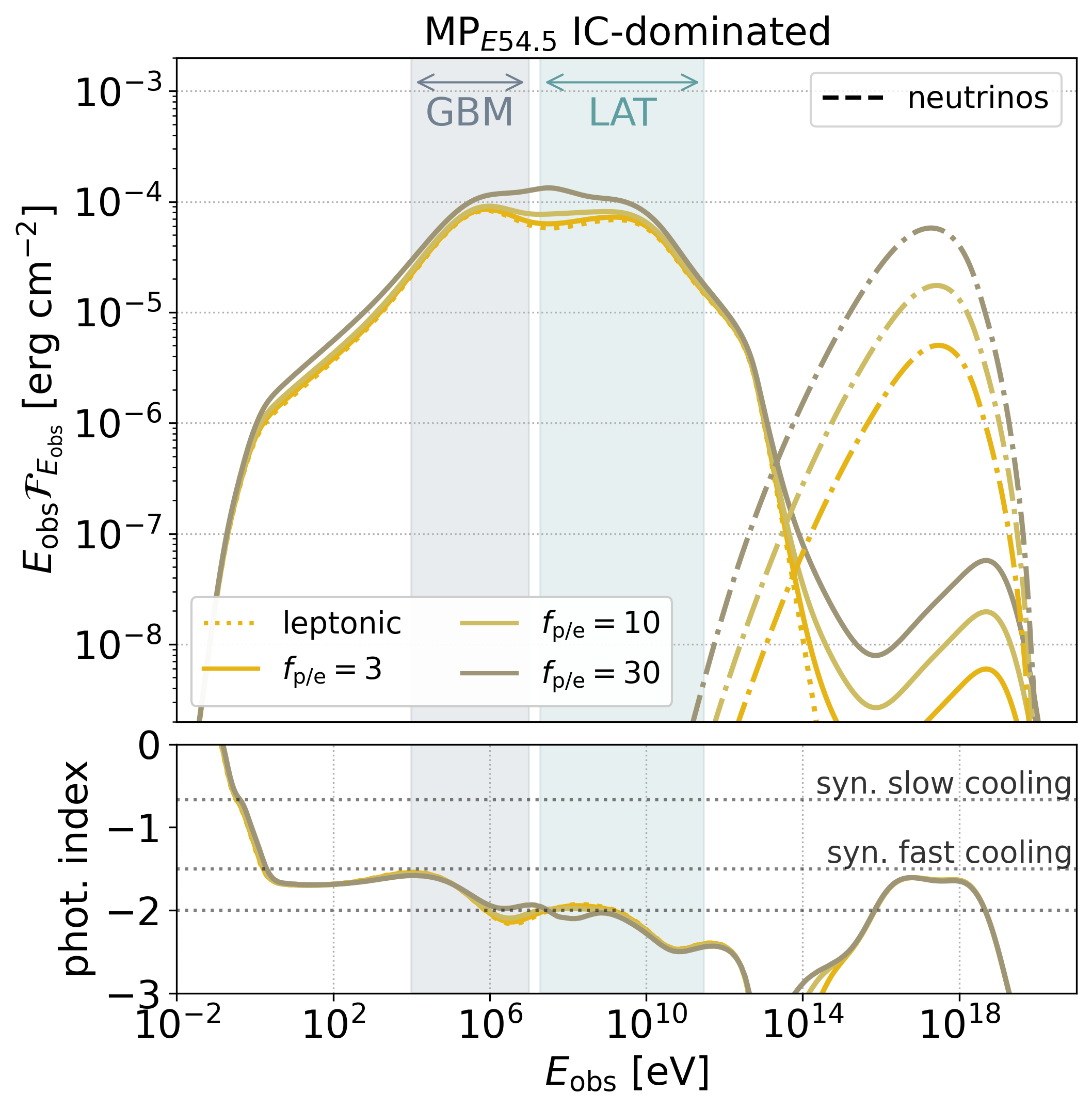}
    \caption{Lepto-hadronic spectra $E_\mathrm{obs} \mathcal{F}_{E_\mathrm{obs}}$ and photon indices for \SP \, (top panel) and \SP \, (bottom panel), examining \textit{(left)} the SYN-dominated scenario and \textit{(right)} the IC-dominated scenario. For all scenarios we show the leptonic case and explore $f_\mrm{p/e} \in \{10, 30, 100\}$ ($f_\mrm{p/e} \in \{3, 10, 30\}$) for \SP \, (\MP). Dash-dotted lines mark the corresponding all-flavour neutrino fluences.
    For the photon indices, we indicate the synchrotron slow- and fast-cooling predictions as dashed lines and a photon index of $-2$ (that marks peaks of $E_\mrm{obs} \mathcal{F}_\mrm{E_{obs}}$) as a solid line.} 
    \label{fig:leptohadronic-baryonicloading}
\end{figure*}

We continue by a systematic study of different baryonic loadings $f_\mrm{p/e}$ for both prototypes in the SYN- and IC-dominated scenarios. 
The spectra and photon indices are displayed in \reffig{leptohadronic-baryonicloading}. For \SP \, we explore $f_\mrm{p/e} \in \{10, 30, 100\}$, and for \MP \, that has a higher isotropic energy $f_\mrm{p/e} \in \{3, 10, 30\}$. For comparison we further show the leptonic modeling results.

We observe that increasing $f_\mrm{p/e}$ leads to similar trends for both prototypes, both in the SYN- and the IC-dominated scenario. We recall that the typical emission radii are similar, however \MP \, has a slightly higher $E_\mrm{iso}$ than \SP. This implies higher energy densities which enhance the efficiency of processes such as photo-pion production and $\gamma \gamma $-annihilation. As a consequence, for the same baryonic loading the signatures of secondary particles are slightly more pronounced for \MP \, than for \SP. As the differences are low we however conclude the more complex structure of \MP \, thus has little effect on the predicted time-integrated spectra. Their properties are instead mostly controlled by the combination of $\langle R_\mrm{Coll} \rangle$ and $E_\mrm{iso}$. 

Let us now discuss selected features of the photon spectra, starting with the SYN-dominated scenarios. As stated earlier, the flat additional component due to secondary synchrotron radiation introduces a wing-like broadening to the sub-MeV peak. This feature is common to both prototypes and increases in intensity for higher $f_\mrm{p/e}$.
This additional component outshines the primary emission for $f_\mrm{p/e} = 100$ for \SP \, ($f_\mrm{p/e} = 30$ for \MP). The flat spectrum is reflected in the photon indices that are smaller than -3/2 (that is the synchrotron fast cooling prediction). For the IC-dominated case the spectra below the peak are similar for all $f_\mrm{p/e}$ explored. For the highest $f_\mrm{p/e}$ an additional peak appears in the LAT range. Overall the spectra for the IC-dominated scenarios remain mostly unaffected by the baryon loading except for the VHE $\pi^0$ peak. \\ 
To put these results in context with observations, we review the shape of the spectra in the GBM band. Measured GRB spectra are typically narrower than what would be expected from the simple synchrotron spectrum, as can be inferred from their spectral width (see \citet{Axelsson:2014ora,Yu:2015jda} but also the critical discussion in \citet{Burgess:2017eek}).
Without explicitly measuring the spectral width in our models, we can by eye identify which models produce narrow or wide spectra.
We compare the leptonic (dotted) to lepto-hadronic (solid) lines in \reffig{leptohadronic-baryonicloading} and find that a leptonic SYN-dominated fast-cooling spectrum is the narrowest spectrum that can be achieved in our models. This is also the model with the highest photon index below the peak and lowest photon index above the peak (see lower panels).

The neutrino flux scales with the baryonic loading for all models, and all peak around $10^{18}$~eV. This is somewhat above the typical sensitivity of IceCube and thus radio neutrino detectors such as GRAND \citep{GRAND_2020} may instead be the adequate observatories for GRB prompt phase neutrinos. 

\subsection{Impact of variability timescale } 
The two prototypes \MP \, and \SP \, have rather similar typical collision radii and energy budget, hence a similar typical energy density. In \refsec{prototypes} we further introduced a modified version of \MP \, which has shorter duration and consequently shorter variability timescale (while using the same Lorentz factor distribution). We remind the reader that in our model, $\delta t_{\mathrm{var}}$ was defined in the \textit{source frame} as the variability timescale of the Lorentz factor distribution and thus tracks the variability time of the engine. Reducing $\delta t_{\mathrm{var}}$ moves the distribution of collisions inwards by a factor 10 and gives a typical collision radius of $\langle R_\mrm{Coll} \rangle \sim 10^{15}$~cm. 

\begin{figure*}
    \centering
\includegraphics[width = 0.8\textwidth]{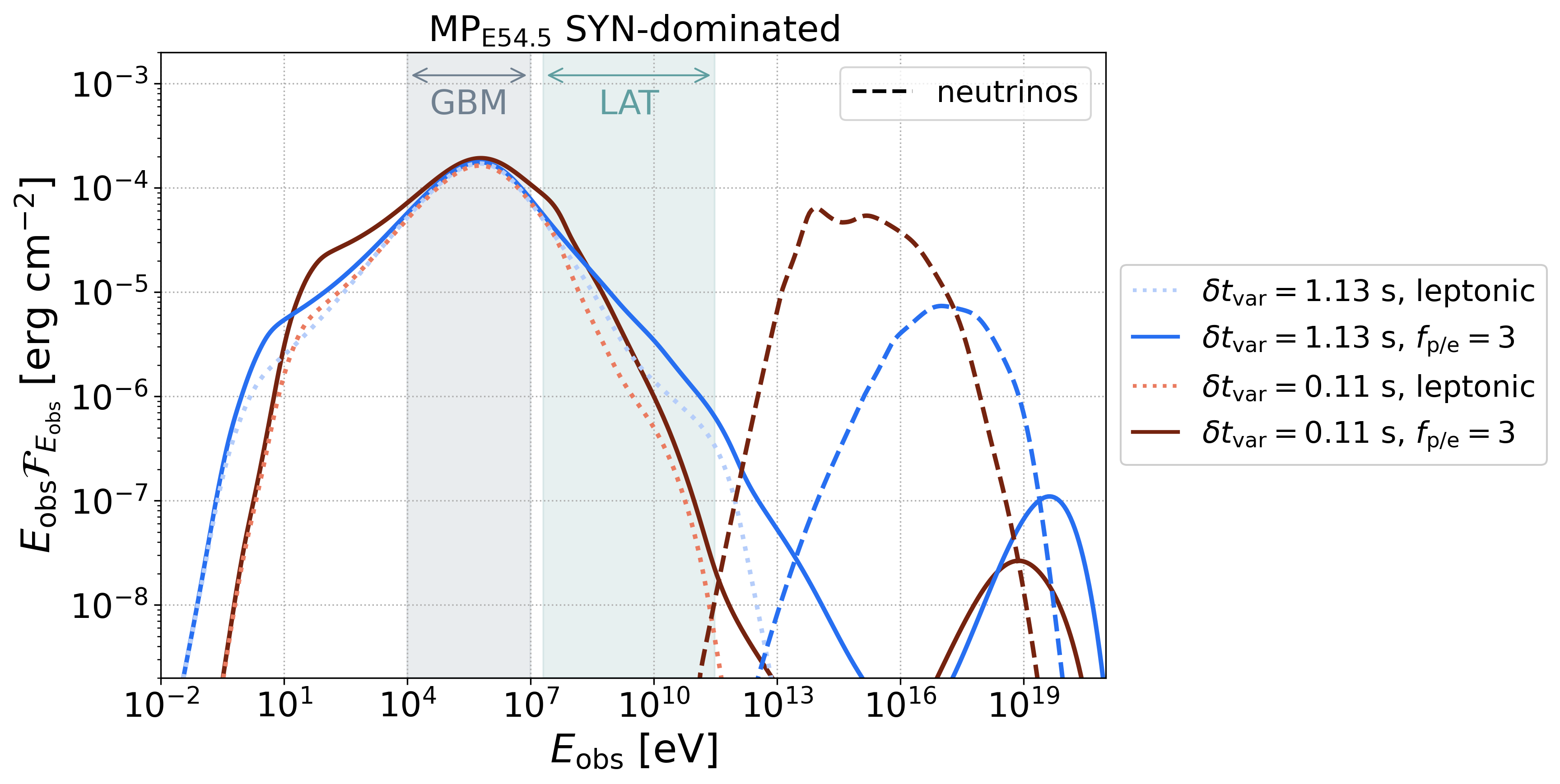}
    \caption{Time-integrated spectra for the \MP \, prototype (SYN-dominated scenario) illustrating the effects of the variability timescale, $\delta t_\mrm{var}$ (see side-panel legend for details).  
    }
    \label{fig:leptohadronic-tvar}
\end{figure*}

In \reffig{leptohadronic-tvar} we show the results for this modified version of \MP\, in orange/brown colors, comparing to the results of \MP \,  shown the last paragraph in blue colors. As before we show leptonic and lepto-hadronic models, for the latter imposing a baryonic loading of $f_\mrm{p/e} = 3$.

Let us first discuss the leptonic results, i.e. comparing the blue and the orange dotted lines in \reffig{leptohadronic-tvar}. Here, two main effects arise with decreasing $\delta t_\mrm{var}$: the synchrotron self-absorption frequency increases, and the suppression in the LAT-range is enhanced. Both effects can be explained by the higher densities because of the lower collision radii that increase the efficiency of internal absorption processes. On the other hand, the spectrum in the GBM band remains unaffected. For lepto-hadronic models we additionally find that the $\pi^0$ decay peak is decreased in peak energy and intensity. This may be understood through the higher pion production efficiency (arising due to the higher densities) for short $\delta t_\mrm{var}$, which reduces the maximal proton energy. This, in turn, lowers the $\pi^0$-decay peak energy. 
The higher pion production efficiency is further reflected in the neutrino fluxes that (for the same baryonic loading) are enhanced almost by a factor of 10 with respect to the longer $\delta t_\mrm{var}$ (compare e. g. solid lines); this can be also shown analytically, see Eq.~\ref{equ:fpi} (second part) later. Moreover, the neutrino spectra contain additional features/breaks which we attribute to cooling effects of primary and secondary particles.

Overall, reducing $\delta t_\mrm{var}$ increases the intensity of the hadronic signatures discussed in the last sections, and a lower baryonic loading may be sustained. We point out that a similar effect can be expected by decreasing the Lorentz factor of the outflow (see also \refapp{modelling-choices} where this is discussed for leptonic scenarios). In this sense, accurate estimates for the Lorentz factor of the outflow and the source-frame $\delta t_\mrm{var}$ are essential when studying IC-dominated or lepto-hadronic models that have large contributions by density-dependent processes to the broadband emission.

\section{Implications for multi-messenger astrophysics}
\label{sec:multimess}

Here we discuss the implications of our model for multi-messenger astrophysics. We examine first the dominant production sites of 
the different messengers (photons, neutrinos, UHECRs) in a GRB jet and check
if these signals are time-correlated, which is relevant to neutrino alert follow-ups (\refsec{location}). One of the most uncertain parameters of lepto-hadronic models is the baryonic loading. We discuss what is the maximum allowed value by considering its impact on the broadband GRB photon spectrum (\refsec{constraints_em_cascade}) and by using constraints from neutrino observations (\refsec{constraints_nu}). Upper bounds on the baryonic loading can also limit the predicted UHECR flux per GRB. We therefore examine the requirements for a sub-population of energetic GRBs, similar to those considered in this work, to be the source of UHECRs (\refsec{UHECRs}).

\subsection{Production regions and observation times of different particle species}\label{sec:location}

\begin{figure}
    \centering
\includegraphics[width = 0.4\textwidth]{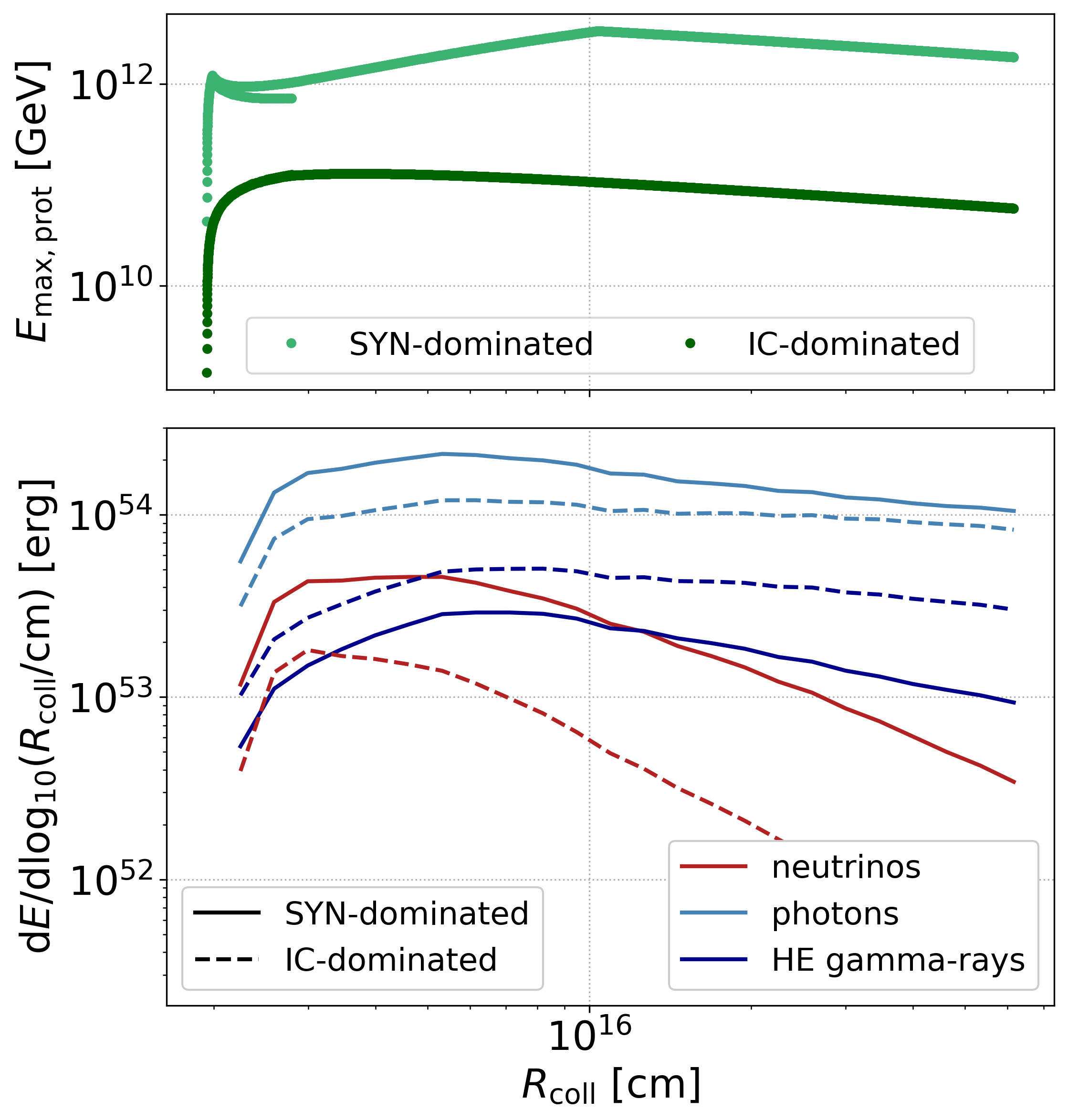}
    \caption{Spatially resolved properties of different messengers in the SYN- and IC-dominated scenarios for the \SP \, burst. 
    \textit{Top panel}: Upper limit on the maximum energy of protons obtained by balancing adiabatic and synchrotron losses with the acceleration time as a function of radius. The true maximal energy may be lower due to photo-hadronic interactions that are neglected here.
    \textit{Bottom panel}: Emitted energy in different particle species as a function of radius for $f_\mrm{p/e} = 30$. Here we define HE gamma-rays as those photons that have $E_\gamma \gtrsim 1$~GeV in the source frame. Recall that the deceleration radius for the hadronic case is $R_\mrm{dec, had} \sim 3 \cdot 10^{17}$~cm. 
    \label{fig:regions}}
\end{figure}

\begin{figure}
    \centering
\includegraphics[width = 0.4\textwidth]{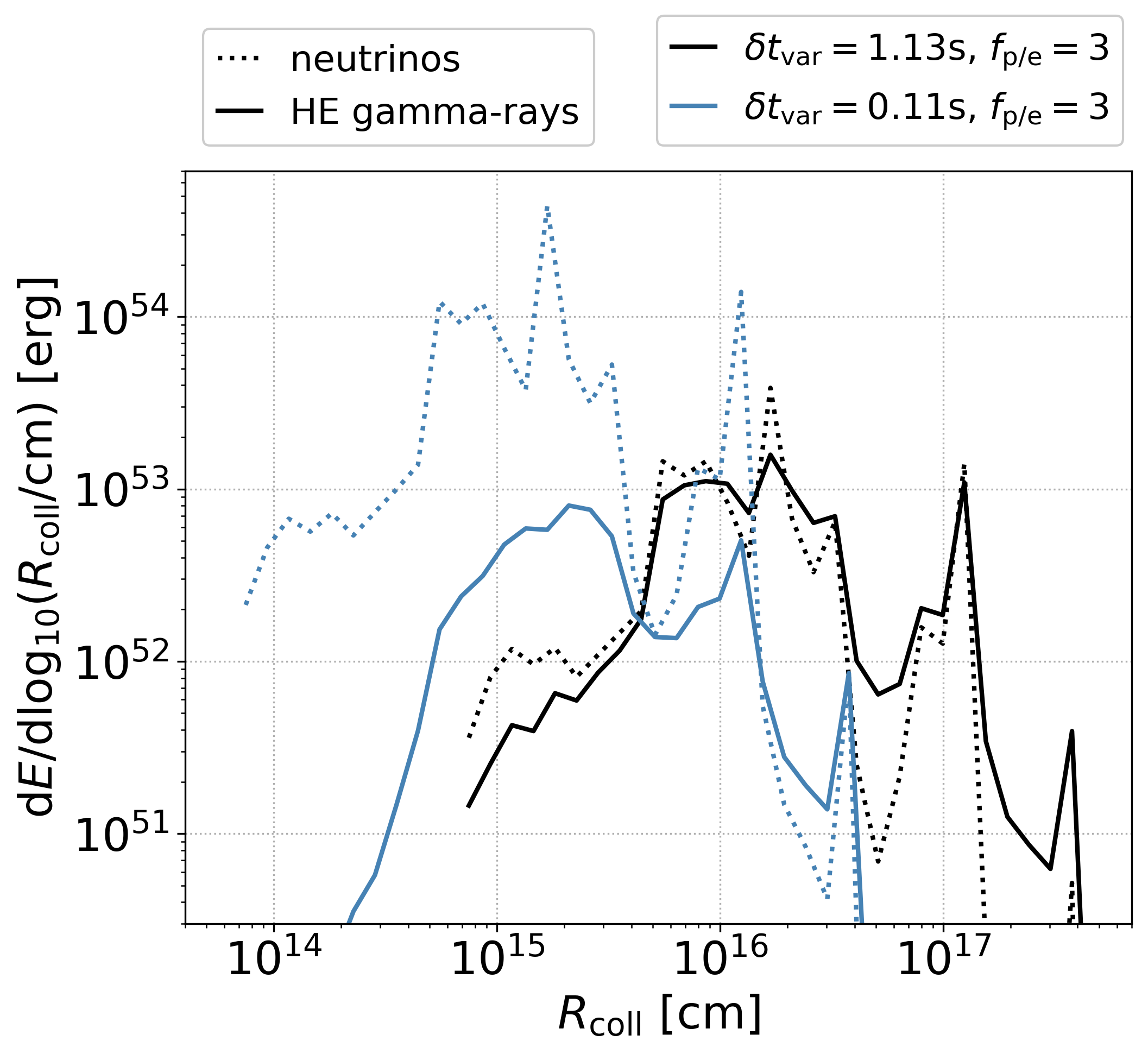}
    \caption{Effects of the variability timescale on the radial profile of emitted energy in neutrinos and HE $\gamma$-rays for the \MP \, burst (SYN-dominated scenario). Different baryonic loadings are used to produce similar peak neutrino energies (see legend). 
    \label{fig:regions_mp}}
\end{figure}

Spatially resolved multi-collision models offer the possibility of investigating the production regions of different particle species \citep[e.g.][]{Bustamante:2016wpu, Rudolph:2019ccl, Heinze:2020zqb}. In \reffig{regions} (lower panel) we show the energy carried by escaping
neutrinos, photons (of all energies) and HE $\gamma$-rays (with energy above 1~GeV in the source frame) as a function of collision radius for the SYN- and IC-dominated scenarios assuming $f_\mrm{p/e} = 30$. Results are shown for the \SP \, burst. The bolometric photon curve is similar to the radial profile of energy dissipation for this burst, which peaks around $R_\mrm{Coll} \sim 10^{16}$~cm (see \reffig{ediss_rcoll}).
In comparison to this,  the production rate of neutrinos is enhanced at the smallest radii. In other words, most of the neutrino energy is emitted at smaller radii where the photon densities are higher, even though the dissipated energy transferred into relativistic particles peaks further out. While in the lepto-hadronic model HE $\gamma$-rays are also produced via photomeson interactions, their energy is suppressed at the smallest radii due to internal $\gamma \gamma$ attenuation. The radial profiles of all messengers are similar for both SYN- and IC-dominated scenarios. Note however that the energy emitted as HE $\gamma$-rays in the IC-dominated scenario is enhanced compared to the SYN-dominated scenario, as demonstrated also in \reffig{leptohadronic-decomposed}.
We point out that the range of collision radii is comparatively small (see for example the results for \MP \,  in \reffig{regions_mp}), hence the differences in the location of different messengers is less pronounced compared to past publications.
The upper panel of \reffig{regions} further shows the maximal proton energy (in the source frame) as a function of radius, as obtained by balancing the acceleration rate with the adiabatic and synchrotron loss rates. The displayed values are strictly speaking an upper bound on $E^\prime_{\rm max, prot}$, since photo-hadronic energy losses, which for simplicity are not included here, may push the maximum energy to even lower values (see Fig.~\ref{fig:pi_mu_coolingtimescales}). The lower magnetic fields involved in the IC-dominated scenario translate to lower acceleration rates which result in lower maximal energies. For the SYN-dominated case, the high synchrotron cooling rate at the smallest radii reduces the maximal proton energy. Overall, our maximal proton energies are of the order of $10^{20}$-$10^{21}$~eV in the source frame which is sufficient to explain UHECRs, as expected for models with large collision radii and large Lorentz factors~\citep[see also][]{Samuelsson:2018fan}.

We further investigate the production regions of neutrinos and HE $\gamma$-rays for two different variability timescales using the \MP\,  burst (see \reffig{regions_mp}). As in \reffig{leptohadronic-tvar}, we choose two engine active times that differ by a factor 10 while keeping the same Lorentz factor distribution. This setup yields variability timescales that differ by a factor of 10 in the source frame. The shorter variability timescale translates to smaller dissipation/emission radii, which can be seen by comparing the neutrino radial profiles (dotted blue and black lines in \reffig{regions_mp}). Besides the shift in radius, the two profiles are very similar because of the same underlying Lorentz factor distribution.  
As discussed before, smaller dissipation radii realised for the shorter variability timescale enhance the pion production rate and, in turn, the energy released in neutrinos. 
As  a result, the energy emitted in neutrinos at the peak  for $\delta t_{\rm var}=0.11$~s is roughly a factor 10 higher than the maximum energy found for the longer variability burst.
The impact of the variability timescale is stronger on the HE $\gamma$-ray radial profile, which is suppressed at all radii for $\delta t_{\rm var}=0.11$~s (compare solid blue and black lines). Differences in the opacity of the emitting region to $\gamma \gamma$ pair production can also be inferred by comparing the neutrino (not attenuated) and HE $\gamma$-ray curves for different variability timescales (compare solid and dotted curves). 

\begin{figure}
    \centering
 \includegraphics[width = 0.45\textwidth]{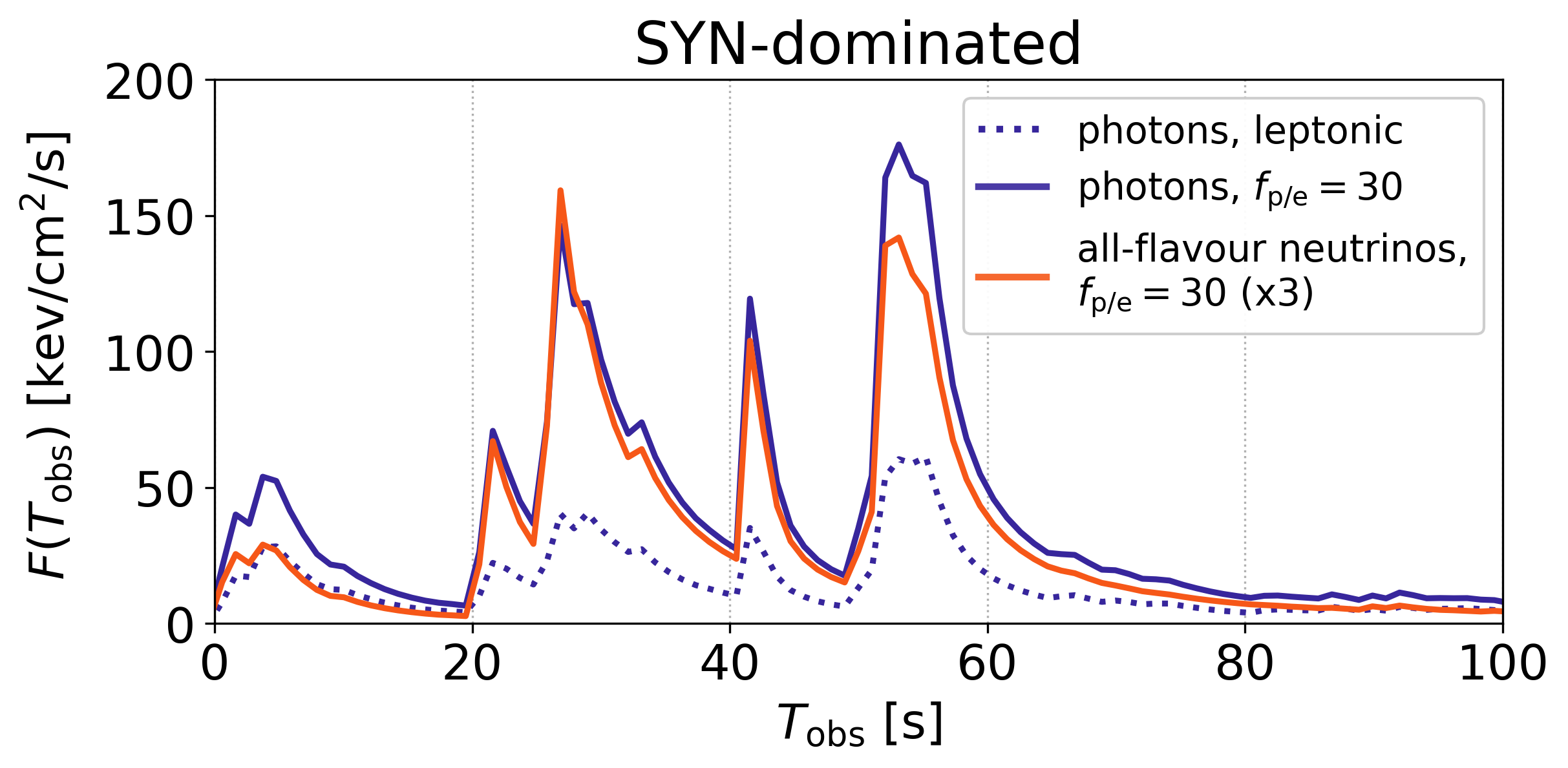}
    \caption{Energy flux for photons (integrated over the all photon energies) and neutrinos (all flavours) as a function of time for the \MP  \, burst. We show the SYN-dom inated scenario with $f_\mrm{p/e} = 30$ (solid lines). and the corresponding leptonic model (dotted line). The neutrino fluxes are scaled up by a factor three to match the same scale.
    \label{fig:nulightcurve}}
\end{figure}

We discuss next the time-dependence of the emission from the different messengers. UHECRs typically cannot be directly associated to transients because of the huge time delays caused by the deflection in intergalactic magnetic fields. We therefore focus on the other two messengers, photons and neutrinos. 
As an indicative example, we show in \reffig{nulightcurve} the energy flux emitted in neutrinos of all flavours and the bolometric photon flux as a function of time (in the observer's frame) for the \MP \, burst in the SYN-dominated scenario. Both messengers have similar light curves and are strongly correlated without evidence of time lags. 
The photon light curve is dominated by the energy range where most of the energy is emitted -- i.e. around the sub-MeV peak. The light curves, however, may look different when computed for different energy bands. For example, low-energy radiation that is produced by secondary lepton pairs will be intense at early times that correspond to small radii, (see also \reffig{leptohadronic-decomposed}), while early HE emission may be suppressed due to high opacity \citep[as discussed in][]{Bustamante:2016wpu}.
Nonetheless, the main detection channel of instruments like \fermi-GBM corresponds to the sub-MeV peak and we expect that the usual follow-up searches between neutrinos and electromagnetic signals in either energy band apply to this model, as differences on the order of seconds to minutes  between photons and neutrinos do not  have any practical impact on the follow-up strategies; the approach would always be to trigger the follow-up immediately.

\subsection{Constraints on the baryonic loading from the electromagnetic cascade}
\label{sec:constraints_em_cascade}

As a next step we assess how the electromagnetic cascade may be employed to infer the maximal baryonic loading that is compatible with observations. 
To identify the energy bands that are potentially relevant in this context we first discuss the impact of EBL absorption. We further compare the fluences in energy bands that are expected to be observationally accessible for different models, and finally test the robustness of our results by studying the impact of the proton power-law slope $p_\mrm{p}$.

\subsubsection*{Effects of EBL absorption}
\label{sec:ebl_absorption}

\begin{figure}
    \centering
\includegraphics[width = 0.45\textwidth]{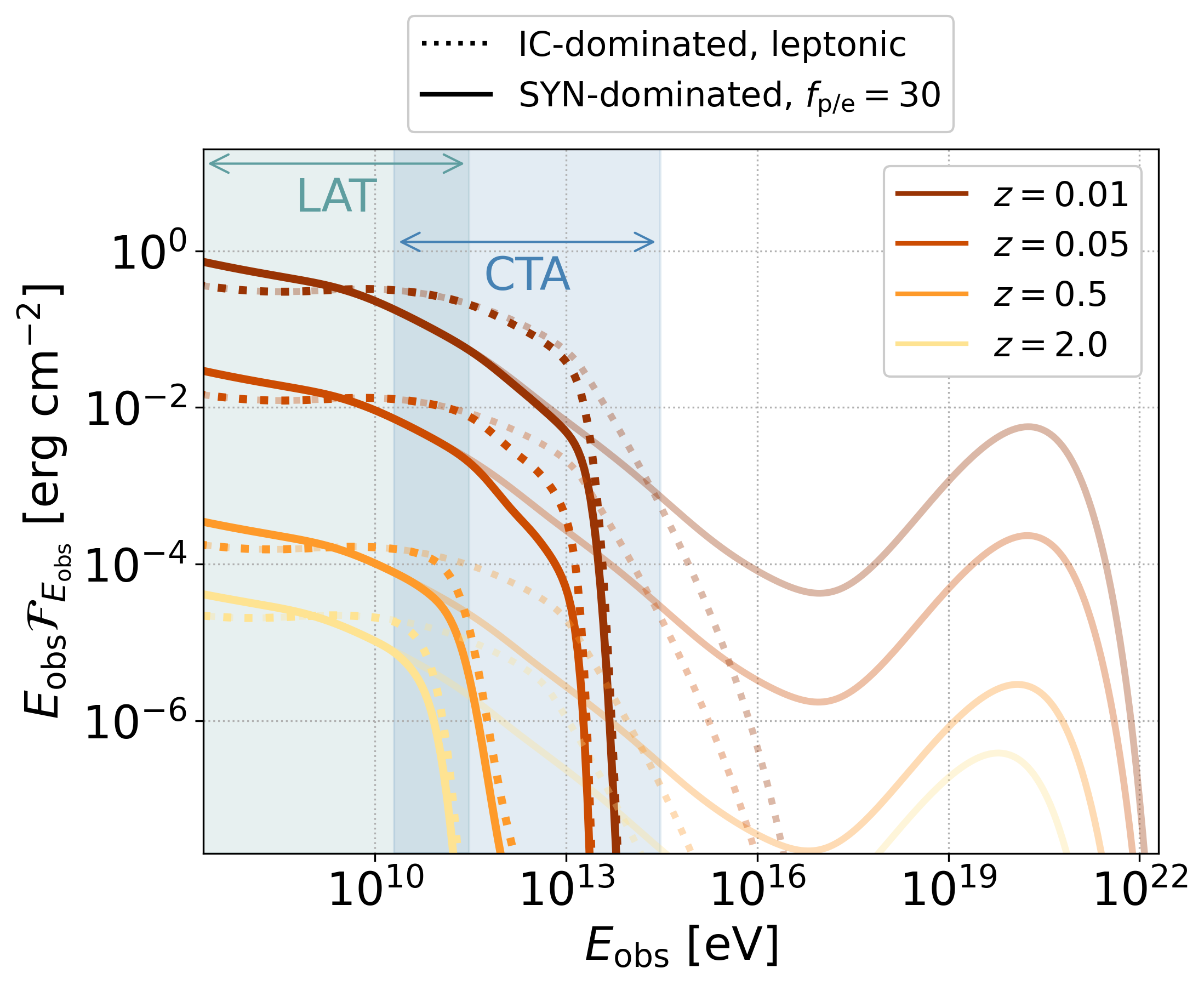}
    \caption{Examining the impact of EBL absorption on energy fluence model spectra by placing \SP at different redshifts (see inset legend). Transparent (non-transparent) curves show the results without (with) EBL absorption. Results are obtained for the EBL model of \protect\cite{Dominguez:2010bv}.}
    \label{fig:redshift_study}
\end{figure}

In \reffig{redshift_study} we commence with a redshift study, placing our prototype \SP \, at different distances to Earth and calculating the spectra with EBL absorption. We only show HE to VHE emission, since lower energies remain unaffected by the EBL absorption. The $\pi^0$-decay peak, which is an unambiguous hadronic signature (compare to the leptonic model, dotted lines), is expected at extremely high energies ($>10$~EeV). Regardless of the degree of EBL attenuation, this feature is beyond the observing window of the Cherenkov Telescope Array (CTA) and hence not accessible. As a result, the baryonic loading of a GRB jet has to be inferred from lower energies which are shaped by secondary cascade emission. For $z \gtrsim 0.5 $ (which are typical redshifts for energetic GRBs, see \citet{Poolakkil:2021jpc}), emission in the CTA range is strongly suppressed. Still, differences in the spectral shape in the LAT and CTA ranges are expected between the leptonic and lepto-hadronic models (depending on parameters like $f_{\rm B/e}$). The prospects of differentiating between a purely leptonic and a lepto-hadronic model with combined \fermi-LAT and CTA analysis is worth investigating further.  

\subsubsection*{Multi-wavelength signatures: Relative fluence in different energy bands}

As these are the observationally accessible energy ranges, we compare the energy fluence of our models in UV, X-rays and HE $\gamma$-rays. For each energy band we normalise with the fluence in the \fermi-GBM band and further give the absolute fluence that would be observed by \fermi-GBM. The values for the \SP \, burst in the SYN- and IC-dominated scenarios are listed in \reftab{fluence_sp_wavelength}. The results again illustrate that the IC-dominated scenario is comparatively dim in low energies (UV and X-rays) while being bright in HE. The HE brightness has for example been analysed in \citet{Nakar:2011qy}, who find $\mathcal{F}_\mrm{LAT}/\mathcal{F}_\mrm{GBM} \lesssim 0.1$, however larger values of up to 1 compatible with our predictions are reported for single events in \citet{Ajello:2019zki}. Further, the table illustrates that for IC-dominated case
leptonic and lepto-hadronic scenarios are unlikely to be distinguishable based solely on energetics (see previous paragraph). In contrast, hadronic contributions result in a very bright UV component for the SYN-dominated scenario. The relative brightness in HE of the IC-dominated scenario cannot be recovered even with hadronic contributions. We thus conclude that a joint observation by a low-energy (optical or UV) and HE instrument would offer good diagnostics to narrow down $f_\mrm{B/e}$ (that controls the SYN- or IC-dominated regimes) and the baryonic loading $f_\mrm{p/e}$.

Let us at this point briefly comment on the efficiency in lepto-hadronic models. The table illustrates that for the scenarios with $f_\mathrm{p/e} = 30 $, the events would be brighter especially in the optical and HE regime. Here it is worth remembering that this comes at the cost of a much higher initial kinetic energy of the jet (see Eq.~\ref{eq:ekin_initial}), and hence the total efficiency of the fireball in converting its initial kinetic energy into radiation is reduced from
$\sim 5 $~\% for leptonic models to $\sim 0.5 $~\% for $f_\mathrm{p/e} = 30 $.

\begin{table*}
\caption{Fluence in the \textit{Fermi}-GBM range between 10~keV and 10~MeV and relative fluence for different instrumental energy ranges with respect to the \textit{Fermi}-GBM for \SP. }
\label{tab:fluence_sp_wavelength}
\centering
\renewcommand{\arraystretch}{1.35}
\begin{threeparttable}
\begin{tabular}{c c c  c c}
\toprule
  & \multicolumn{2}{c}{SYN-dominated} & \multicolumn{2}{c}{IC-dominated} \\ 
 & leptonic & $f_\mathrm{p/e} = 30 $  & leptonic & $f_\mathrm{p/e} = 30 $ \\ \hline
 \textit{Fermi}-GBM [$10^{-5}$erg/cm$^{-2}$]& 9.4 & 12.7 & 5.1 & 5.5 \\ \hline
  \textsc{ULTRASAT} /\textit{Fermi}-GBM [$10^{-4}$] & 4.3 & 18.3 & 3.0 & 3.6\\ 
\textit{Swift}-XRT /\textit{Fermi}-GBM [$10^{-1}$]& 1.2 & 2.1 & 0.85 & 0.87\\
 \textit{Fermi}-LAT  /\textit{Fermi}-GBM [$10^{-1}$]& 2.4 & 4.6 & 12.0 & 14.1 \\
\hline
\end{tabular}
\begin{tablenotes}
    \item \textit{Notes.--} We use the energy ranges of \textsc{ULTRASAT} (as an example for a UV-observatory, 220 - 280 nm), \textit{Swift}-XRT (as an example for an X-ray observatory, 0.2 - 10 keV) and \textit{Fermi}-LAT (as an example for a $\gamma$-ray observatory, 20~MeV - 300~GeV). 
\end{tablenotes}    
\end{threeparttable}
\end{table*}

\subsubsection*{Relative intensity of cascade emission}

\begin{figure}
    \centering
\includegraphics[width = 0.45\textwidth]{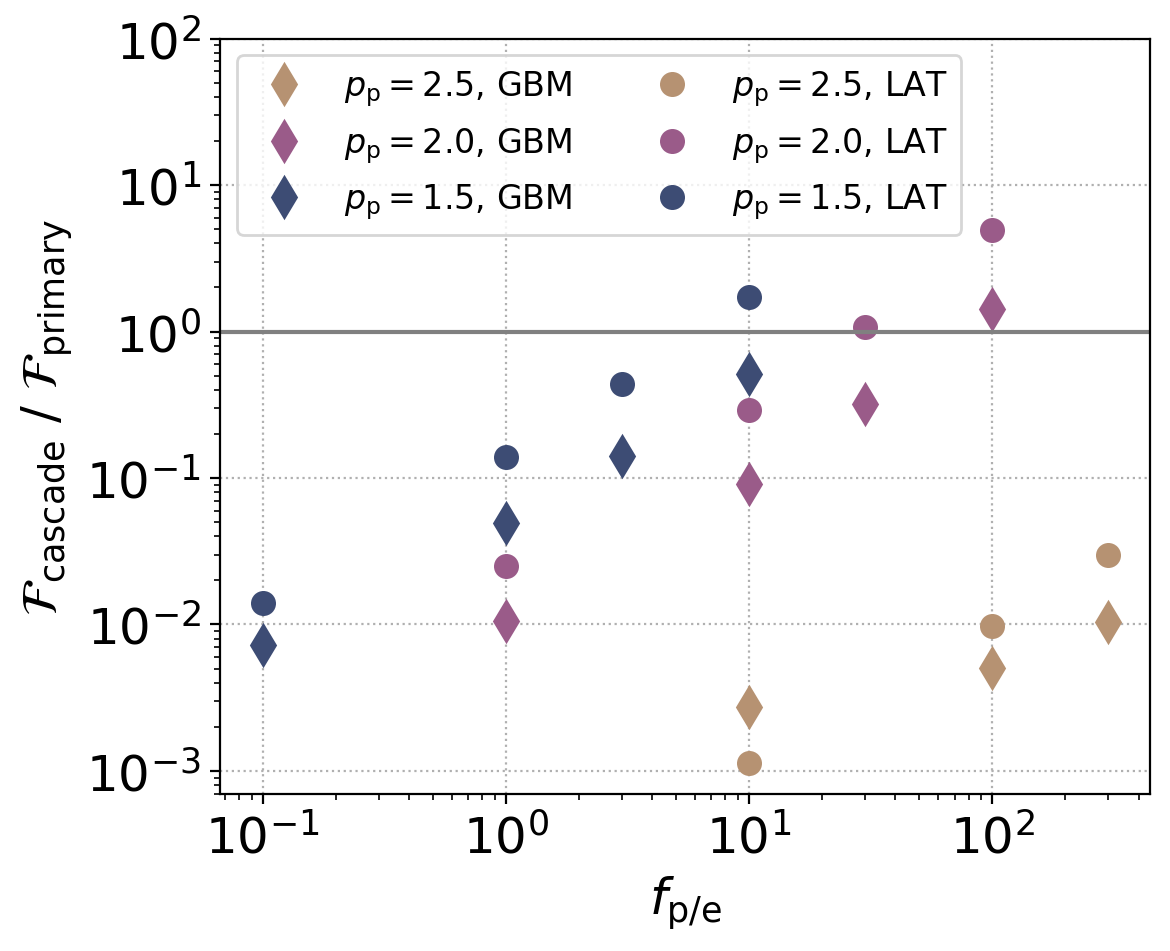}
    \caption{Ratio $\mathcal{F}_\mrm{cascade}$ and $\mathcal{F}_\mrm{primary}$ for \SP \, (the integrated fluence produced by cascade/ primary leptons, see \refeq{fluence_instrument}) in the GBM and LAT band for different values of $f_\mrm{p/e}$ and primary proton power-law slopes $p_\mrm{p}$. All calculations are performed in the SYN-dominated scenario, the solid line indicates where the fluence of the cascade equals the one of primary leptons.}
    \label{fig:hl_primary_vs_secondary_fluxes}
\end{figure}

A critical question in the context of lepto-hadronic models is for which baryonic loading $f_\mrm{p/e}$ the cascade emission outshines the one of primary electrons. We investigate this aspect using the SYN-dominated scenario of \SP\ as an example. The ratio of fluences produced by primary electrons $\mathcal{F}_\mrm{primary}$ (sum of light and dark red lines in \reffig{leptohadronic-decomposed}) and cascade leptons produced by $\gamma \gamma$-annihilation $\mathcal{F}_\mrm{cascade}$ (sum of light and dark blue lines in \reffig{leptohadronic-decomposed}) in the \textit{Fermi}-GBM and LAT range 
are shown in \reffig{hl_primary_vs_secondary_fluxes}
for different pairs of  $(f_\mrm{p/e}, p_\mrm{p})$ values. EBL absorption is neglected. The fluences are computed as 
\begin{equation}
\mathcal{F}  = \int_{E_\mathrm{min}}^{E_\mathrm{max}}\mathcal{F}_{E_\mrm{obs}} \mrm{d}E_\mrm{obs} \, .
\label{eq:fluence_instrument}
\end{equation} 
where $\mathcal{F}_{E_\mrm{obs}}$ 
is the differential fluence, $E_\mathrm{min}=10$~keV (20 MeV), and $E_\mathrm{max}=10$~MeV (300 GeV) for the \textit{Fermi}-GBM (LAT) instruments. The horizontal line indicates where the primary and cascade emission are equally intense ($\mathcal{F}_{\mathrm{cascade}}$= $ \mathcal{F}_{\rm primary}$), but noticeable distortions of the primary synchrotron spectrum may be expected for $\mathcal{F}_\mathrm{cascade} \gtrsim 0.5 \, \mathcal{F}_\mrm{primary} $. 

In purple we show the results for the classical proton power-law index of $p_\mrm{p} = 2.0$ (recall that the injected proton distribution follows ${\rm d}n'_{\rm p}/{\rm d}\gamma'_{\rm p}\propto \gamma_{\rm p}^{'-p_{\rm p}}$). In addition we show $p_\mrm{p} = 1.5$ and $p_\mrm{p} = 2.5$ with yellow and blue symbols, respectively. For equal energy budget of non-thermal protons and minimum proton Lorentz factor $\gamma_{\rm p, \mrm{min}}$, $p_\mrm{p}$ determines the energy transferred to protons at the highest energies; the lower $p_\mrm{p}$ is, the more energy is transferred to high-energy protons. 
Due to the larger relative luminosity of high-energy protons for $p_\mrm{p} = 1.5$, the secondary emission outshines the primary emission already for $f_\mrm{p/e} \gtrsim 10$. With increasing $p_\mrm{p}$, larger baryonic loadings can be allowed. For instance, for soft proton power-law spectra ($p_{\rm p}=2.5$), the cascade emission does not distort the primary emission in the GBM/LAT bands even for $f_{\rm p/e}=300$. In other words, a lot of energy is carried by protons that do not undergo photo-hadronic interactions. The linear increase (in log-log space) is common to all choices of $p_\mrm{p}$ and may therefore be interpreted as a general feature. For the usually adopted value of $p_\mrm{p} = 2$, the secondary emission outshines the primary one in the GBM band for baryonic loadings of $f_\mrm{p/e} \gtrsim 100$ in this SYN-dominated scenario. We again stress that this will not be true in an IC-dominated case where the spectra in GBM and LAT energy bands are not much affected unless $f_\mrm{p/e} \gg 100$. \\

\subsection{Constraints on the baryonic loading from neutrinos}\label{sec:constraints_nu}

\subsubsection*{Point source and multiplet constraints}

No neutrinos associated with individual GRBs have been detected so far \citep{IceCube:2012qza,Aartsen:2017wea,IceCube:2022rlk}. Meanwhile, the energetic GRBs considered here are very luminous out in $\gamma$-rays, i.e., they could have been detected in most cases, see e.g. Fig.~1 in \cite{Kistler:2009mv}. Depending on the baryonic loading factor, they could also be bright neutrino emitters (see e.g. our \reffig{leptohadronic-baryonicloading}). Therefore, we can place constraints on $f_\mrm{p/e}$ for an individual GRB by requiring that significantly less than one event is expected, similar to constraints derived for GRB~130427A in \citet{Gao:2013fra}, and for GRB~160625B in \citet{Fraija:2017mlx}. We compute the number of expected neutrinos for each example integrating over the full energy range of neutrinos and with the ``most favorable'' point-source effective area, $A_\mrm{eff}$, between 0 and 2.29 degrees \citep[as presented in][]{IceCube:2021xar}.

We find that the number of expected neutrinos per flavour for \MP \, (SYN-dominated scenario) is  $N_\nu \sim 3.7 \cdot 10^{-2}$ for $f_\mrm{p/e} = 10$, and for \SP \, (SYN-dominated scenario) is $N_\nu \sim 3.4 \cdot 10^{-3}$ for $f_\mrm{p/e} = 10$, which is significantly below one in both cases. Therefore, one would not expect these GRBs to stick out in neutrino point source analyses based on neutrino data alone (for the assumed redshift of $z = 2$). For \MP \, and short $\delta t_\mrm{var}$ we calculate the number of expected events as $N_\nu \sim 0.14$ for $f_\mrm{p/e} = 1$; here, the lower peak energy increases the detection prospects.

In some rare cases multiplets are expected even if the predicted event rate is smaller than one. For example, taking into account the whole population of sources including redshift distribution, the maximal possible neutrino output for sources with GRB-like redshift evolution per transient has been limited to $E_\nu \lesssim 10^{53} \, \mathrm{erg}$ in \cite{Aartsen:2018fpd} (Fig.~2) for a local source rate $\dot n_0 \simeq 0.1 \, \mathrm{Gpc^{-3}} \, \mathrm{yr}^{-1}$ expected here (see below).  We expect a total energy output in neutrinos of $8.4 \cdot 10^{52}$ erg for \SP \, and $f_\mrm{p/e} = 10$ and and $5.4 \cdot 10^{53}$ erg for \MP \, and $f_\mrm{p/e} = 10$, which means that the baryonic loading $f_\mrm{p/e} \lesssim 10$ is roughly compatible with these transient searches. For \MP \, and $\delta t_\mrm{var}= 0.11$s, we find that the energy emitted in neutrinos is $3.8 \cdot 10^{53}$ erg for $f_\mrm{p/e} = 1$, which means that for these sources even a baryonic loading $f_\mrm{p/e} = 1$ would lead to a detectable neutrino signal. This highlights the role of the collision radius, which is related to the variability timescale, in the production rate of neutrinos from a GRB. 

\subsubsection*{Stacking searches}

\begin{figure*}
    \captionsetup[subfigure]{labelformat=empty}
    \centering
    \includegraphics[width = 0.42\textwidth]{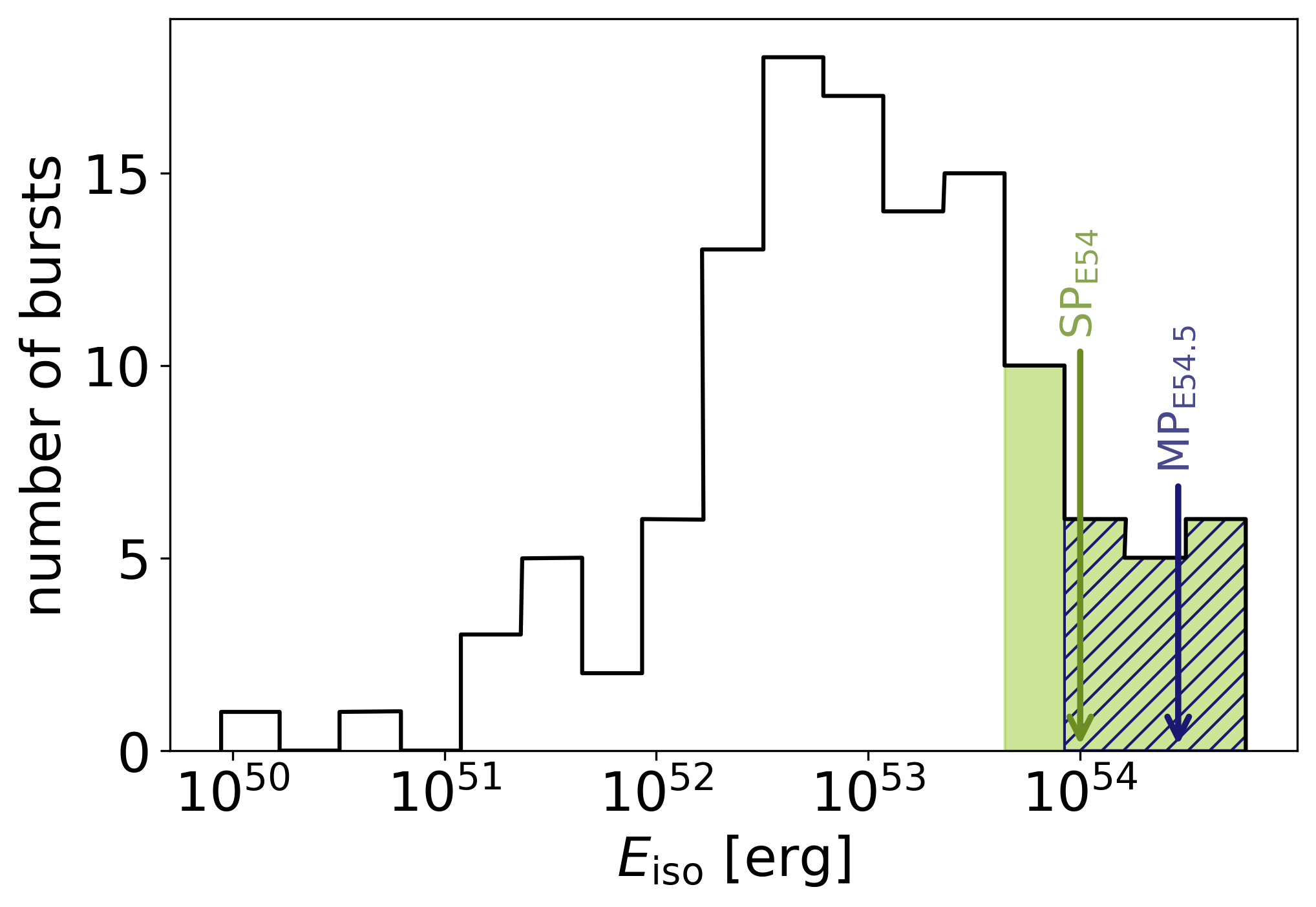}
    \includegraphics[width = 0.47\textwidth]{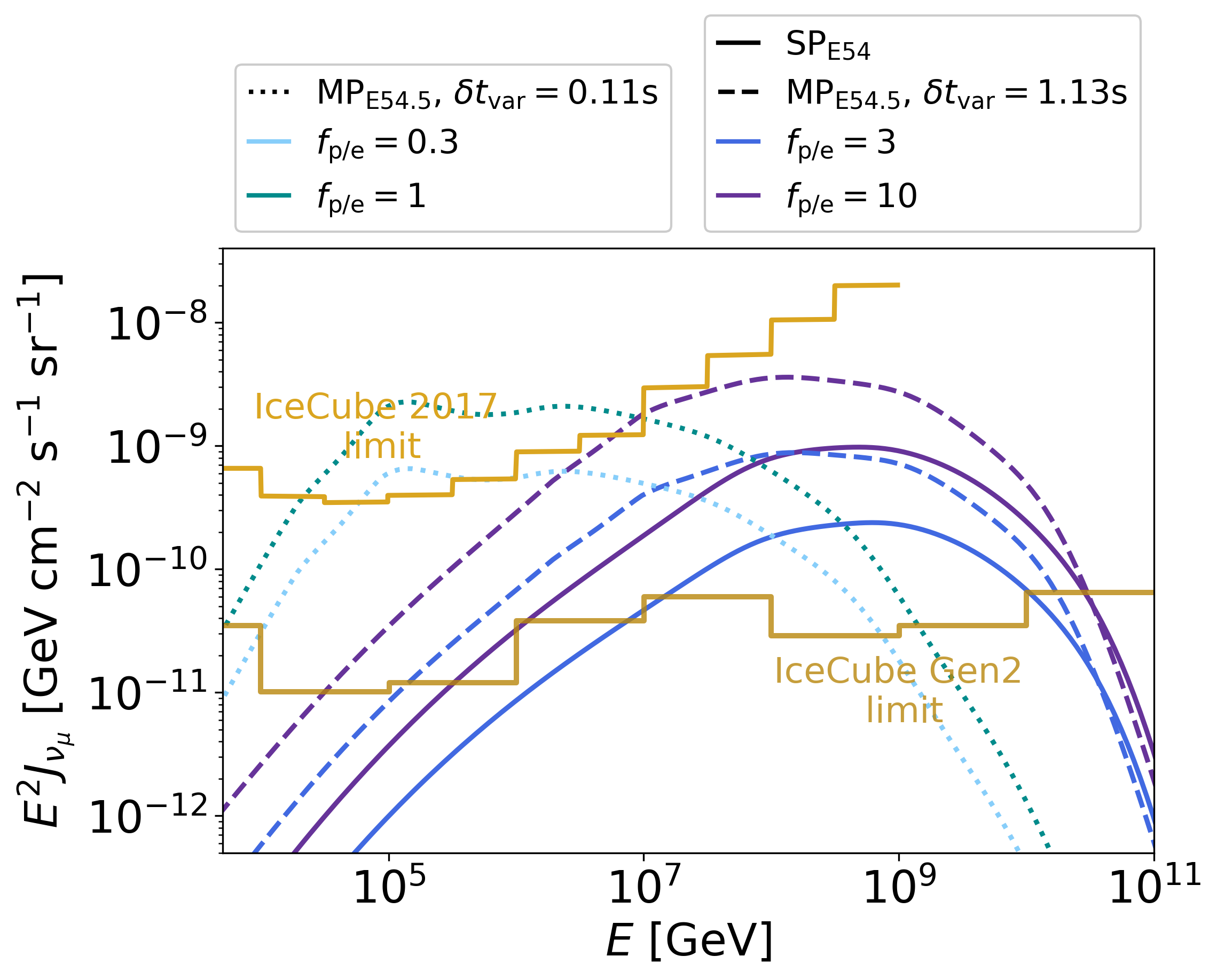}
    \caption{\textit{Left:} Distribution of bursts as a function of source-frame $E_\mathrm{iso}$ for the 122 long GRBs with measured redshift of the \textit{Fermi}-GBM catalogue presented in \citet{Poolakkil:2021jpc}. Arrows indicate the $E_\mrm{iso}$ of our two prototypes.
    The green shaded (hashed blue) areas correspond to the number of bursts that have $E_\mathrm{iso}$ comparable to \SP \, (\MP).
    \textit{Right:} All-sky quasi diffuse flux (per flavour) for the SYN-dominated scenarios for both prototype bursts. $J_\mathrm{\nu_{\mu}}$ is calculated from the neutrino fluence $\mathcal{F}_\mathrm{\nu_{\mu}}$ from one GRB through $J_\mathrm{\nu_{\mu}} =\mathcal{F}_\mathrm{\nu_{\mu}} \dot{N}/(4 \pi) $
    assuming a redshift of $z=2$ and $\dot{N} = 148$ yr$^{-1}$ for \SP \, and $\dot{N} = 93$ yr$^{-1}$ for \MP. 
    These rates were obtained by reducing the conventional number of observable GRBs $\dot{N} = 667$ yr$^{-1}$ accounting for the $E_\mathrm{iso}$ distribution.
    For comparison we show the current IceCube limits from \citet{Aartsen:2017wea} as well as the projected IceCube Gen2 limits (for 5000 bursts) from \citet{IceCube-Gen2:2020qha}, which assume that all GRBs contribute to the neutrino emission. \label{fig:neutrino}}
\end{figure*}

Very strong constraints on GRBs come from stacking searches \citep{IceCube:2012qza,Aartsen:2017wea}, which limit the prompt phase GRB contribution to $\lesssim 1\%$ of the diffuse astrophysical neutrino flux~\citep{IceCube:2022rlk}. Since we hypothesize that only the energetic GRBs produce neutrinos and UHECRs, the corresponding limits cannot be taken literally, but are a good indicator. The first question is how large the population of energetic GRBs is, which we address in \reffig{neutrino}, left panel, for the  \textit{Fermi}-GBM catalogue presented in \citet{Poolakkil:2021jpc}. From that distribution, we can estimate that about 22\% and 14\% of the GRBs are similar (in terms of energetics) to the \SP  (green shaded) and \MP\ (hashed blue) prototypes, respectively. These correspond to a local energetic GRB rate $\dot n_0 \simeq 0.02 - 0.1 \, \mathrm{Gpc^{-3}} \, \mathrm{yr}^{-1}$, as used in earlier studies (see e.g. Table~1 in \citet{Baerwald:2014zga}). 

We compute the expected diffuse fluxes from energetic GRBs for different baryonic loadings with similar methods as those used for the stacking limits, see \reffig{neutrino}, right panel. We can read off the figure that $f_\mrm{p/e} \lesssim 10$ for \SP \, and $f_\mrm{p/e} \lesssim 3$ for \MP \, are perfectly compatible with the current stacking limits, where the difference in neutrino flux mostly comes from the different $E_{\gamma,\mathrm{iso}}$ of the prototypes. For \MP \, with a shorter variability timescale, already $f_\mrm{p/e} = 0.3$ violates the stacking limit. This is also due to the lower peak energies of the neutrino spectra (that arise due to strong cooling effects of secondaries). The next-generation instrument IceCube-Gen2 \citep{IceCube-Gen2:2020qha} may test all scenarios in stacking searches. However, we note that the observational stacking limits are derived under the assumption that all GRBs produce neutrinos. We expect that the corresponding limits obtained selecting only energetic GRBs would be a factor of a few weaker, which does not affect our qualitative conclusions. Such a selection is, however, not trivial, as the redshift of the GRBs is not always known in the stacking samples, which means that $E_{\gamma,\mathrm{iso}}$ cannot be unambiguously derived.

\subsection{Requirements for UHECR energetics}\label{sec:UHECRs}

Following \citet{Baerwald:2014zga}, the required (isotropic-equivalent) energy ejection per GRB in UHECRs between $10^{10}$~GeV and $10^{12} \, \mathrm{GeV}$ to power the UHECR flux is
\begin{equation}
 E_{\mathrm{UHECR}} \simeq 10^{53} \, \mathrm{erg} \cdot \frac{\dot \varepsilon_{\mathrm{UHECR}}}{10^{44} \, \mathrm{erg \, Mpc^{-3} \, yr}}    \cdot \frac{\mathrm{Gpc^{-3} \, yr^{-1}}}{\dot n_0} \, ,
 \label{eq:euhecr}
\end{equation}
where $\dot \varepsilon_{\mathrm{UHECR}}$ is the required local energy injection density to describe data in that energy range.\footnote{Note that $\dot n_0$ is the apparent local GRB rate; since for UHECRs, not only GRBs emitting into our direction matter, the actual GRB rate is much higher. By using the isotropic-equivalent energy for UHECRs, that beaming effect cancels. } Using the baryonic loading, one can relate this energy per GRB to $E_{\gamma,\mathrm{iso}}$ by
\begin{equation}
E_{\mathrm{UHECR}} = E_{\gamma,\mathrm{iso}} \, f_\mrm{p/e} \, f_{\mathrm{bol}} \, f_{\mathrm{esc}}   \, ,
\label{eq:euhecr2}
\end{equation}
where $f_{\mathrm{bol}} \simeq 0.2$ (for an $E^{-2}$ spectrum) describes how much of the non-thermal proton energy falls into the UHECR range\footnote{For an $E^{-2.2}$ ($E^{-1.8}$) spectrum, one finds $f_{\mathrm{bol}} \simeq 0.03$ ($0.6$) for the parameters chosen in this paper. The required baryonic loading in  Eq.~\ref{eq:barload} is higher (lower) by the corresponding factor seven (three). However, note that this correction critically depends on the minimal proton energy for spectra softer than $E^{-2}$.}, and $f_{\mathrm{esc}}$ describes how many of the UHECR protons can escape; we use $f_{\mathrm{esc}} \simeq 0.5$ inspired by the values obtained in \citet{Heinze:2020zqb}. Combining \refeq{euhecr} with \refeq{euhecr2}, we can estimate the required baryonic loading for our population as
\begin{equation}
 f_\mrm{p/e}  \simeq 10 \cdot \frac{10^{54} \, \mathrm{erg}}{E_{\gamma,\mathrm{iso}}} \cdot \frac{\dot \varepsilon_{\mathrm{UHECR}}}{10^{44} \, \mathrm{erg \, Mpc^{-3} \, yr}} \cdot \frac{0.1 \, \mathrm{Gpc^{-3} \, yr^{-1}}}{\dot n_0} \, .
 \label{eq:barload}
\end{equation}
This means that the thick curves shown in \reffig{neutrino}, right panel, actually correspond to the estimated baryonic loadings from the UHECR paradigm: $f_\mrm{p/e}  \simeq 10$ for \SP, and $f_\mrm{p/e}  \simeq 3$ for \MP. On the other hand, the required 
$f_\mrm{p/e}$ for powering the UHECR does not depend on $\delta t_{\rm var}$. Hence,  bursts like \MP \, with short time variability cannot power the UHECRs if the stacking bounds are taken into account. This implies that the collision radius of the energetic GRBs has to be large enough to avoid efficient neutrino production. In conclusion, the population of energetic GRBs with large enough collision radii might power the UHECRs and at the same time produce a fraction of the diffuse astrophysical neutrino flux. The hypothesis is testable with the next generation of experiments.

\section{Discussion}
\label{sec:discussion}

\subsection*{Photon spectra and compatibility with observations}
Let us start by comparing the simulated photon spectra to the ones of \fermi-LAT detected events. For GRB~170214A, which served as inspiration for our \MP \, prototype, \citet{Tang:2017mmr} report that the photon spectral index in the LAT energy range is similar to or even steeper than the high-energy index $\beta$ in the GBM range. This would favor equipartition between electrons and the magnetic field (i.e. our SYN-dominated scenario) over an IC-dominated scenario. We point out that higher values of $f_\mrm{B/e}$ are in principle also possible; however in this case the magnetisation of the outflow may be considerable and as pointed out in \citet{2011MNRAS.416.2193N}, models based on magnetic energy dissipation should be considered instead of the internal shock scenario, such as \citet{Spruit:2000zm, Giannios:2007yj, Zhang:2010jt}. Further, the steep power-law index in the LAT range may be explained by a suppression of HE emission due to internal $\gamma \gamma$-absorption in case of small variability timescales/ low dissipation radii. 
Lepto-hadronic models have also been proposed in context of other LAT-detected bursts, e.g. in \citet{Asano:2009ta} for GRB~090510, and generally for a delayed HE emission in \citet{Asano:2012jr}. Especially interesting is GRB~090902B, which has excess at both low and high energies \citep{FermiGBM:2009ubu} -- thus, a flat underlying spectrum like in our lepto-hadronic SYN-dominated scenario. Independent of the spectral shape of the sub-MeV peak supposedly incompatible with a synchrotron origin \citep[see][]{Ryde:2010} it was pointed out that this flat component may be of hadronic origin in \citet{Wang:2018xkp}. Soft power-law components that extend from low to high energies are however also present in other bursts  \citep[see][]{Guiriec:2015ppa}. Generally, a flat additional component was equally predicted by a range of optically thin lepto-hadronic models such as \citet{Asano:2009ta, Murase:2011cx}, and we thus confirm their findings.

A major criticism of optically thin synchrotron models for GRBs arises from the typical slopes below the sub-MeV peak. For example, \citet{Poolakkil:2021jpc} reported value $\alpha \sim  - 1.1 $ for fluence spectra in the recent {\it Fermi} GBM catalog, while \citet{Yu:2016epf} found $\alpha \sim  - 0.8 $ for the time resolved spectra of brightest GBM bursts. For GRB~170214A, that inspired our \MP prototype, a Band-function fit with $\alpha = - 0.98 $ was reported in the GBM catalogue. Such a steep slope could not be reproduced in any of our models, generally the synchrotron fast-cooling prediction of $\alpha = - 1.5 $ was the highest low-energy slope found. We comment on an potential additional contribution from thermal electrons: Since one of the assumptions of our model is that only a small fraction $\zeta_\mrm{e}$, typically $\sim$ 0.1\%, of electrons are accelerated in a non-thermal power-law distribution, the low energy slopes can be further modified by the contribution of the remaining quasi-thermal population of electrons which were not considered in this work. We point out that, in order to compare e.g. to the GBM catalogue, data should be forward-folded with the detector response \citep[see e.g.][]{Burgess:2018dhc}. Additionally, the accuracy of reported spectra depends on the position of $E_\mrm{peak}$ in the detector energy range and whether the asymptotic slopes can be reached within the fitting range. 
As the spectral slopes also tend to be sensitive to the fitted function (\textit{e.g.} Band-function or smooth broken powe-law) we further discuss the spectral width or narrowness \citep{Axelsson:2014ora, Burgess:2017eek, Yu:2015jda} as a measure for spectra. 
The spectral width is related to the difference between the low- and high-energy spectral slopes (typically the photon indices $\alpha$ and $\beta$); simply put, if the difference between the two spectral slopes (ie., $\beta - \alpha$) is large, the spectrum is narrow. 
GRB spectra are generelly very narrow, for example when compared to typical spectra of Active Galactic Nuclei. In our model, the narrowest spectra that we find are the leptonic SYN-dominated ones. For IC-dominated models or for models with large contributions from secondary leptons we find larger high-energy slopes that induce wider spectra. 

Next, we comment on low-energy signatures in the eV range and compare with the respective conclusions of other works. The potential of constraining models and parameters by optical measurements was pointed out in \citet{Oganesyan:2019fpa}, who identified an additional break in the keV-range \citep[see also][]{Toffano:2021awx, Ravasio:2019kiw}. The photon indices below and above the break ($-2/3$ and $-3/2$ respectively) are consistent with an interpretation of marginally fast cooling particles
\footnote{Defined as the regime where the electron critical Lorentz factor $\gamma_\mathrm{e, c} \lesssim \gamma_\mathrm{e, min}$.}.
The possibility of a proton synchrotron interpretation  
was explored by \citet{Ghisellini:2019lgz, Florou:2021grp, Begue:2021pzr}. While the observed spectra can be reproduced in both scenarios, the inferred physical conditions might be in tension with observables like the short variability timescale and/or the energetics of GRB jets.
In the models we examined primary electrons were always in the fast-cooling regime, hence a cooling break in the keV range is not found. Moreover, the synchrotron emission of secondaries, whenever prominent, leads to photon indices close to $-2$ below a keV break energy.

Finally, we discuss the maximal baryonic loading compatible with observed photon spectra derived in the literature\footnote{Compatibility with observations basically means that the contribution of secondary particle emission does not outshine the primary photon emission.} \citep[e.g.][]{Asano:2008tc,Asano:2014nba,Petropoulou:2014awa}.
A summary of their findings can be found in Fig.~11 of \cite{Petropoulou:2014awa}, where the maximum baryonic loading is plotted against the  photon compactness of the emitting region, there defined as $\ell_\gamma = \sigma_\mrm{T} L_\gamma / (4 \pi R_\mrm{Coll} \Gamma_\mrm{em}^3 m_e c^3)$. This quantity depends on all parameters that affect the comoving photon density and, in turn, the photomeson interaction rate (i.e. the observed luminosity, the collision radius, and the Lorentz factor). Overall they conclude that for low compactness ($\ell_\gamma \sim 0.1$) the baryonic loading should be below $f_\mrm{p/e}\lesssim 30$, whereas in high-compactness scenarios ($\ell_\gamma \sim 100$) the baryonic loading is limited to much lower values, i.e. $f_\mrm{p/e} \lesssim 10^{0.25} \simeq 1.8$. To put our work in context of these results
we first compute the compactness of the representative collision of \SP, which is $\ell_\gamma = 1.3$. Throughout the full fireball evolution we find $\mrm{max}(\ell_\gamma) \simeq 7 $ for \SP \, and $\mrm{max}(\ell_\gamma) \simeq 41 $  for \MP. These high values of $\ell_\gamma$ are however achieved for collisions that do not dissipate most of the energy, and thus are not most relevant for constraining $f_\mrm{p/e}$. Indeed, it is the product of both the dissipated energy and $\ell_\gamma$ that is relevant for constraining the baryonic loading in a multi-collision scenario like ours. 

\subsection*{Baryonic loading in light of neutrino constraints}
The absence of neutrinos from GRBs~\citep{IceCube:2012qza,IceCube:2016ipa,Aartsen:2017wea,IceCube:2022rlk} has been a big mystery, especially if these are sources of UHECRs. In the simple one-zone model, where all internal shocks occur at the same $R_{\mathrm{Coll}}$, ``typical'' GRB parameters, i.e. $\Gamma \simeq 300$ and $\delta t_{\mathrm{var}} \simeq 0.01 \, \mathrm{s}$ (corresponding to $R_{\mathrm{Coll}} \simeq 5 \cdot 10^{13} \, \mathrm{cm}$) can be ruled out for the frequently used baryonic loading $f_\mrm{p/e} \simeq 10$, see e.g.~\citet{IceCube:2016ipa}. 

The baryonic loading used in neutrino studies is, however, an {\em ad hoc} assumption only roughly motivated by the UHECR connection, whereas it can be self-consistently derived in source-propagation models describing both cosmic rays and neutrinos. Systematic parameter space studies for protons~\citep{Baerwald:2014zga} and nuclei~\citep{Biehl:2017zlw} describing UHECR observations point towards either large $R_{\mathrm{Coll}}$ or low luminosities (see e.g. low-luminosity GRBs in~\citet{Zhang:2017moz, Rudolph:2021cvn}) as possible solutions to the tension with the neutrino GRB stacking searches, where the pion production efficiency is lower. Furthermore, it has been pointed out that the assumption of a single $R_{\mathrm{Coll}}$ may be too strong, as different messengers are preferably produced in different regions~\citep{Bustamante:2014oka,Bustamante:2016wpu, Rudolph:2019ccl}: neutrinos may be most efficiently produced close to the photosphere (the energy density scales $\propto R^{-2}$), whereas UHECRs prefer larger $R_{\mathrm{Coll}}$ where the energy density (to prevent nuclear disintegration) and the magnetic field (to prevent synchrotron losses limiting the maximal energy) are lower. 

GRB multi-collision models can produce lower neutrino fluxes, which allow for a combined description of UHECRs and neutrinos \citep{Globus:2014fka,Heinze:2020zqb}. For example, \citet{Heinze:2020zqb} obtain a baryonic loading $f_\mrm{p/e} \simeq 60-110$ to power the UHECRs with $R_\mrm{Coll} \simeq 10^{15}$ to $10^{16} \, \mathrm{cm}$, using  $E_{\gamma,\mathrm{iso}} \simeq  (4-8) \cdot \, 10^{52} \, \mathrm{erg}$ as the typical GRB isotropic energy.  In the multi-collision model for energetic GRBs considered here we find that $f_{\rm p/e}\sim 3-10$ is consistent with the energetic requirements for powering UHECRs without violating neutrino stacking limits. Our baryonic loading is lower than the one found by \citet{Heinze:2020zqb} due to the much higher energy budget per burst that compensates for small local GRB rate for bursts with $E_{\gamma, \rm iso} \gtrsim 10^{54}$~erg. Note that spectral effects can act in the other direction and enhance the neutrino flux, as discussed in Sec.~\ref{sec:leptohadronicmodel}.

\subsection*{UHECR composition}
To fully describe UHECR data one also needs to take into account the information about composition. We do not explicitly describe UHECR composition in this work, as this requires a more sophisticated numerical treatment of the radiation processes in the source, including nuclear disintegration and emission processes of all by-products. This is still beyond the state-of-the-art for self-consistent hadronic radiation source models. Nevertheless, qualitative changes are not expected because our production radii are large enough that the nuclear cascade cannot develop~\citep{Biehl:2017zlw} (Figs.~7 and 8) and the energetics requirement for nuclei is similar to that of protons~\citep{Jiang:2020arb}; however, the composition should be described by a mix of injection isotopes similar to \citet{Globus:2014fka,Heinze:2020zqb}. As shown in~\citet{Biehl:2017zlw}, the prompt GRB neutrino fluence exhibits very little dependence on the injection composition; one reason is that the secondary cooling (which depends on $B'$ but not on the composition) is important for the peak. The electromagnetic signatures of nuclei can also be speculated to be similar for an $E^{-2}$ injection spectrum, as each nucleus with energy $E_A$ can be (to first order) regarded as a superposition of $A$ nucleons of energy $E_A/A$. Our qualitative conclusions are thus expected to hold roughly even if heavier nuclei are accelerated in the source. This has been demonstrated in \citet{Globus:2014fka,Heinze:2020zqb}, but still lacks the self-consistent treatment of the electromagnetic signatures of nuclei at the level used for protons in this study.

\subsection*{Why should energetic GRBs be more favourable for UHECR production than ``ordinary'' GRBs?}

In this study we argued that energetic GRBs are sufficient for powering UHECRs without violating current neutrino constraints. An interesting question that arises then is whether these energetic GRBs are special in terms of proton/nuclei acceleration compared to the more numerous and less energetic GRBs. The pion production efficiency to photohadronic interactions can be analytically estimated from observables as (see Eq.~A16 in ~\citet{Guetta:2003wi})
\begin{equation}
f_\pi \propto \frac{L_{\gamma,\mathrm{iso}} \, \delta t_{\mathrm{var}} }{ R_{\mrm{Coll}}^2 \, \epsilon_{\gamma, \mathrm{eff}}}
\propto \frac{ L_{\gamma,\mathrm{iso}}}{\Gamma^4 \, \delta t_{\mathrm{var}} \, \epsilon_{\gamma,\mathrm{eff}} }  \,.
\label{equ:fpi}
\end{equation} 
Here, the co-moving photon number density is estimated from the photon energy divided by co-moving volume, $\epsilon_{\gamma, \mathrm{eff}}$ is the photon energy where the number of photons is maximal, 
and the shell width is estimated from $\delta t_{\mathrm{var}}$. This means that similar hadronic signatures on the GRB prompt emission are expected if the relationships $L_{\gamma,\mrm{iso}} \propto R_{\mrm{Coll}}^2$, or $L_{\gamma,\mrm{iso}} \propto  \Gamma^4$ are obeyed, neglecting the dependence on $\delta t_{\mathrm{var}}$ for the sake of simplicity; see e.g. Fig.~8 in \citet{Biehl:2017zlw} for nuclear disintegration, where this degeneracy can be clearly seen. In that sense, GRBs with very different parameters (e.g. $L_{\gamma,\mathrm{iso}}$, $\Gamma$, $R$, $\epsilon_{\gamma,\mathrm{eff}}$) can have similar properties in terms of the photon density, such as the low luminosity GRBs and large $R_{\mathrm{Coll}}$ regions mentioned earlier.

On the other hand, an empirical correlation between the isotropic energy and bulk Lorentz factor has been reported, namely $E_{\gamma,\mrm{iso}} \propto \Gamma^{2.7}$ for a wind-like circumburst medium or $\propto \Gamma^{2.2}$ for the homogeneous case -- see Eq.~20 in \citet{Ghirlanda:2017opl}. This empirical relation was obtained by using a sample of GRBs for which the peak time of the afterglow light curve (and the redshift) were measured. The bulk Lorentz factor before the deceleration can be then inferred from the peak time, assuming a density profile for the circumburst medium, and the isotropic energy of the afterglow after the prompt phase. It is clear that the $E_{\gamma,\mrm{iso}}-\Gamma$ relationship scales differently from the above relationship and therefore breaks that degeneracy. This means that less energetic GRBs, which have lower Lorentz factors according to the $E_{\gamma,\mrm{iso}}-\Gamma$ relationship, have larger photon densities and therefore qualitatively different properties. 
Here efficient photohadronic interactions (such as nuclear disintegration) might affect the acceleration of nuclei, and these GRBs may consequently have lower baryonic loadings (describing the non-thermal population).  

\subsection*{Potential future directions}
Possible extensions of the presented work may concern theoretical aspects (e.g. improvement of the modeling) or connection to new experimental results. First of all, a refined outflow model informed by magnetohydrodynamic simulations \citep[as presented for example in][]{Gottlieb:2021avb, Gottlieb:2022sis, Gottlieb:2022tkb} might provide a more realistic picture of the physical conditions in the plasma, such as locations of energy dissipation and magnetization. Alternatively, one could lift the assumption of a top-hat jet model made here, and apply our work to a structured jet, similar to the one inferred 
by observations of GRB~170814A \citep{Troja:2018uns}. One might further consider the feedback between different radiation zones in the jet, which might affect the photohadronic interaction rate. 
In our work the properties of accelerated particle distributions were motivated by observations instead of being informed by theories of particle acceleration. For instance, the default value for the power-law slope of the accelerated protons ($p_{\rm p}=2$) corresponds to the one usually assumed in an UHECR context. On the contrary, a steeper power law was assumed for the accelerated electrons in order to reproduce the typical photon spectrum in the GBM band.  
Particle acceleration processes can be studied self-consistently using Particle-In-Cell (PIC) simulations. It is noteworthy that the parameter space of mildly relativistic shocks relevant for the GRB prompt emission was explored with PIC simulations only recently 
\citep[for a review see][]{Marcowith:2020vho}. 
For example in the context of highly magnetised jets, where magnetic reconnection is the main energy dissipation mechanism, similar power-law indices for electrons and protons are found \citep[e.g. ][]{Guo:2015ydj, Petropoulou:2019bse}. For non-magnetised relativistic collisionless shocks, \citet{Sironi:2013ri} indeed find an electron power-law of $p_\mrm{e} = 2.5 $ differing from the proton index $p_\mrm{p} = 3.0$. Weakly magnetised mildly relativistic shocks (that correspond to the conditions of the outflow assumed in our model) have been explored in \citet{Crumley:2018kvf}, who report an electron index of $p_\mrm{e} \sim 2.2 $ but no good constraints on the ion indices are found. For magnetised mildly relativistic shocks, \citet{Ligorini:2021lbj} find a small supra-thermal population of electrons with  $ p_\mrm{e} \sim  2 $.  
If the acceleration process produces similar power-law indices, then a steeper proton power-law index than the one used here will affect the neutrino spectra. First and foremost, since less energy is deposited in energetic protons, the overall neutrino flux will be lower. Additionally, the neutrino spectral shape might differ, depending on the shapes of the photon and proton spectra.
Finally we point out that in order to describe UHECR data, nuclei heavier than protons should be included in the radiation modeling. 
In the past, \citet{Globus:2014fka, Heinze:2020zqb} have treated the nuclear cascade of elements up to iron, however without taking the feedback of nuclei on photon spectra into account. In a fully self-consistent picture, disintegration photons may produce photon cascades and the systems of nuclei and photons should not be treated separately. 
Radiation models can be improved further by making a better connection to observational data. For electromagnetic spectra, an adequate procedure would be to forward-fold predictions with detector responses in order to assess the feasibility of models \citep[as shown in][]{Burgess:2018dhc}. 
For multi-messenger searches, separate analysis for different sub-classes of GRBs could be included in future instruments like IceCube Gen2~\citep{IceCube-Gen2:2020qha} and KM3NeT~\citep{KM3Net:2016zxf}. Alternatively, the contribution of different GRBs could be weighted by their luminosity~\citep[see \textit{eg.}][]{Lucarelli:2022ush}.

\section{Summary \& Conclusion}
\label{sec:summary}
In this paper we investigated lepto-hadronic radiation models for the prompt phase of energetic GRBs with $E_\mrm{iso} \gtrsim 10^{54}$~erg. 
Our study was motivated by the fact that a sub-class of energetic events may power the UHECR energy flux without requiring excessive  
baryonic loadings. Moreover, strong constraints may be placed on the baryonic loading from the non-observation of associated neutrinos from such energetic bursts. Finally, lepto-hadronic models have been proposed for the HE emission detected in some energetic GRBs by the \fermi-LAT.

Our aim was to understand whether the baryonic loading necessary to power the UHECR energy flux is feasible in light of multi-messenger constraints. For this we modelled the prompt phase photon and neutrino emission of two prototype GRBs (inspired by LAT-detected events) within a multi-collision internal shock model that accounts for different emission regions along the jet. 
In contrast to past multi-collision approaches we self-consistently accounted for the feedback between photon and proton spectra by a
a time-dependent radiation model treating the coupled evolution of hadrons, leptons (including neutrinos) and photons. allowed us to examine the contribution of different radiative processes at different energies and times, including synchrotron radiation and synchrotron self-absorption, inverse Compton scattering, photo-pair and photo-pion production, and  $\gamma\gamma$-annihilation.
For each prototype we evaluated synchrotron (SYN)- and inverse Compton (IC)-dominated scenarios and discussed the maximal baryon loading compatible with observations (for which the secondary cascade emission would not outshine the emission of primary electrons in the GBM band). Beyond single-source spectra and light curves, we compared the diffuse neutrino flux expected for a sub-population of energetic events to current IceCube and predicted IceCube-Gen2 limits. Below we summarise our main findings:
\begin{enumerate}
\item \textit{Maximum baryon loading.} This was determined by electromagnetic observations (i.e. spectral modifications in the GBM/LAT band) and neutrino observations (i.e. GRB stacking limits) and was found to be about 3--10 (depending on $E_{\gamma,{\mathrm{iso}}}$), assuming a $E^{-2}$ proton spectrum for average dissipation radii around $10^{16}$~cm. This value is consistent with the baryonic loading needed to power the UHECRs with energetic GRBs only.  Note that empirical $E_{\gamma,\mrm{iso}}-\Gamma$ relationships points towards large dissipation radii for energetic GRBs.
\item \textit{Neutrino signatures}. For large enough collision radii, the non-observation of single energetic events in high-energy neutrinos and the current IceCube stacking limits are consistent with the same baryonic loadings of about 3--10. The corresponding neutrino spectra peak at higher energies than previously anticipated, making energetic GRBs interesting targets for radio detection of neutrinos \citep[with future detectors such as GRAND,][]{GRAND_2020}. 
For smaller collision radii (corresponding to shorter variability timescales), pion and muon cooling reduces the neutrino peak energies, bringing them within the range where current neutrino observatories are more sensitive. Consequently, $f_\mrm{p/e}$ can be limited to smaller values. Note that the neutrino fluence can be enhanced by spectral effects, such as the spectral index in the synchrotron fast-cooling regime or electromagnetic cascade contributions -- which we self-consistently include.  
\item \textit{Signatures in photon spectra}. 
We limit ourselves to photon energies below $\sim$100~GeV that will survive propagation.
For SYN-dominated scenarios -- even for the moderate baryonic loading of $3-10$ -- the synchrotron emission from secondaries introduces an additional broadband power-law component with photon index $-2$. This implies correlated increase in fluence in the optical-UV, soft X-ray and HE regimes. 
Hence, to identify hadronic signatures in these energy ranges, prompt low-energy observations combined with HE observations by \fermi-LAT would be crucial. The lack of spectral breaks at low energies (UV/X-rays) and/or at high energies (LAT) would exclude low values of $\epsilon_{\rm B}/\epsilon_{\rm e}$ and constrain even more the baryonic loading.
For IC-dominated scenarios and baryonic loading $< 100$, the broadband photon spectra obtained in leptonic and lepto-hadronic models are indistinguishable. For these cases, neutrinos are thus the only messengers that could pinpoint hadronic interactions.
Generally, a IC-dominated scenario may be identified by a low relative X-Ray/UV flux, and a high HE flux. 
\item \textit{The role of different collision radii}.
The typical collision radius, which is usually determined by the variability timescale and Lorentz factor of the outflow, is important for models with density-dependent processes (e.g. IC and photo-pion production). Only purely leptonic SYN-dominated models are found to be robust against variations of the collision radius. 
While electromagnetic spectral features at low energies (below a few keV) and high energies (above $\sim 100$ MeV) are produced in collisions taking place in high-density regions of the jet, the peak (sub-MeV to MeV part) of the spectrum is mainly determined by collisions that dissipate most of the energy. Consequently, the region where most of the energy is dissipated is representative for the complete burst in that energy range; on the other hand, it is not representative for the whole GRB's neutrino emission, as it would overestimate the neutrino fluence and underestimate the peak energy. 
\end{enumerate}
We conclude that energetic GRBs are promising candidates for multi-messenger searchers.
Baryonic loadings as high as necessary to power UHECRs probably do not modify the spectral energy distribution as much as to rule out the UHECR paradigm -- in IC-dominated scenarios, these cannot even be uniquely identified. However, interesting insights can be obtained in SYN-dominated scenarios from correlated flux enhancements between the optical-UV to soft X-rays, and the GeV to TeV gamma-ray bands. Finally, the discovery of a neutrino from an energetic GRB, such as GRB~221009A, would be the smoking gun signature for our hypothesis that a class of energetic GRBs could be the sources of UHECRs.

\textit{Note. --} During completion of this work GRB~221009A was detected. As this is likely to become the most prominent example for this class of energetic GRBs, we discuss the implications of our model for this burst in a companion letter~\citep{Rudolph:2022dky}. 
\acknowledgments 
We would like to thank Iftach Sadeh and Marc Klinger for reading the manuscript and useful comments. 
M.P. acknowledges support from the MERAC Fondation through the project THRILL and from the Hellenic Foundation for Research and Innovation (H.F.R.I.) under the ``2nd call for H.F.R.I. Research Projects to support Faculty members and Researchers'' through the project UNTRAPHOB (Project ID 3013). A.R. was supported by the International Helmholtz-Weizmann Research School for Multimessenger Astronomy, largely funded through the Initiative and Networking Fund of the Helmholtz Association and received funding from the Carlsberg Foundation (CF18-0183).

\software{\soft{astropy}\citep{Astropy:2013muo, Astropy:2018wqo}, \soft{gammapy} \citep{Deil:2017yey, Nigro:2019hqf} and Python v3.9.}

\vspace{5mm}

\bibliography{references}

\appendix

\section{Time evolution in the representative collision}
\label{app:buildup_spectra} 
\begin{figure*}
    \centering
    \includegraphics[width = 0.5 \textwidth]{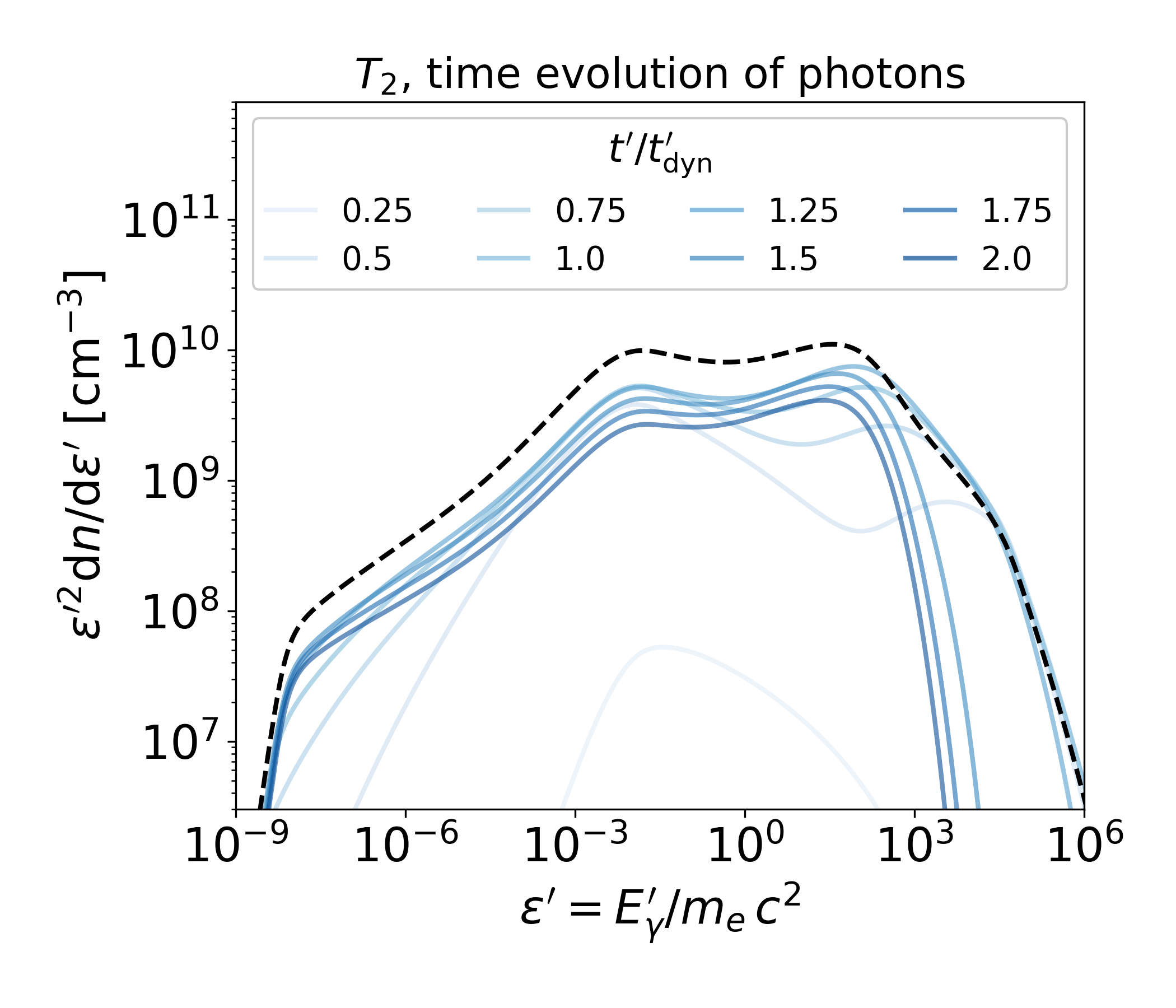}
    \includegraphics[width = 0.48 \textwidth]{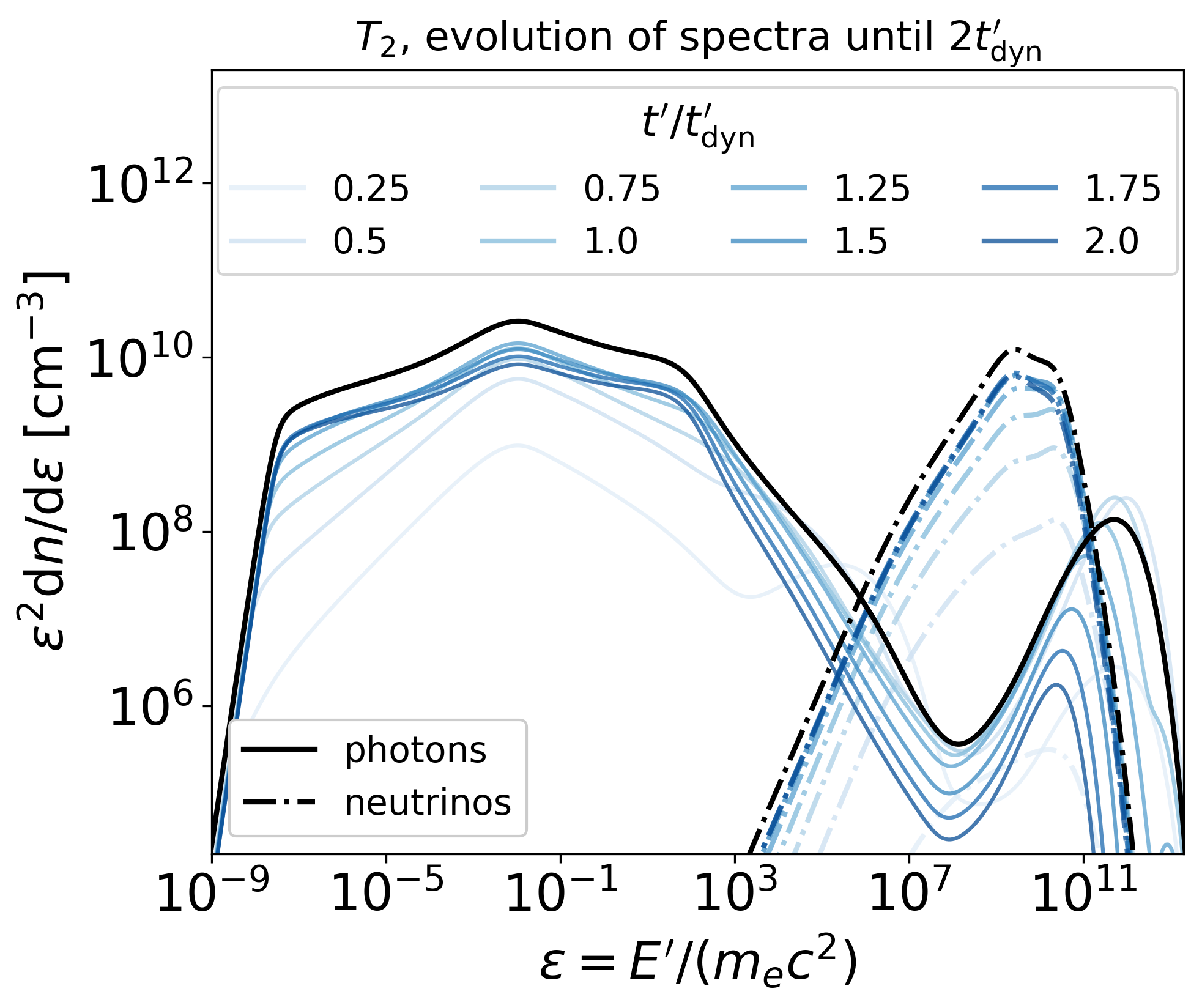}
    \caption{Temporal evolution of comoving photon spectra until 2$t^\prime_\mathrm{dyn}$ for our representative collision, assuming $t^\p_\mrm{inj} = t^\p_\mrm{dyn}$. 
    We show the result for the \textit{(left)} IC-dominated leptonic case and \textit{(right)} SYN-dominated case with $f_{p/e} = 30$. Black lines indicate the emitted spectra.}
    \label{fig:evolution_comoving_spectra}
\end{figure*}
For each collision we evolve the system until $t^\prime = t^\prime_\mrm{inj} + t^\prime_{dyn}$ and calculate the emitted spectrum integrating over the escaping the photon spectrum. As this approach differs from the one of past publications \citep[e.g.][]{Bosnjak:2008bd, Rudolph:2021cvn}, we illustrate its effect in \reffig{evolution_comoving_spectra} for the representative collision of \SP, showing the IC-dominated leptonic scenario on the left and the SYN-dominated lepto-hadronic scenario on the right. 

Solid lines represent the comoving photon spectrum at different simulation times $t^\prime$, where early (late) times correspond to light (dark) blue colors. The emitted spectra are shown in black.
We first discuss the leptonic IC-dominated case (left plot). For $t^\prime < 0.75 t^\prime_{dyn}$ the synchrotron peak around $E^\prime \simeq 10^{-2} m_e c^2$ and the inverse Compton peak at higher energies are clearly visible. At later times, synchrotron and inverse Compton radiation of secondary leptons enhances the photon fields at low and intermediate energies. 
The emitted spectrum encodes all this information. In principle, the same behaviour is observed for the photon fields of the lepto-hadronic case (right plot). Here we further notice how internal $\gamma \gamma$-annihilation suppresses VHE $\pi^0$-decay with increasing simulation time. 

The neutrino spectra (that are composed of $\nu_\mu$, $\bar{\nu}_\mu$, $\nu_e$, and $\bar{\nu}_e$) also show an evolution with time. For example, the peak of the spectrum evolves to lower energies with time. This may be understood as follows: early on, the spectrum is dominated by $\nu_\mu$ from pion decays. Later, the decay products of muons dominate. However, due to the muons being subject to synchrotron cooling, those secondaries have lower peak energies. 

Overall our approach equally captures features that are produced early on or late in the simulation. This may not be true  e.g. for a steady-state solution.

\section{Impact of modeling choices}\label{app:modelling-choices}
\begin{figure*}
    \centering
    \includegraphics[width = 0.49 \textwidth]{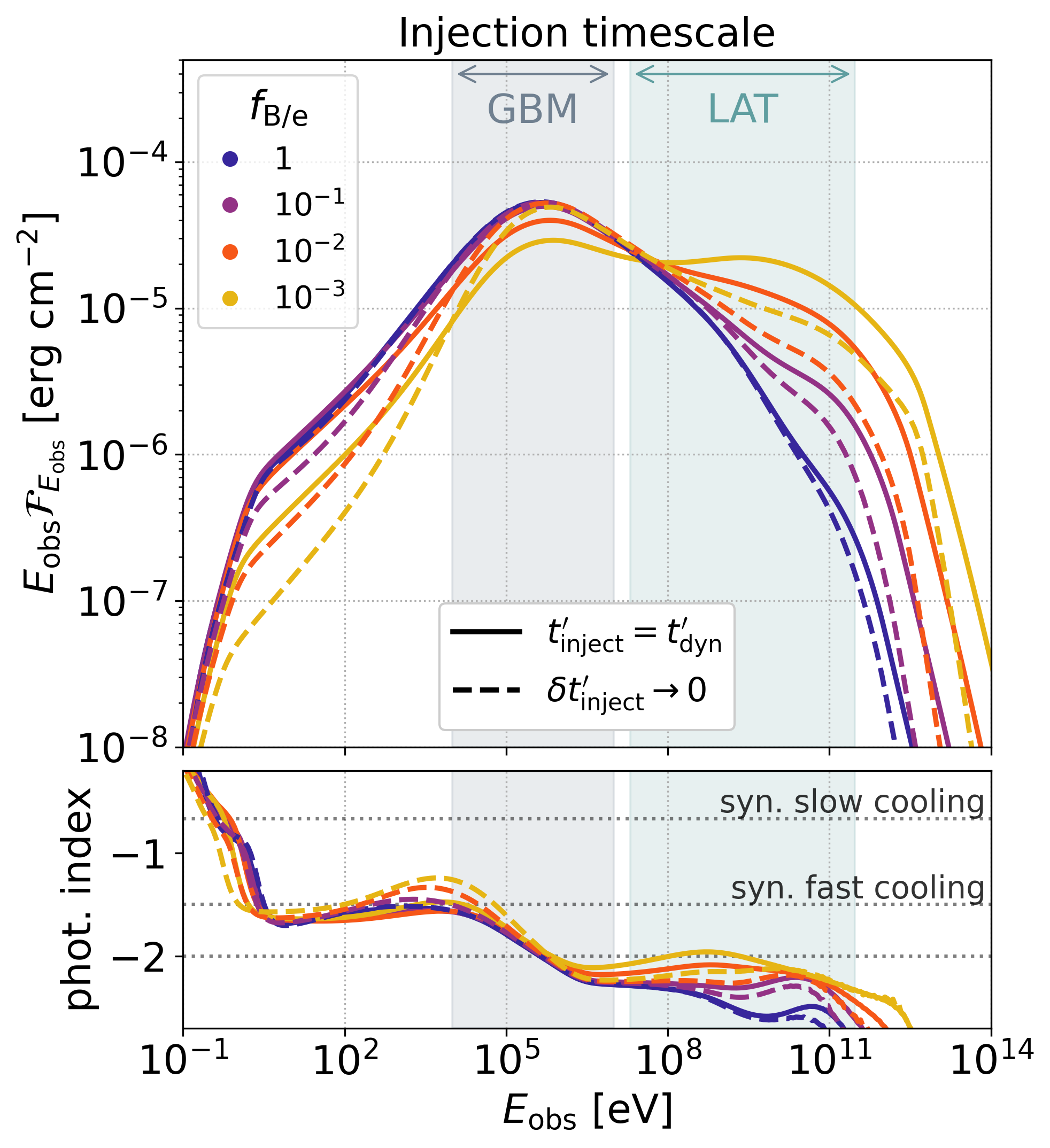}
    \includegraphics[width = 0.47 \textwidth]{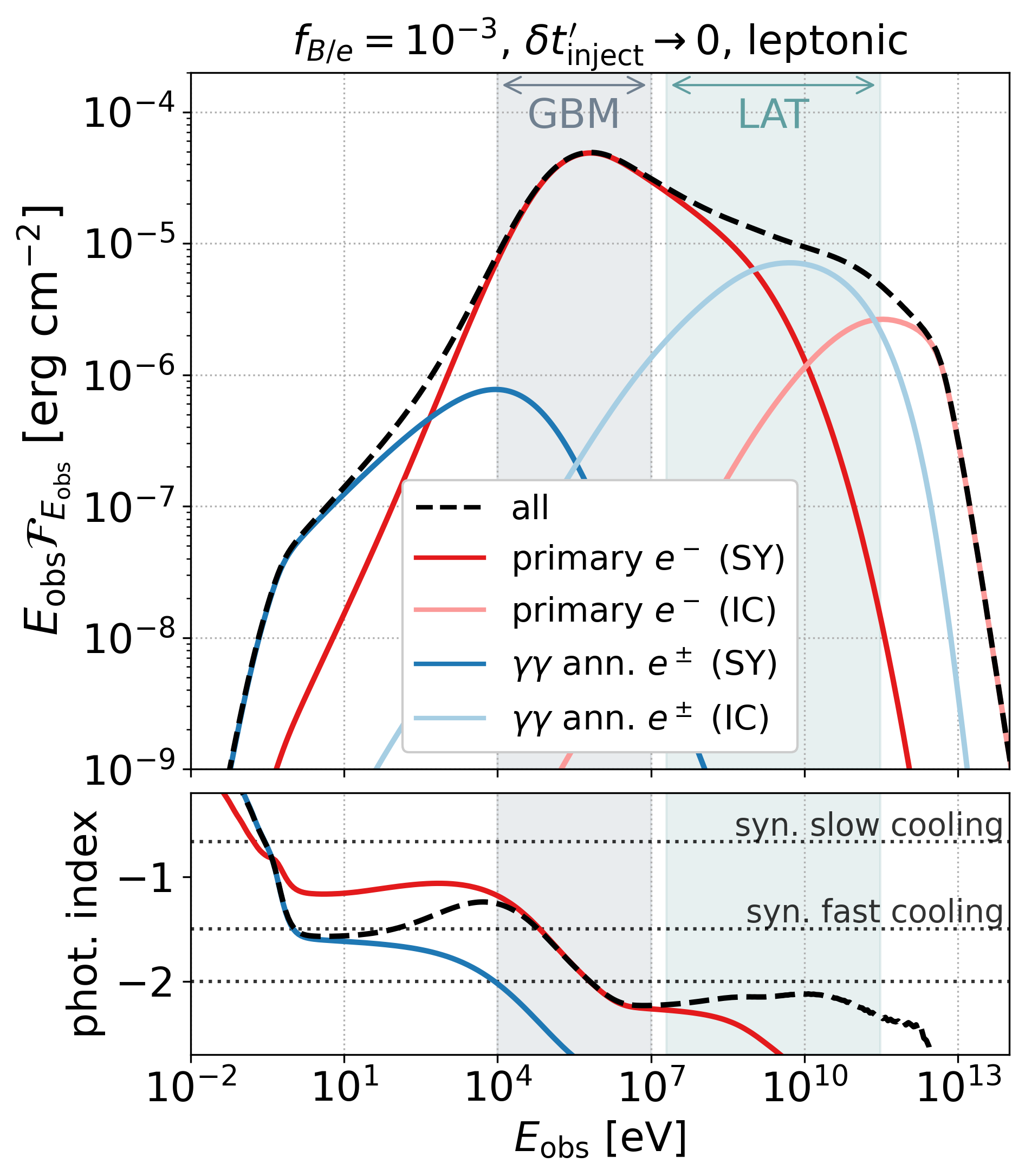}
    \\
    \vspace{3mm}
   \includegraphics[width = 0.49 \textwidth]{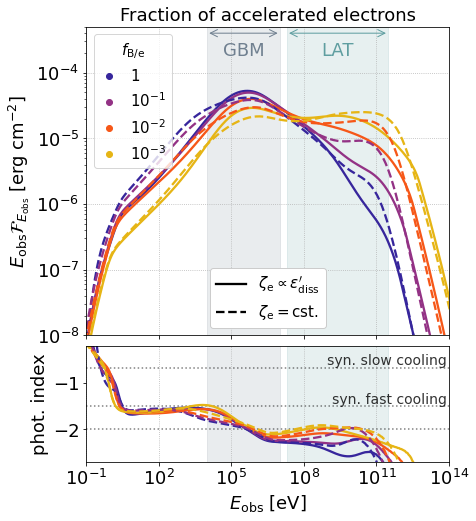}
    \includegraphics[width = 0.49 \textwidth]{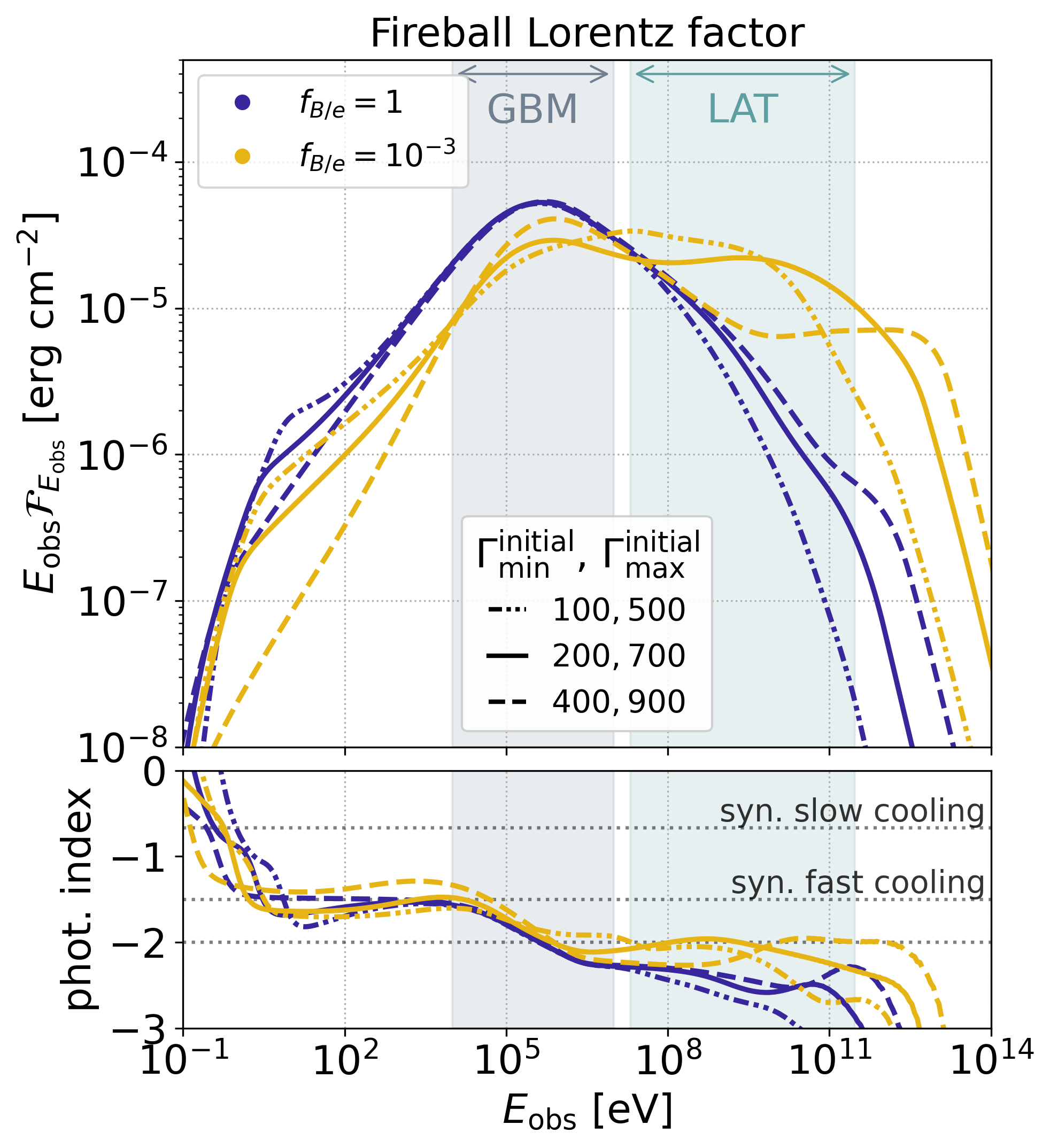}
    \caption{Examining the impact of different modeling choices in a leptonic modeling setup for the single-pulse scenario at the time-integrated spectra $E_\mathrm{obs} \mathcal{F}_{E_\mathrm{obs}}$. We show comparisons of \textit{(top left)} different injection schemes, \textit{(bottom left)} different choices for the fraction of accelerated electrons and \textit{(bottom right)} different initial Lorentz factors of the outflow. In all figures, we explore different $f_\mrm{B/e}$.
    In the \textit{(top right)} panel we further show the decomposed spectrum for $\delta t^\p_\mrm{inj} \rightarrow 0$ and $f_\mrm{B/e} = 10^{-3}$. 
   For the photon indices in the lower panels we indicate the synchrotron slow- and fast-cooling predictions as well as a photon index of $-2$ (that marks local maxima/minima of $E_\mathrm{obs} \mathcal{F}_{E_\mathrm{obs}}$) as dotted lines.}
    \label{fig:hl_spectra_leptonic}
\end{figure*}

To complement the main results, we systematically explore the impact of the following parameter assumptions and modeling choices: \textit{(a)} The injection timescale of accelerated particles, \textit{(b)} the scaling of the number fraction of accelerated electrons throughout the fireball evolution, \textit{(c)} the maximum and minimum (initial) Lorentz factor of the outflow and \textit{(d)} the magnetic field strength (set through $f_\mrm{B/e}$). For $f_\mrm{B/e}$ the maximum (minimum) value considered corresponds to the SYN-dominated (IC-dominated) scenario presented in the main text.
For simplicity we limit ourselves to leptonic models and show the simulated full-burst spectra examining these aspects in \reffig{hl_spectra_leptonic}. 
We will discuss the results in context to the parameter and model choices generally applied in GRB radiation models and necessary to reproduce spectra with properties similar to observed ones. 

\subsection*{(a) Injection timescale of accelerated particles}
In \cite{Bosnjak:2008bd, Daigne:2010fb, Bosnjak:2014hya}, 
the injection timescale of primary electrons (for each single-collision radiation calculation) was considered to be much smaller than the dynamical timescale. 
To scrutinise the implications of different choices of the injection timescale, we show the simulated spectra for $t^\p_\mrm{inj} = t^\p_\mrm{dyn}$ and $t^\p_\mrm{inj} = 0.01 t^\p_\mrm{dyn}$ (labelled as $\delta t^\p_\mrm{inj} \rightarrow 0$) in the top left plot \reffig{hl_spectra_leptonic}. 

We find that the results for the two injection timescale agree well for strong magnetic fields (high $f_\mrm{B/e}$). For low $f_\mrm{B/e}$, the results for $\delta t^\p_\mrm{inj} \rightarrow 0$ differ in three main aspects:  \textit{(1)} The fluence of at the sub-MeV peak does not change with respect to the one for high $f_\mrm{B/e}$, \textit{(2)} the (V)HE component and low-energy fluences are weaker than for $t^\p_\mrm{inj} = t^\p_\mrm{dyn}$. 
and \textit{(3)} the low-energy photon index depends systematically on $f_\mathrm{B/e}$ (where high $f_\mathrm{B/e}$ yield low photon indices and vice versa, see also \cite{Rudolph:2021cvn}). On the other hand, the low-energy photon index is almost independent of $f_\mrm{B/e}$ for $t^\p_\mrm{inj} = t^\p_\mrm{dyn}$. As these effects appear for low $f_\mrm{B/e}$, we attribute them to inverse Compton scatterings that are more efficient in case weak magnetic fields. 
We relate this to the three mains points identified before: 
\textit{(1)} The higher fluence at the sub-MeV peak may be explained through the dominating energy loss mechanism of primary electrons at $\gamma^\p_\mrm{e, min}$: for $\delta t^\p_\mrm{inj} \rightarrow 0$ energetic primaries are present only at early $t^\p$. However, at this point the photon spectrum is still building up and cannot serve as target for inverse Compton scatterings. Consequently, the efficiency of inverse Compton scatterings is low(er) and primary electrons at $\gamma^\p_\mrm{e, min}$ predominantly cool through synchrotron emission also in low $f_\mathrm{B/e}$ scenarios. 
This, in turn, also means that inverse Compton scatterings are generally less efficient, which yields lower (V)HE photons and also less contributions from the secondary cascade. This is further illustrated in the top right plot of \reffig{hl_spectra_leptonic}, where we show the decomposed spectrum for $\delta t^\prime_\mrm{inj} \rightarrow 0 $ and $f_\mrm{B/e} = 10^{-3}$: Despite the fact that the parameter choices correspond to the IC-dominated scenario, primary synchrotron emission still dominates the observed spectrum. This implies that the low- and high-energy fluences which are produced by cascade emission are lower than for $t^\p_\mrm{inj} = t^\p_\mrm{dyn}$ (point \textit{(2)} mentioned before).
The lower contribution of the secondary cascade also explains why harder low-energy photon indices can be achieved: as detailed above, inverse Compton scatterings in Klein-Nishina regime modify the low-energy slope of the primary synchrotron such that values up to $-1$ can be achieved. This was also observed in \citet{Daigne:2010fb}, but could not be reproduced by our results in the main text due to the strong contribution of synchrotron emission of secondary lepton pairs. Now choosing the same setup as \citet{Daigne:2010fb} we can however confirm their findings. 

We conclude that  SYN-dominated models (i.e. with high $f_\mrm{B/e}$) are robust against variations of the injection timescale of energetic particles, whereas low-$f_\mrm{B/e}$ scenarios are sensitive to the choice of $t^\prime_\mrm{inj}$. Meanwhile, the findings of past publications of \citet{Daigne:2010fb} can only be reproduced for the same settings of the injection timescale.
 
\subsection*{(b) Evolution of the number fraction of accelerated electrons $\zeta_\mrm{e}$}
GRB synchrotron models within the internal shock scenario commonly set the number fraction of accelerated electrons proportional to the dissipated energy per unit mass ($\zeta_\mrm{e} \propto \varepsilon^\p_\mrm{diss}$). The intention is to keep the minimum Lorentz factor of electrons $\gamma^\p_\mrm{e, min}$ constant throughout the fireball evolution to reproduce the correct spectral evolution. Additionally, it results in narrow(er) spectra. 
In the lower left panel of \reffig{hl_spectra_leptonic}, we show the results for $\zeta_e \propto \varepsilon^\p_\mrm{diss}$ and $\zeta_\mrm{e} = \mrm{cst.}$ (with our benchmark choice of $t^\p_\mrm{inj} = t^\p_\mrm{dyn}$ and setting the constant $\zeta_e$ such that the same peak energy is reproduced). Independent of $f_\mrm{B/e}$ the simulations with $\zeta_\mrm{e} = \mrm{cst.}$ are different in two main points. First, the synchrotron peak is broadened and reduced in energy fluence. Second, the low-energy and high-energy fluences are enhanced. The broad synchrotron peak is reflected in the photon indexes, which are relatively flat around the peak. With values that are overall below $-1.5$, it is questionable whether  $\zeta_\mrm{e} = \mrm{cst.}$ can produce narrow spectra compatible with observations. 

We conclude that $\zeta_\mrm{e} \propto \varepsilon^\p_\mrm{diss}$ (which yields a constant minimum Lorentz factor of electrons $\gamma^\p_\mrm{e, min}$ throughout the fireball evolution) is necessary to reproduce spectra compatible with observed ones within a multi-collision model. We point out that this is a somewhat fine-tuned choice and should be confirmed by simulations of particle acceleration in (mildly) relativistic shocks for a broad range of shock strengths and comoving energy densities.

\subsection*{(c) Initial jet Lorentz factor}
Next, we examine the impact of the jet Lorentz factor. We choose three different initial Lorentz factor distributions, characterised by their minimal and maximal $\Gamma$:
\begin{equation*}
(\Gamma_\mrm{min}^\mrm{inital}, \Gamma_\mrm{max}^\mrm{inital}) \in \{ (100, 500), (200, 700), (400, 900) \} \, . 
\end{equation*}
For the sake of simplicity, the lower right panel of \reffig{hl_spectra_leptonic} only shows the results for $f_\mrm{B/e} = 1$ and $f_\mrm{B/e} = 10^{-3}$ (i.e. the SYN- and IC-dominated scenario introduced in the main text). For all cases we adjusted $\zeta_\mrm{e,0}$ such that the peak energy remains the same.
We find that for the SYN-dominated scenario ($f_\mrm{B/e} = 1$), the results are robust against variations of the initial Lorentz factor distribution. 
However, the IC-dominated scenario of $f_\mrm{B/e} = 10^{-3}$ shows strong dependence on the Lorentz factor distribution: the lower the Lorentz factor is, the stronger the distortions of the spectrum are. A natural explanation for this behaviour is the impact of lepton pairs produced in $\gamma \gamma$-annihilation, which are produced in dense environments and distort the spectrum by synchrotron radiation at low energies and inverse Compton radiation at high energies. For low Lorentz factors, the comoving densities are higher (due to smaller collision radii and a smaller effect of boosting). Consequently, the $\gamma \gamma$-annihilation efficiency is increased and a large number of pairs reshape the spectrum. 

We point out that the initial jet Lorentz factors are only one way of altering the comoving (energy) densities. Other parameters would be the overall duration of the burst (short durations yield small collision radii and hence large densities) or the isotropic equivalent energy (large $E_\mrm{\gamma, iso}$ induce large densities). 

\subsection*{(d) Magnetic field strength}
We proceed by reviewing the impact of the magnetic field strength (set through $f_\mrm{B/e}$), which we varied for all scenarios presented above. 
The most extreme choices presented as blue and yellow lines in \reffig{hl_spectra_leptonic} correspond to the SYN- and IC-dominated scenario presented in the main text. 
Varying $f_\mrm{B/e}$ systematically between these two values we observe a smooth transition between the two scenarios: the features of the IC-dominated scenario such as an increase in low- and high-energy fluence become more pronounced as $f_\mrm{B/e}$ is decreased.
As outlined above, for $\delta t^\prime_\mrm{inject} \rightarrow 0 $ also the low-energy photon index depends systematically on $f_\mrm{B/e}$, where low (high) $f_\mrm{B/e}$ yield hard (soft) photon indices. This however cannot be confirmed in scenarios with a longer injection timescale.

\section{Pion and muon cooling and impact on neutrino spectra}
\label{app:pion_muon_cooling_neutrinos}
In this section we study the cooling of intermediate pions and muons and its impact on neutrino spectra. 
We account for synchrotron and adiabatic cooling and decay. For a particle of Lorentz factor $\gamma^\p$ and species $i$ (with corresponding mass $m_i$, charge number $Z_i$ and rest-frame lifetime $\tau_{0,i}$ ) the corresponding cooling/decay times in the plasma rest frame are
\begin{eqnarray}
t^\p_\mrm{sy, i} (\gamma^\p) &=& \frac{6 \pi}{Z_i^4 \sigma_t}\frac{m_i c^2}{c} \left( \frac{m_i}{m_e} \right)^2 \frac{1}{B^{\p 2} \gamma^\p} 
\label{eq:tprime_sy}\\
t^\p_\mrm{ad, i} &=& t^\p_\mrm{ad}  \\
t^\p_\mrm{dec, i}(\gamma^\p)  &=& \tau_{0,i} \gamma^\p \, , 
\label{eq:tprime_decay}
\end{eqnarray}
with corresponding cooling rates $t^{\p -1}_\mrm{sy, i} (\gamma^\p) $, $t^{\p -1}_\mrm{ad, i}$ and  $t^{\p -1}_\mrm{dec, i} (\gamma^\p) $. 

\subsection{Pion and muon cooling in the representative collision and identification of cooling regimes}

\begin{figure}
    \centering
    \includegraphics[width = 0.43 \textwidth]{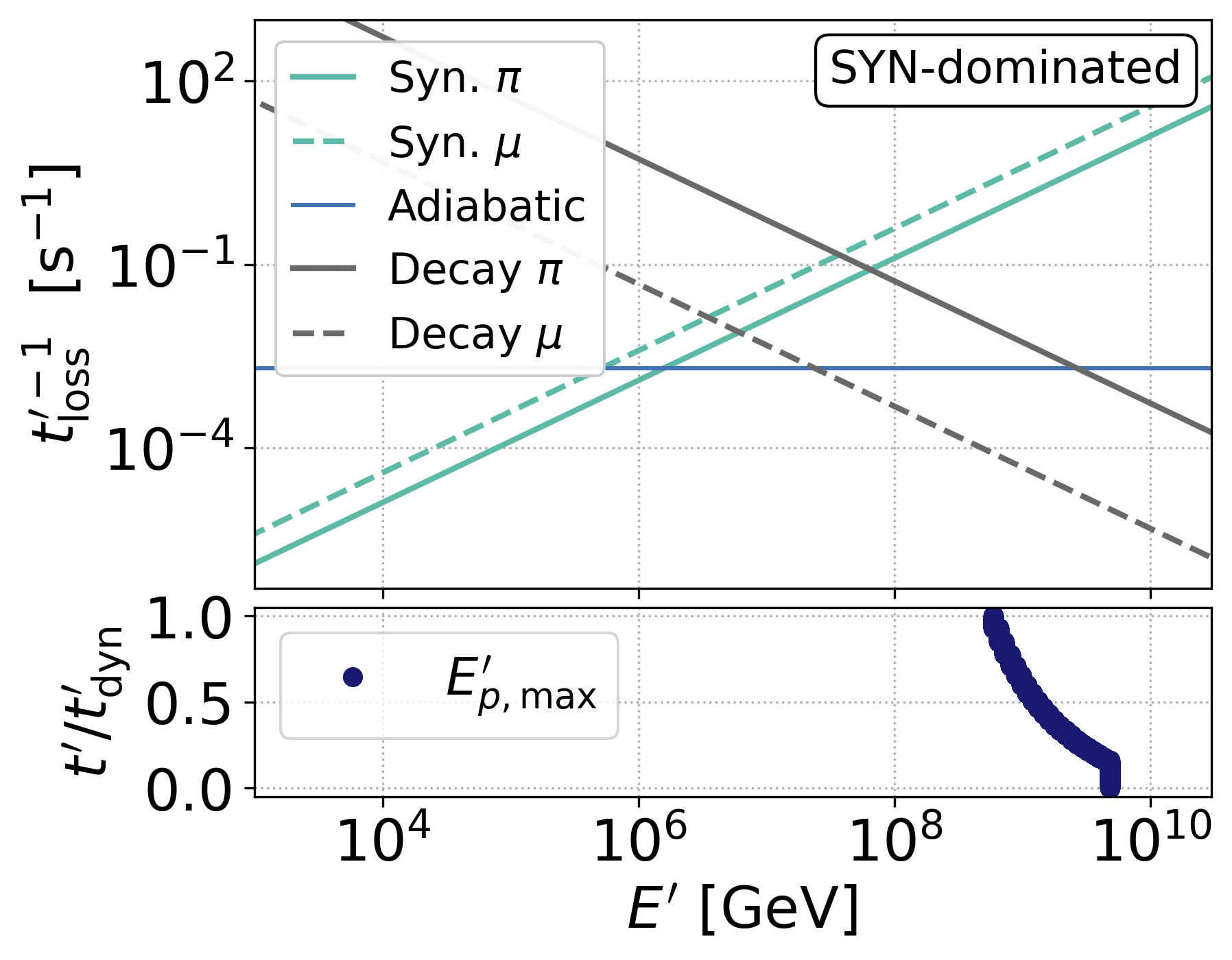}
    \hspace{3mm}
    \includegraphics[width = 0.43 \textwidth]{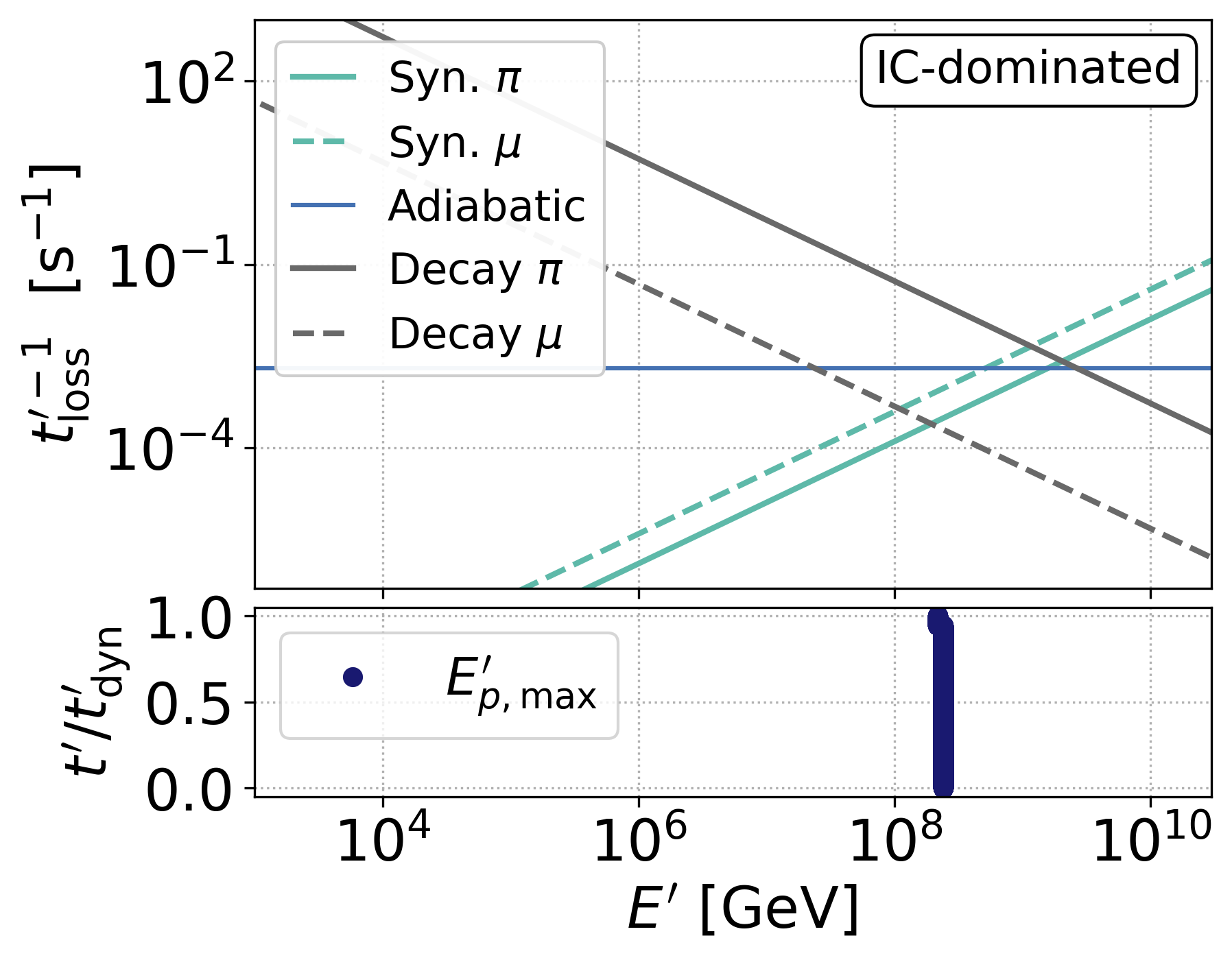}
    \caption{Pion and muon loss rates $t^{\prime -1}_\mrm{cool}$ for the representative collision, for \textit{(top)} the SYN-dominated scenario and \textit{(bottom)} the IC-dominated scenario; Complemented by the corresponding evolution of (comoving) maximal proton energies obtained for the radiative calculations until $t^\p_\mrm{dyn}$. } 
 \label{fig:pi_mu_coolingtimescales}
\end{figure}

We study the cooling of pions and muons first at the example of representative collision. While adiabatic cooling and decay rates are independent of the magnetic field, synchrotron cooling scales with $B^{\p 2}$. Similar to the results in the main text we again study the SYN- and IC-dominated scenarios. 

The cooling timescales for pions and muons are shown in the upper panel of \reffig{pi_mu_coolingtimescales}, calculated from \refeq{tprime_sy} - \refeq{tprime_decay}.
For the SYN-dominated scenario, adiabatic cooling is sub-dominant at all energies.
The longer decay time of muons shifts their critical energy $E_\mrm{c, sy/dec}$ to lower energies with respect to pions. 
For the IC-dominated scenario, synchrotron cooling is less efficient. For pions however either synchrotron cooling or decay domiantes while for muons adiabatic cooling dominates at intermediate energies. 

The figure further displays the maximum proton energies as a function of $t^\prime / t^\prime_\mrm{dyn}$ (recall that the injection time was set to $t^\prime_\mrm{inj} = t^\prime_\mrm{dyn}$. For the SYN-dominated scenario the proton energies are initially limited by adiabatic cooling and by photo-pion production at later $t^\p$. 
This implies that the maximum proton energies consecutively move to lower values with evolving $t^\prime$ as photo-pion production gets more efficient with the building up of the photon fields. 
For the IC-dominated case, the maximum proton energy is dominated by adiabatic cooling until almost $t^\p_\mrm{dyn}$ and remains almost constant.
We compare the maximum proton energies to the pion and muon cooling rates: For the SYN-dominated scenario, secondaries are produced at high enough energies to be able to cool prior to their decay. In contrast to this, the pions produced in the IC-dominated scenario decay without prior cooling. The secondary muons cool adiabatically before decaying.

We label the energies at which the timescales of two different processes are equal as \textit{critical} energies.  These mark the intersections of timescale/ cooling rates in the respective plots. Three processes imply three critical energies: $E_\mrm{c, ad/dec}$ (where $t^\p_\mrm{ad} = t^\p_\mrm{dec}$), $E_\mrm{c, sy/dec}$ (where $t^\p_\mrm{sy} = t^\p_\mrm{dec}$) and $E_\mrm{c, ad/sy}$ (where $t^\p_\mrm{ad} = t^\p_\mrm{sy}$).

As the decay rate decreases with particle energy, it can be assumed to be the dominating loss process at the lowest energies. Depending on the relative importance of the three processes, two scenarios are possible and can be identified in \reffig{pi_mu_coolingtimescales}:

\begin{enumerate}
    \item[(1)]  $E_\mrm{c, ad/dec} < E_\mrm{c, ad/sy} < E_\mrm{c, sy/dec}$ : At the highest energies synchrotron cooling dominates, at intermediate ones adiabatic cooling and at the lowest decay. It is realised for muons in the IC-dominated (right) plot of \reffig{pi_mu_coolingtimescales}.
    \item[(2)]  $E_\mrm{c, ad/sy} < E_\mrm{c, sy/dec} < E_\mrm{c, ad/dec}$: Adiabatic cooling is subdominant for all energies. At high energies synchrotron losses dominate, at low ones decay. It is nicely illustrated for muons and pions in the SYN-dominated (left) plot of \reffig{pi_mu_coolingtimescales}.
\end{enumerate}

\subsection{Cooling regimes throughout the fireball dynamical evolution}

\begin{figure}
\centering
    \includegraphics[width = 0.43 \textwidth]{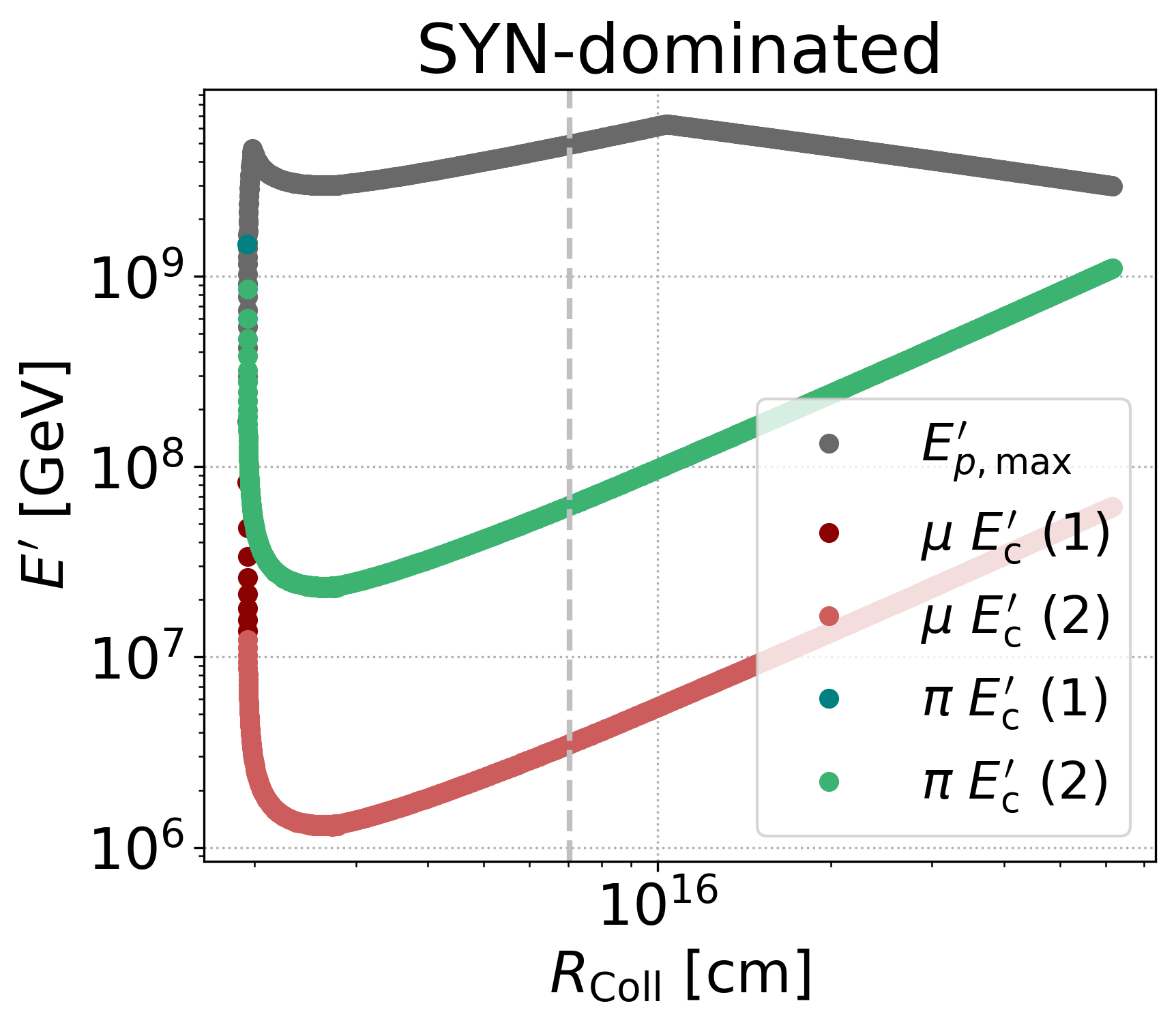}
    \hspace{3mm}
    \includegraphics[width = 0.43
    \textwidth]{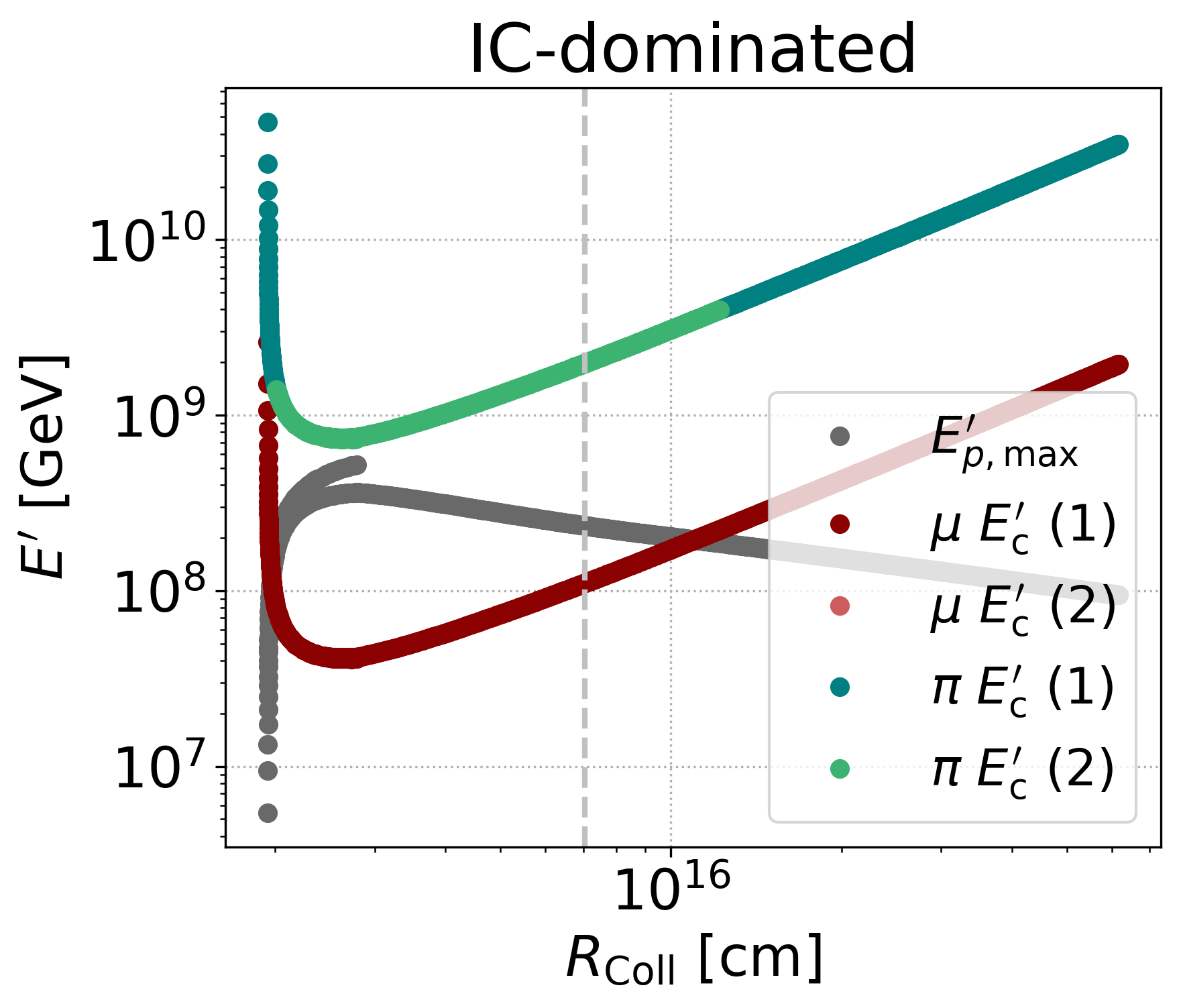}
 \caption{Critical energies $E^\p_\mrm{c}$ (see \refeq{critical_energy_pimu}) for pions and muons. The number indicates whether above the critical energy \textit{(1)} adiabatic or \textit{(2)} synchrotron cooling dominates. 
 For reference we further indicate the maximal proton energies obtained by balancing acceleration with adiabatic and synchrotron losses (in contrast to \reffig{pi_mu_coolingtimescales} where also photo-pion and photo-pair production were taken into account). 
 The dashed line indicates the radius of the representative collision. }
 \label{fig:pi_mu_cooling_fireball}
\end{figure}

From the representative collision we move to the complete fireball dynamical evolution. We define the critical energy as
\begin{equation}
E^\p_\mrm{c} = \mrm{min} \{ E^\p_\mrm{c, sy/dec} , E^\p_\mrm{c, ad/dec} \} \, .
\label{eq:critical_energy_pimu}
\end{equation}
It may be re-phrased as energy below which particles do not cool prior to their decay. 
The evolution of critical energy as a function of collision radius for our educative example is shown in \reffig{pi_mu_cooling_fireball}. The color-coding further indicates if above the critical energy, adiabatic or synchrotron cooling dominate: Dark red/green means that adiabatic cooling dominates, for light red/green synchrotron cooling is dominant above the critical energy. Thus, light (dark) colors correspond to scenario (1) ((2)) introduced before. For comparison we further indicate the maximal proton energy calculated by balancing synchrotron and adiabatic losses with acceleration. We point out that this is rather an upper limit on $E^\prime_\mrm{prot, max}$, since photo-pair or photo-pion production may further reduce the maximal energy attainable. 

For the SYN-dominated scenario, synchrotron cooling dominates for both pions and muons throughout the full fireball evolution. Further, the critical energies lie several orders of magnitude below $E_\mrm{p, max}^\prime$. In this scenario we thus expect strong impact of pion and muon cooling.
For the IC-dominated case, pion critical energies are dominated partially by adiabatic, partially by synchrotron cooling. Muon critical energies are dominated by adiabatic cooling throughout the complete fireball evolution.
In this scenario, the maximum proton energies lie below the pion critical energies for all collisions. Pions will thus not cool before decaying. Muon critical energies are below $E_\mrm{p, max}^\prime$ for some part of the fireball evolution, the difference in energy is however one order of magnitude at most.

We conclude that synchrotron cooling of secondary pions and muons will reduce their energies prior to their decay for high magnetic fields (high $f_\mrm{B/e}$). As this introduces a spectral break, it re-shapes the secondaries such as lepton pairs and neutrinos.
For low magnetic fields (low $f_\mrm{B/e}$) on the other hand, the maximal proton energies are too low to enable pion cooling before their decay. Muons may cool adiabatically before their decay. Overall, the cooling effects of pions and muons will much less impact lepton and neutrino spectra.

\subsection{Impact on neutrino spectra}
\label{app:pion_muon_cooling_neutrinos_singlespec}
\begin{figure}
\centering
\includegraphics[width = 0.45 \textwidth]{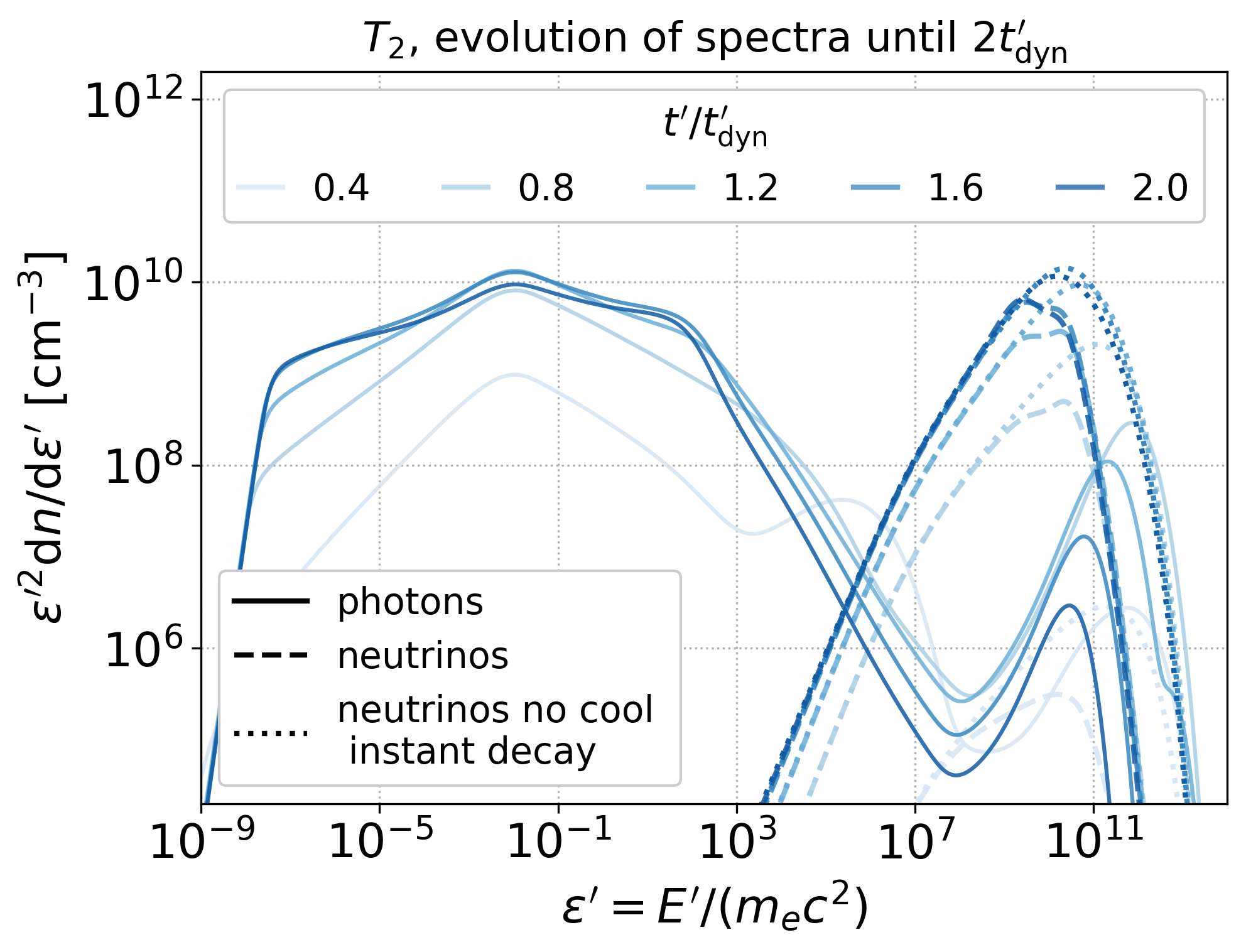}
 \caption{Evolution of comoving photon and neutrino spectra until $2 t^\p_\mrm{dyn}$ for the representative collision and the SYN-dominated scenario. }
 \label{fig:cooling_impact_on_neutrinos}
\end{figure}

Finally, we review the impact of muon and pion cooling on neutrino spectra. 
We found that for the SYN-dominated scenario the neutrino spectra peak at much lower energies compared to the photons produced in $\pi^0$ decays. For the IC-dominated scenario, neutrino and $\pi^0$-decay spectra extended to approximately the same energies.
In the last section we deduced that for the SYN-dominated scenario, intermediate pions and muons are subject to synchrotron cooling prior to their decay. To validate that this explains the difference between neutrino and photon spectra, we perform simulations with/without pion and muon cooling for the representative collision in the SYN-dominated scenario.

We show the evolution of comoving photon and neutrino spectra in the left plot of \reffig{cooling_impact_on_neutrinos}. Throughout the complete evolution, the neutrino spectra peak at lower energies than the HE photons from $\pi^0$ decays. The right plot shows the evolution of neutrino spectra \textit{(1)} with full treatment of pions/muon, \textit{(2)} without pion and muon cooling and \textit{(3)} assuming an infinitesimally small decay time of muons and pions (which implicitly means that pion and muon cooling are not accounted for).

For the first snapshot at $t^\prime = 0.2 t^\p_\mrm{dyn}$, the scenarios with/ without pion and muon cooling are little different and do not extend to the highest energies. 
If however, pions and muons are assumed to decay instantaneously (dotted lines in the right plot), energies comparable to the photons from $\pi^0$-decays are reached. 
This means that at early times $t^\prime$ the maximal energies of neutrinos are limited by the longer decay times of high-energy pions and muons (recall the dependency of the decay time on particle Lorentz factor, \refeq{tprime_decay}). As the system evolves, the decay time even of high-energy particles gets smaller than $t^\prime$ and the dash-dotted and dotted lines in \reffig{cooling_impact_on_neutrinos} extend to similarly high energies. 
We point out that the effect of the decay timescale is not as relevant if pion and muon cooling are accounted for: By cooling down, particles move to shorter decay times.

Overall, \reffig{cooling_impact_on_neutrinos} confirms that the lower peak energies of neutrino spectra for a SYN-dominated scenario can be attributed to cooling effects of intermediate pions and muons. 

\end{document}